\documentclass[twocolumn,english,showpacs,superscriptaddress,prl]{revtex4-1}
\usepackage[latin9]{inputenc}
\usepackage{hyperref}
\usepackage{color}
\usepackage{float}
\usepackage{textcomp}
\usepackage{amstext}
\usepackage{amsmath} 
\usepackage{graphicx}
\usepackage{gensymb}
\makeatletter

\usepackage{graphics}\usepackage{subfigure}\usepackage{longtable}
\usepackage{pstricks}
\usepackage{dcolumn}\usepackage{bm}
\usepackage{babel}
\makeatother
\newcommand{\eq}[1]{Eq.~(\ref{#1})}
\newcommand{\fig}[1]{Fig.~\ref{#1}}
\def\bea{\begin{eqnarray}}
\def\eea{\end{eqnarray}}

\def\vq{{\bf q}}

\def\vk{{\bf k}}

\begin{document}
\title{Gapped collective charge excitations and interlayer hopping in cuprate superconductors}

\author{M.~Hepting}
\email[]{hepting@fkf.mpg.de}
\affiliation{Max-Planck-Institute for Solid State Research, Heisenbergstra{\ss}e 1, 70569 Stuttgart, Germany}
\author{M.~Bejas}
\affiliation{Facultad de Ciencias Exactas, Ingenier\'{\i}a y Agrimensura and Instituto de F\'{\i}sica de Rosario (UNR-CONICET), Avenida Pellegrini 250, 2000 Rosario, Argentina}
\author{A.~Nag}
\affiliation{Diamond Light Source, Harwell Campus, Didcot OX11 0DE, United Kingdom}
\author{H.~Yamase}
\affiliation{International Center of Materials Nanoarchitectonics, National Institute for Materials Science, Tsukuba 305-0047, Japan}
\affiliation{Department of Condensed Matter Physics, Graduate School of Science, Hokkaido University, Sapporo 060-0810, Japan}
\author{N.~Coppola}
\affiliation{Dipartimento di Ingegneria Industriale, Universit\`{a} di Salerno, I-84084 Fisciano (SA), Italy}
\author{D.~Betto}
\affiliation{Max-Planck-Institute for Solid State Research, Heisenbergstra{\ss}e 1, 70569 Stuttgart, Germany}
\author{C.~Falter}
\affiliation{Institut f\"{u}r Festk\"{o}rpertheorie, Westf\"{a}lische Wilhelms-Universit\"{a}t, Wilhelm-Klemm-Str. 10, 48149 M\"{u}nster, Germany}
\author{M.~Garcia-Fernandez}
\affiliation{Diamond Light Source, Harwell Campus, Didcot OX11 0DE, United Kingdom}
\author{S.~Agrestini}
\affiliation{Diamond Light Source, Harwell Campus, Didcot OX11 0DE, United Kingdom}
\author{K.-J.~Zhou}
\affiliation{Diamond Light Source, Harwell Campus, Didcot OX11 0DE, United Kingdom}
\author{M.~Minola}
\affiliation{Max-Planck-Institute for Solid State Research, Heisenbergstra{\ss}e 1, 70569 Stuttgart, Germany}
\author{C.~Sacco}
\affiliation{Dipartimento di Ingegneria Industriale, Universit\`{a} di Salerno, I-84084 Fisciano (SA), Italy}
\author{L.~Maritato}
\affiliation{Dipartimento di Ingegneria Industriale, Universit\`{a} di Salerno, I-84084 Fisciano (SA), Italy}
\affiliation{CNR-SPIN Salerno, Universit\`{a} di Salerno, I-84084 Fisciano (SA), Italy}
\author{P.~Orgiani}
\affiliation{CNR-SPIN Salerno, Universit\`{a} di Salerno, I-84084 Fisciano (SA), Italy}
\affiliation{CNR-IOM, TASC Laboratory in Area Science Park, 34139 Trieste, Italy}
\author{H.~I.~Wei}
\affiliation{LASSP, Department of Physics, Cornell University, Ithaca, NY 14853, USA}
\author{K.~M.~Shen}
\affiliation{LASSP, Department of Physics, Cornell University, Ithaca, NY 14853, USA}
\author{D.~G.~Schlom}
\affiliation{Department of Materials Science and Engineering, Cornell University, Ithaca, NY 14853, USA}
\affiliation{Kavli Institute at Cornell for Nanoscale Science, Ithaca, NY 14853, USA}
\affiliation{Leibniz-Institut f\"{u}r Kristallz\"{u}chtung, Max-Born-Str. 2, 12489 Berlin, Germany}
\author{A.~Galdi}
\affiliation{Dipartimento di Ingegneria Industriale, Universit\`{a} di Salerno, I-84084 Fisciano (SA), Italy}
\affiliation{Cornell Laboratory for Accelerator Based Sciences and Education, Cornell University, Ithaca NY 14853, USA}
\author{A.~Greco}
\email[]{agreco@fceia.unr.edu.ar}
\affiliation{Facultad de Ciencias Exactas, Ingenier\'{\i}a y Agrimensura and Instituto de F\'{\i}sica de Rosario (UNR-CONICET), Avenida Pellegrini 250, 2000 Rosario, Argentina}
\author{B.~Keimer}
\affiliation{Max-Planck-Institute for Solid State Research, Heisenbergstra{\ss}e 1, 70569 Stuttgart, Germany}

\begin{abstract}
We use resonant inelastic x-ray scattering (RIXS) to probe the propagation of plasmons in the electron-doped cuprate superconductor Sr$_{0.9}$La$_{0.1}$CuO$_2$ (SLCO). We detect a plasmon gap of $\sim$~120 meV at the two-dimensional Brillouin zone center, indicating that low-energy plasmons in SLCO are not strictly acoustic. The plasmon dispersion, including the gap, is accurately captured by layered $t$-$J$-$V$ model calculations. A similar analysis performed on recent RIXS data from other cuprates suggests that the plasmon gap is generic and its size is related to the magnitude of the interlayer hopping $t_z$. Our work signifies the three-dimensionality of the charge dynamics in layered cuprates and provides a new method to determine $t_z$.

\end{abstract}

\maketitle

\begin{figure}
 \begin{centering}
\includegraphics[width=1.\columnwidth]{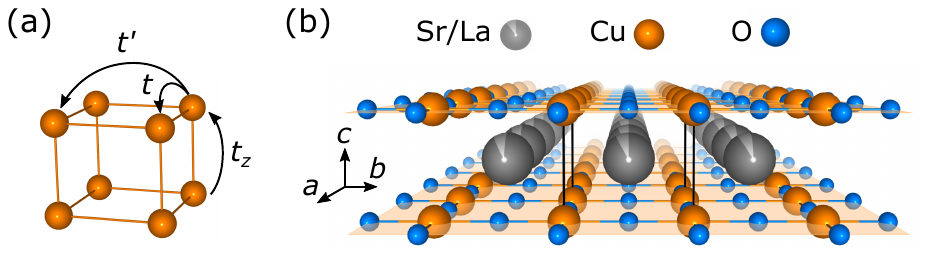}
\par\end{centering}
\caption{\textcolor{black}{(a) Schematic of the hopping integrals $t, t'$, and $t_z$ on a stacked square lattice. (b) Schematic of the crystal structure of Sr$_{0.9}$La$_{0.1}$CuO$_2$ (SLCO). Solid black lines correspond to the crystallographic unit cell. CuO$_2$ planes are indicated in orange.}}  
\label{experiment}
\end{figure}

A variety of enigmatic states emerge in layered cuprates upon hole- or electron-doping, such as the pseudogap, charge order, strange metal, and --- most prominently --- high-temperature superconductivity \cite{keimer15,armitage10}. While it is widely believed that antiferromagnetic spin fluctuations play a key role in the superconducting pairing \cite{lee06,scalapino12,dai01,letacon11,dean13,vilardi19}, it is not yet established whether the spin channel alone is responsible for the superconductivity in cuprates. For instance, the relevance of electron-phonon coupling is still under debate \cite{franck94,cuk04,heid09,giustino08,reznik08,maksimov10,johnston10}, 
and theoretical studies propose that low-energy plasmons \cite{ruvalds87,kresin88,cui91,ishii93,malozovsky93,varshney95,pashitskii08} or plasmon-phonon modes \cite{falter94,bauer09} mediate superconductivity in cuprates, or contribute constructively to the high superconducting transition temperature $T_c$ \cite{bill03}. Irrespective of the specific type of pairing glue, Anderson and coworkers suggested that $T_c$ is not a single-plane property \cite{anderson92} and interlayer Josephson tunneling of Cooper pairs strongly amplifies the $T_c$ of cuprates \cite{chakravarty93,anderson95}, which was discussed controversially in subsequent studies \cite{leggett96,tsvetkov98,moler98,anderson98,kirtley98,chakravary99,basov99}. The interlayer tunneling mechanism is most effective when coherent single particle hopping between adjacent CuO$_2$ planes is inhibited in the normal-state \cite{chakravarty93}. Nevertheless, coherent normal-state $c$-axis transport properties were detected in various experiments on overdoped cuprates \cite{uchida96,hussey03,homes05}, while in lightly doped cuprates it is challenging to assess whether interlayer hopping is small or absent \cite{uchida96,yamase21b}. In fact, the extraction of accurate values of the interlayer hopping integral $t_z$ from experimental data has proven difficult not only for lightly but also for overdoped cuprates \cite{takeuchi05,horio18,matt18,zha96,hussey03,grissonnanche21}. Hence, the $t_z$ determined from first-principle calculations \cite{andersen95,markiewicz05} is frequently employed, which was suggested to be as large as $t'$ or 0.1$t$ in some cuprates \cite{markiewicz05}, with $t'$ and $t$ denoting the in-plane next-nearest and nearest neighbor hopping, respectively [Fig.~\ref{experiment}(a)]. In contrast, other studies assume that $t_z$ is negligibly small compared to $t'$ and $t$, which is in line with a smaller interlayer hopping evaluated from experiments \cite{yamase21b,horio18}. Thus, new methods for a reliable determination of the contentious parameter $t_z$ are highly desirable. 

Notably, recent theoretical work has emphasized that the interlayer hopping is also encoded in the plasmon spectrum of cuprates and should manifest as a gap at the two-dimensional (2D) Brillouin zone (BZ) center \cite{greco16}. More specifically, in layered systems, such as the cuprates, the plasmon dispersion (for small $q$) neither follows a $\sqrt{q}$ dependence that is typical for 2D metals, nor the $q^2$ behavior of isotropic 3D metals \cite{grecu73,fetter74,grecu75}. Instead, poorly screened interlayer Coulomb interaction between the CuO$_2$ planes 
gives rise to a plasmon spectrum containing a set of acoustic branches, which disperse linearly for small $q$, and one optical branch \cite{kresin88}. In the presence of interlayer charge transfer, however, the former plasmon branches are not strictly acoustic, but gapped at the 2D BZ center.
Yet, while seemingly acoustic plasmons were identified in recent resonant inelastic x-ray scattering (RIXS) experiments on various electron- and hole-doped cuprates \cite{hepting18,jlin20,nag20,singh21} including ${\rm La_{1.825}Ce_{0.175}CuO_4}$ (LCCO) and ${\rm La_{1.84}Sr_{0.16}CuO_4}$ (LSCO), a gap has not been observed unambiguously.

In this work, we study plasmon excitations in the electron-doped cuprate Sr$_{0.9}$La$_{0.1}$CuO$_2$ (SLCO), which exhibits the infinite-layer crystal structure. Using RIXS, we detect an energy gap of the acoustic-like modes at the in-plane BZ center. 
The observed plasmon dispersion, including the gap, is accurately captured by $t$-$J$-$V$ model calculations, and characteristic microscopic parameters, such as $t_z$, are determined. The application of our analysis scheme to previously published RIXS data of other cuprates suggests considerably smaller plasmon gaps and interlayer hoppings in LCCO and LSCO.

The RIXS measurements were performed on a SLCO thin film with a superconducting transition temperature $T_c \sim 30 $ K and a thickness of 294 {\AA} grown by molecular-beam epitaxy (MBE) on a (110) oriented TbScO$_3$ substrate ~\cite{galdi18}. Figure~\ref{experiment}(b) shows the crystal structure of SLCO, which exhibits CuO$_2$ planes stacked along the $c$-axis direction with La/Sr spacer layers, which corresponds to the infinite-layer structure
\cite{fournier15,tomaschko12}. The lattice constants $a, b = 3.960$ {\AA} and $c = 3.405$ {\AA} were determined by x-ray diffraction (XRD). Note that for SLCO the interlayer distance is equivalent to the $c$-axis lattice constant, whereas in LCCO the interlayer distance corresponds to $c/2 \sim 6.05$ {\AA} \cite{hepting18}. 

All RIXS spectra were collected at the Cu $L_3$-edge with high energy resolution ($\Delta$E $\approx$ 40 meV) at $T = 20$ K at the I21-RIXS beamline of the Diamond Light Source, UK \cite{zhou22}. The momentum-resolution was $\Delta q$ $\approx$ 0.01 {\AA}$^{-1}$ \cite{zhou22}. A similar scattering geometry as in Ref.~\onlinecite{nag20} was employed, with the $a/b$-axis and the $c$-axis of SLCO lying in the scattering plane and incident photons linearly polarized perpendicular to the scattering plane ($\sigma$-polarization). Importantly, the continuous rotation of the RIXS spectrometer arm allowed for a variation of the scattering angle, and the corresponding rotation of the sample enabled the independent variation of the in-plane ($q_{\parallel}$) and out-of-plane momentum transfer ($q_{z}$). In the following we denote the momentum transfer by $(H,K,L)$ in reciprocal lattice units $(2\pi/a, 2\pi/b, 2\pi/c)$. 

\begin{figure*}
 \begin{centering}
\includegraphics[width=2\columnwidth]{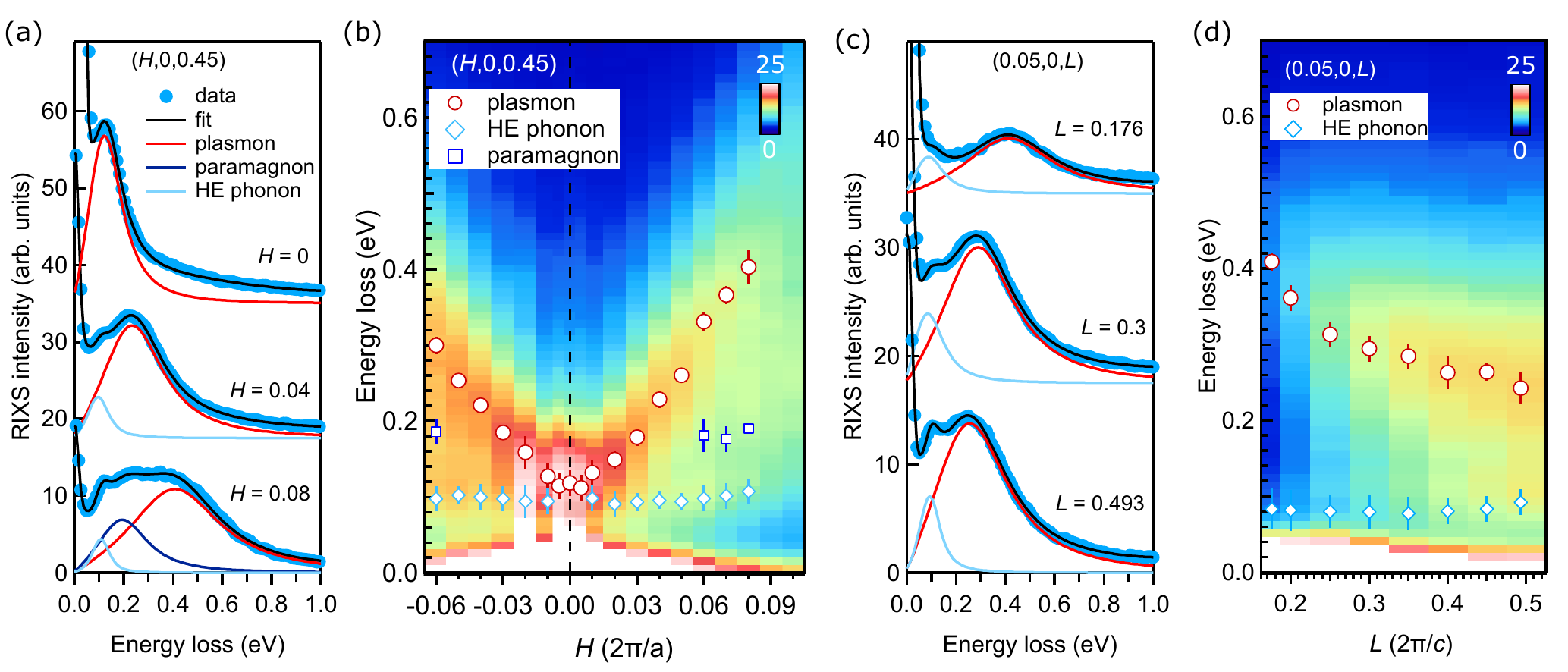}
\par\end{centering}
\caption{\textcolor{black}{(a) Vertically stacked RIXS spectra for representative momenta along the $(H,0,0.45)$ direction. Solid black lines are fits to the spectra. Fitted peak profiles of the plasmon, paramagnon, and high-energy (HE) phonon are shown, while the other contributions to the fit are omitted for clarity (for details of the fitting procedure see Supplemental Material \cite{suppmat}). 
(b) RIXS intensity map for momenta along the $(H,0,0.45)$ direction. Open symbols are peak positions extracted from fits. The color scale of the map is capped at 25 arb. units (white color). (c) Vertically stacked RIXS spectra for representative momenta along the $(0.05,0,L)$ direction. (d) RIXS intensity map for momenta along the $(0.05,0,L)$ direction.}}
\label{RIXS}
\end{figure*}

Figure~\ref{RIXS}(a) shows a representative set of RIXS spectra for different in-plane momenta $H$. The spectra were fitted by the sum of the elastic line at zero-energy loss and several damped harmonic oscillator functions accounting for inelastic features. Details about the fitting procedure, assignment of the features, and the complete set of spectra are given in the Supplemental Material \cite{suppmat}. As the most relevant features, we identify (i) an essentially non-dispersive peak around 95 meV, (ii) a paramagnon peak around 190 meV for $\vert H \vert \geq 0.06$, and (iii) a fast dispersive peak with an energy higher than 120 meV [Figs.~\ref{RIXS}(a),(b)]. We attribute the 95 meV feature to a high-energy (HE) phonon. While phonons in cuprates are typically observed below 85 meV \cite{gunnarsson08,koval96,mun01}, zone-boundary phonons with energies as high as 91.4 meV were predicted for the infinite-layer cuprate SrCuO$_2$ \cite{klenner94}. The further increase of phonon energy in our SLCO film could be due to the epitaxial strain induced by the substrate or the La-doping.
The assignment of the paramagnon peak (see Supplemental Material \cite{suppmat}) is in line with Ref.~\onlinecite{dellea17}, which investigated the paramagnon dispersion in a similar SLCO film, but focused on large in-plane momenta and employed a lower energy resolution of $\Delta$E = 265 meV. In contrast to the paramagnon, the fast dispersing peak exhibits substantial spectral intensity at the zone center $H = 0$ [Figs.~\ref{RIXS}(a),(b)]. Close inspection of the spectra with small $H$ reveals that the peak likely also contains contributions from the HE phonon, which precludes unambiguous fitting for $\vert H \vert \leq 0.005$. 
Nevertheless, we emphasize that the center of gravity of the superposed peak (labelled as plasmon in Figs.~\ref{RIXS}(a),(b)) exceeds 120 meV even at $H = 0$, indicating the presence of an energy gap for this mode. Notably, a charge origin of the fast dispersing mode was already proposed in the low-resolution measurements in Ref.~\onlinecite{dellea17}, while a gap around $H = 0$ was not resolved and the investigation of a possible dispersion along $L$ was lacking. As shown in Figs~\ref{RIXS}(c),(d), our high-resolution measurement reveals that the peak exhibits a distinctive dispersion as a function of $L$. Along the lines of Refs.~\onlinecite{hepting18,nag20}, such $L$-dependence with a minimum around $L = 0.5$ identifies the mode as an acoustic plasmon excitation. On the other hand, the $H$-dependence in Fig.~\ref{RIXS}(b) shows that the mode exhibits a gap of $\sim$ 120 meV at $H = 0$. Thus, the seemingly acoustic plasmons in SLCO are not strictly acoustic, but gapped. 
This observation calls for a thorough theoretical investigation of the emergence and the energy scale of the gap, which we present next. 

\begin{figure*}
 \begin{centering}
\includegraphics[width=2\columnwidth]{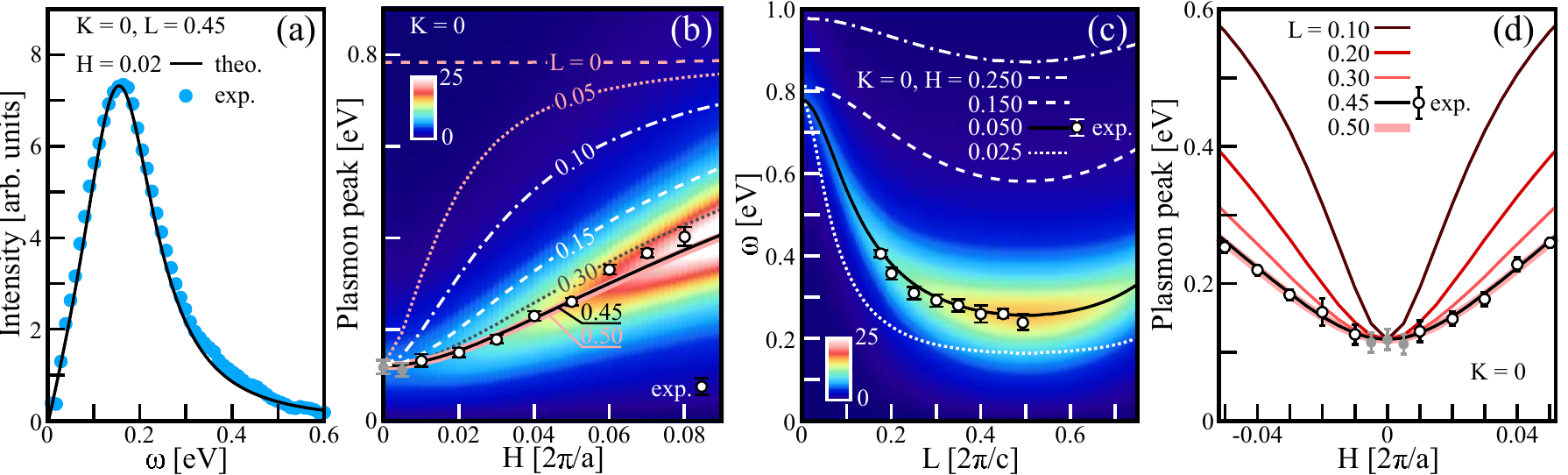}
\par\end{centering}
\caption{\textcolor{black}{
(a) Imaginary part of the charge susceptibility $\chi_c''({\bf q},\omega)$ for momentum $(0.02, 0, 0.45)$ (solid black line) computed in the layered $t$-$J$-$V$ model. Superimposed are experimental data (blue symbols), which correspond to the plasmon component in the RIXS raw data. 
The intensity of $\chi_c''({\bf q},\omega)$ is scaled such that it fits to the maximum of the RIXS data.
(b) Computed intensity map of $\chi_c''({\bf q},\omega)$ for momenta along the $(H,0,0.45)$ direction. The solid black line corresponds to the maxima of $\chi_c''({\bf q},\omega)$. The other lines indicate the maxima of $\chi_c''({\bf q},\omega)$ computed for different $L$. Experimental plasmon peak positions for momenta along the $(H,0,0.45)$ direction are superimposed as white and gray symbols. The former symbols correspond to peak positions used as an input for the fitting procedure for the $t$-$J$-$V$ model, while the latter were not included (see text). (c) Computed intensity map and maxima for different $H$ along the $(0.05,0,L)$ direction. (d) Computed plasmon dispersion around the 2D BZ center at $H = 0$ for different $L$. 
}}

\label{spectrum}
\end{figure*}

Previously, typical properties of the plasmons in cuprates were studied by random phase approximation (RPA) calculations \cite{kresin88,grecu73,bill03,markiewicz08,singh21}, a combination of determinant quantum Monte Carlo (DQMC) and RPA in a layered Hubbard model \cite{hepting18}, an extended variational wave function (VWF) approach VWF+1/$N_f$ \cite{fidrysiak21}, and a large-$N$ theory of the layered $t$-$J$-$V$ model \cite{greco16,greco19,greco20}. In the following, we turn to the latter theory, which emphasized in Ref.~\onlinecite{greco16} that acoustic plasmons in cuprates are not strictly acoustic, but exhibit an energy gap at the 2D BZ center. The $t$-$J$ model is widely employed as an effective model for cuprates \cite{fczhang88} and accounts for strong correlations. Importantly, the $t$-$J$-$V$ model includes not only first-nearest-neighbor ($t$), second-nearest-neighbor $(t')$, and interlayer ($t_z$) hoppings [Fig.~\ref{experiment}(a)], but also the long-range Coulomb interaction $V(\textbf{q})$ (see Supplemental Material \cite{suppmat}), which is crucial given the three-dimensional character of plasmons in layered cuprates \cite{hepting18}. Figure~\ref{spectrum}(a) shows the imaginary part of the charge susceptibility $\chi_c''({\bf q},\omega)$ computed in the framework of a  large-$N$ theory of the $t$-$J$-$V$ model for doping $\delta=0.1$ and a broadening parameter $\Gamma/t=0.1$,  which is required to account for the experimental resolution and a possible broadening due to correlations \cite{prelovsek99}. All other fit components (see Supplemental Material \cite{suppmat}) were subtracted from the RIXS spectrum in
Fig.~\ref{spectrum}(a) to make a direct comparison with $\chi_c''({\bf q},\omega)$. To capture the full plasmon dispersion in SLCO, we have applied an error minimization fitting procedure for the $t$-$J$-$V$ model (see Supplemental Material \cite{suppmat}), using the experimentally determined plasmon peak positions [Figs.~\ref{RIXS}(b),(d)] as an input. Figures~\ref{spectrum}(b),(c) show the computed plasmon branches as a function of momenta $H$ and $L$, respectively, together with corresponding experimental data. Note that the experimentally determined peak positions for $\vert H \vert \leq 0.005$ (gray symbols in Figs.~\ref{spectrum}(b),(d)) with a strong overlap with the HE phonon were excluded from the fitting procedure for the $t$-$J$-$V$ model, to avoid any bias in the determination of a gap at the 2D BZ center. Besides the measured plasmon branches along the $(H,0,0.45)$ and $(0.05,0,L)$ directions, additional calculated branches for unmeasured $H$ and $L$ values are shown in Figs.~\ref{spectrum}(b) and (c), respectively.  
The obtained spectrum of plasmon modes, including the optical branch for $L=0$, is qualitatively reminiscent of previous calculations for cuprates \cite{kresin88,bill03,markiewicz08,hepting18}, but additionally features a distinct energy gap at $H,K = 0$ [Fig.~\ref{spectrum}(b)], which is along the lines of calculations in Refs.~\onlinecite{greco16,nag20,greco19,greco20,fidrysiak21}.

Figure~\ref{spectrum}(d) focuses on small in-plane momenta around the observed plasmon gap at the 2D BZ center and illustrates the excellent agreement between experiment and theory. As the key result of our study, the fitting with the $t$-$J$-$V$ model yields $t_z/t=0.055$ (corresponding to $t_z=55$ meV, see Supplemental Material) for the interlayer hopping. This is in line with $t_z/t=0.06$, determined by first-principles calculations for CaCuO$_2$ \cite{botana20}, which is a closely related infinite-layer cuprate. Moreover, we obtain the in-plane and out-of-plane dielectric constants $\epsilon_\parallel/\epsilon_0 =5.89$ and $\epsilon_\perp/\epsilon_0=1.06$ from the $t$-$J$-$V$ model fitting, which are similar to theoretical predictions for infinite-layer cuprates \cite{klenner94}. We emphasize that a gap with a magnitude of $\sim 120$ meV is a robust feature, which is rooted in the presence of a finite $t_z$ \cite{greco16} and cannot be attributed exclusively to other effects, such as the broadening $\Gamma$. In fact, in absence of interlayer hopping, $\Gamma$ can induce only a relatively small gap in SLCO (see Supplemental Material \cite{suppmat}). 

As a next step, we apply the present fitting procedure for the $t$-$J$-$V$ model to other systems and revisit previous RIXS data on LCCO and LSCO reported in Refs.~\onlinecite{hepting18,nag20}, respectively. Note that in the case of LCCO the gap was estimated to be approximately zero \cite{hepting18}, while in LSCO the previous analysis with the $t$-$J$-$V$ model indicated 75 meV as an upper limit \cite{nag20}. The present analysis (see Supplemental Material \cite{suppmat}) indicates an upper limit of 82 meV for the gap in LCCO and 55 meV for LSCO, which are both substantially smaller than the gap size in SLCO. While the strength of the Coulomb interaction is comparable in the three cuprates, a particularly strong interlayer hopping can be expected in the latter compound due to its distinct infinite-layer crystal structure with narrowly spaced adjacent CuO$_2$ planes [Fig.~\ref{experiment}(b)], thus rationalizing the large gap value of more than one hundred meV, which in turn enabled our first conclusive observation of a plasmon gap with RIXS. 

The corresponding hoppings $t_z/t$ for the upper limits of the plasmon gaps in LCCO and LSCO are 0.03 ($t_z=30$ meV) and 0.01 ($t_z=7$ meV), respectively. 
Employing a larger $t_z/t$ of 0.1 for LSCO \cite{markiewicz05} would yield a large gap of 344 meV with the present methodology. 
This suggests that the interlayer hopping in LSCO is indeed very small and motivates future RIXS studies with higher resolution to determine the value of the gap in LSCO experimentally. Furthermore, we estimate the lower bound of the plasmon gap by assuming a hypothetical $t_z = 0$, leading to 58 meV and 41 meV for LCCO and LSCO, respectively. These lower bounds arise purely from the broadening $\Gamma$ (see Supplemental Material \cite{suppmat}). 

Having established the plasmon gap at the 2D BZ center in different cuprates, we next focus on the plasmon properties for nonzero in-plane momentum transfer. A close inspection of the $H$-dependence of the dispersion in the vicinity of $H = 0$ in LCCO and LSCO \cite{hepting18,nag20} reveals that the intensity of the plasmon peak decreases when approaching the 2D BZ center. This behavior is also expected from the $t$-$J$-$V$ model calculations [Fig.~\ref{spectrum}(b)]. For SLCO, a similar trend might not be obvious in the RIXS intensity map in Fig.~\ref{RIXS}(b), due to the overlap with the paramagnon, phonons, and the elastic line. Nevertheless, a plot of the fitted integrated intensity of the plasmon peak as a function of $H$ reveals that also SLCO exhibits a comparable trend (see Supplemental Material \cite{suppmat}) --- except for momenta $\vert H \vert \leq 0.005$ where a sharp increase of the intensity occurs. This increase is likely a result of the superposition of the plasmon and the HE phonon peak, but cannot be disentangled unambiguously in the present RIXS data owing to an insufficient energy resolution. Future higher-resolution RIXS experiments might be capable to resolve the different spectral components around $H = 0$ in SLCO, and are also desirable for LSCO, where a coupling between the plasmon and a $c$-axis polarized phonon mode of apical oxygen ions was predicted for small momenta, with an expected gap of the mixed plasmon-phonon mode of $\sim 60$ meV \cite{bauer09}. Note that in SLCO, however, plasmon-phonon coupling can be ruled out (see Supplemental Material). 

In summary, our observation and theoretical description of the plasmon gap provide the missing piece of the puzzle alongside the $q_z$-dependence \cite{hepting18,nag20,singh21} to consolidate the 3D character of plasmons in cuprates. This gap was neither evidenced in previous optical spectroscopy \cite{bozovic90} nor electron-energy loss spectroscopy (EELS) studies \cite{fink01}, yet its unambiguous presence underscores the importance of explicit inclusion of the interlayer hopping $t_z$ for a comprehensive description of the charge dynamics of cuprates. On a fundamental level, the charge degrees of freedom and the dynamics of the normal state are considered as a prerequisite for understanding the superconducting state \cite{basov99}. Hence, in a broader context, the presence of a substantial plasmon gap at the 2D BZ center calls for a reassessment of the theories proposing that acoustic plasmons mediate superconductivity \cite{ruvalds87,kresin88,cui91,ishii93,malozovsky93,varshney95,pashitskii08} and raise the $T_c$ of cuprates as much as 20\% \cite{bill03}. In particular, it should be evaluated whether the gap energy has a positive or negative effect in the suggested pairing scenarios.   
Along the lines of previous discussions about a correlation between $T_c$ and the magnitude of $t'$ \cite{pavarini01}, future applications of our methodology may provide new insights into a possible relation between $T_c$ and $t_z$, as well as the putative scaling of $T_c$ with the number of CuO$_2$ planes per unit cell \cite{iyo07,vincini19}. In this context, we note that the large-$N$ theory for the $t$-$J$-$V$ model indicates that the doping-dependence of the plasmon gap exhibits a dome-like shape \cite{greco16}, similar to the $T_c$ dome of cuprates \cite{keimer15}.  Nevertheless, detailed calculations are required to assess the impact of a gap on the value of $T_c$, considering not only the electron self-energy, but also vertex corrections, as the relatively large energy scale of the plasmon may invalidate Migdal's theorem \cite{mahan}.

More specifically, our approach combining RIXS and $t$-$J$-$V$ model calculations enables the extraction of robust values of $t_z$ in cuprates --- possibly even in lightly doped cuprates which has not been accomplished with other experimental methods \cite{takeuchi05,horio18,matt18,zha96,yamase21b}.
Moreover, we anticipate that our methodology will be applicable to other materials, including layered 2D materials and van der Waals heterostructures \cite{cudazzo12,groenewald16,thygesen17}, as well as the newly discovered infinite-layer nickelate superconductors \cite{Li19,hepting20,zeng22}, which might possess even larger interlayer hoppings than SLCO \cite{botana20}.

 We thank A.~V. Boris and P.~Horsch for fruitful discussions, and W. Metzner for a critical reading of the manuscript. A.~Greco acknowledges the Max-Planck-Institute for Solid State Research in Stuttgart for hospitality and financial support. H.~Y. was supported by JSPS KAKENHI Grant No.~JP20H01856, Japan. K.M.S. and H.I.W. acknowledge support from the Air Force Office of Scientific Research through Grant No. FA9550-21-1-0168. We acknowledge Diamond Light Source for providing the beamtime under the proposal MM23933.

\clearpage

\renewcommand{\eq}[1]{Eq.~(\ref{#1})}
\renewcommand{\fig}[1]{Fig.~\ref{#1}}
\setcounter{figure}{0}  
\renewcommand{\thefigure}{S\arabic{figure}}

\section*{Supplemental Material for ``Gapped collective charge excitations and interlayer hopping in cuprate superconductors"}


\section*{RIXS raw data and fits}
\label{app:RIXS}

\begin{figure}
 \begin{centering}
\includegraphics[width=.9\columnwidth]{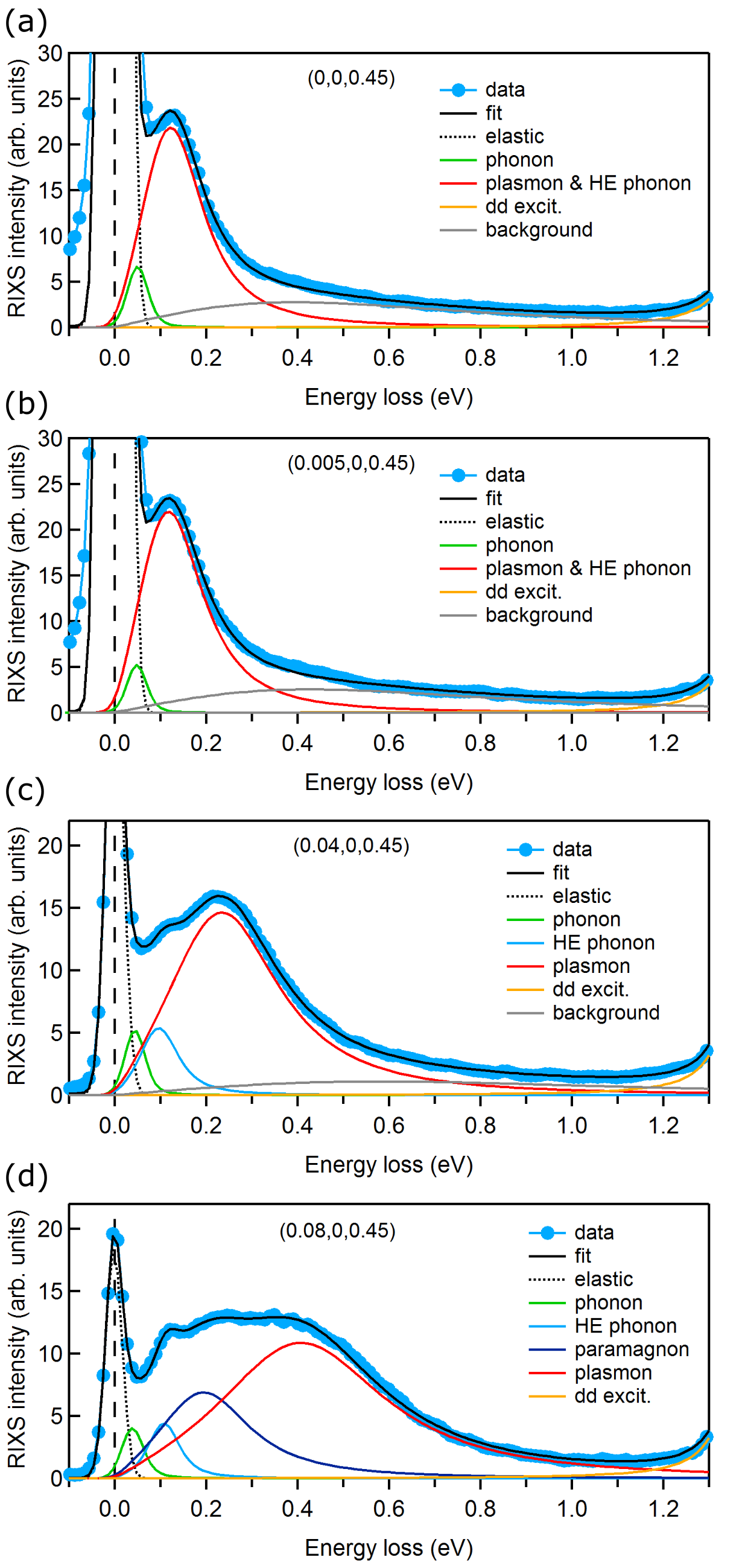}
\par\end{centering}
\caption{(a)-(d) Fits of the RIXS spectra at representative momenta (0, 0, 0.45), (0.005, 0, 0.45), (0.04, 0, 0.45) and (0.08, 0, 0.45), respectively. The model uses a Gaussian for the elastic peak (dashed black line) and anti-symmetrized Lorentzians for all other contributions in the spectrum, convoluted with the energy resolution of 40 meV via Gaussian convolution. The individual contributions are described in the text.}
\label{fig:fits}
\end{figure}

\begin{figure}
 \begin{centering}
\includegraphics[width=1.\columnwidth]{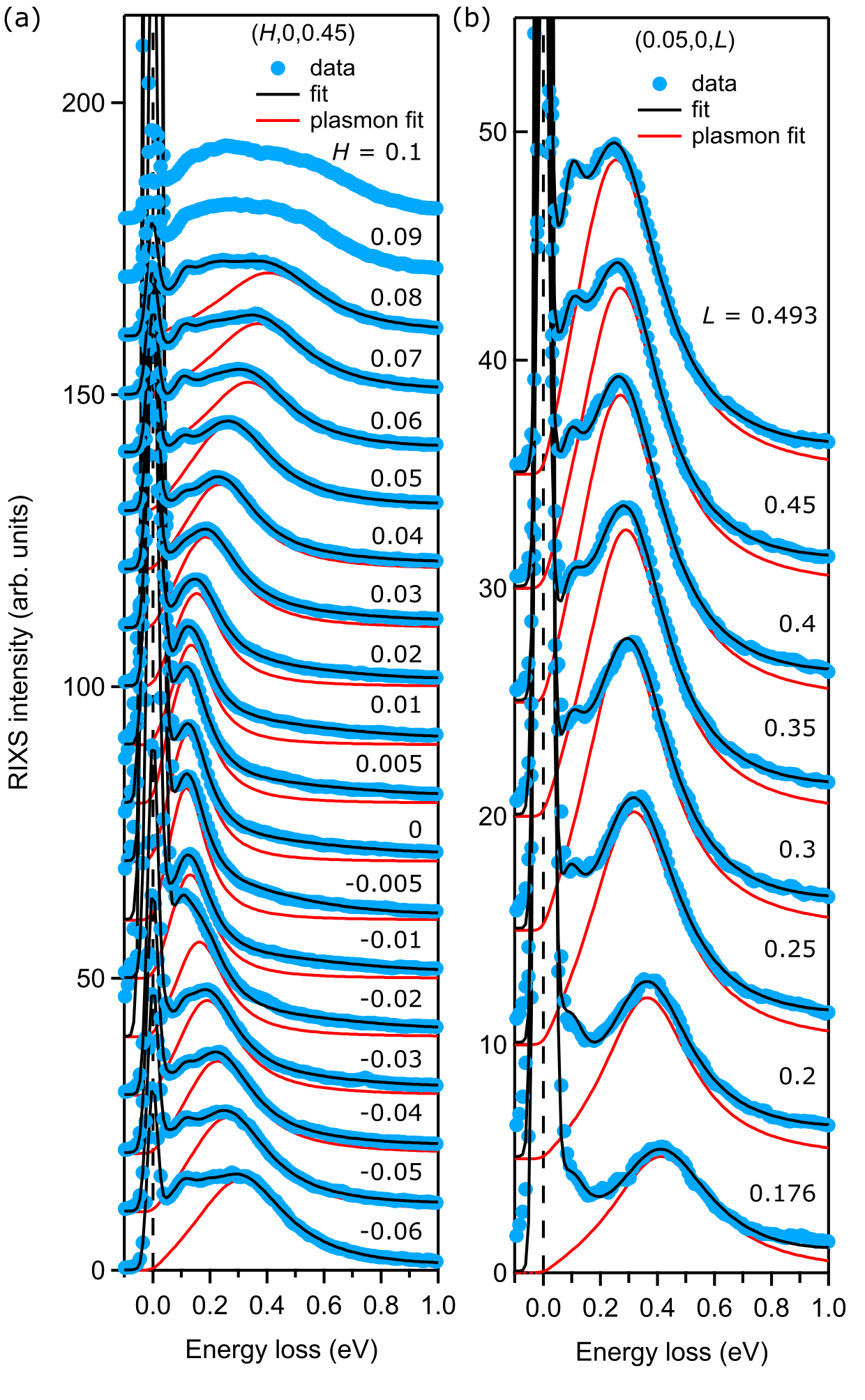}
\par\end{centering}
\caption{Raw RIXS spectra (blue symbols) of SLCO together with fits (solid
lines) for momentum transfer along the $H$ direction (a) and $L$ direction (b). Red solid lines are the anti-symmetrized Lorentzian fit profiles of the plasmon peak. Black solid lines are the sum of all contributions fitted to the spectra. Spectra are offset in vertical direction for clarity. Note that the overlap and broadness of the features in the $H = 0.1$ and 0.09 spectra in (a) precluded a reliable fit, hence no fitted curve is shown for these momenta.}
\label{fig:raw}
\end{figure}

Figures~\ref{fig:fits}(a)-(d) show all components fitted to the RIXS spectra for the representative momenta (0, 0, 0.45), (0.005, 0, 0.45), (0.04, 0, 0.45) and (0.08, 0, 0.45), respectively. The elastic peak was modeled by a Gaussian and the other contributions in the spectra by anti-symmetrized Lorentzians \cite{hepting18a,nag20a}, convoluted with the energy resolution of 40 meV via Gaussian convolution. The anti-symmetrized Lorentzian profiles ensure zero intensity at zero energy loss (prior to convolution) for the inelastic features. In detail, the inelastic features are assigned to a phonon, a high-energy (HE) phonon, a paramagnon, a plasmon, and $dd$-excitations. For the $dd$-excitations, only the low-energy tail below 1.3 eV energy loss was fitted. For spectra with $\vert H \vert \leq 0.05$, an additional phenomenological background contribution was fitted. This broad and almost featureless background is reminiscent of the background observed in the low-energy region of Cu $L_3$-edge RIXS spectra of other cuprates \cite{hepting18a,fumagalli19} and possibly originates from charge excitations with (partly) incoherent character and/or a bimagnon continuum \cite{minola15}. Figures~\ref{fig:raw}(a),(b) show the full set of RIXS spectra measured along the $H$ and $L$ direction, respectively, with the plasmon peak profiles and the sum of all fitted contributions indicated.

\begin{figure}
 \begin{centering}
\includegraphics[width=.9\columnwidth]{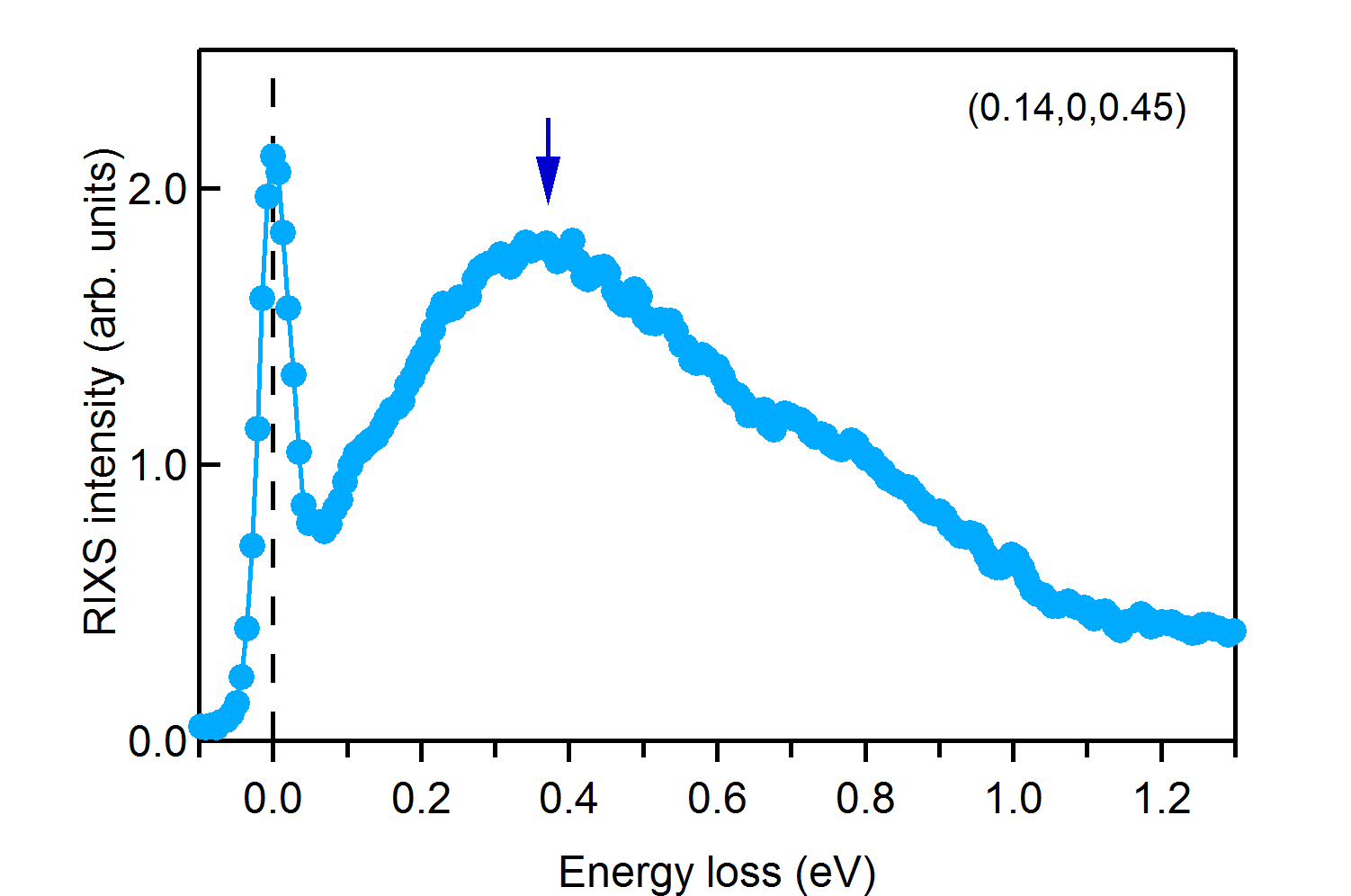}
\par\end{centering}
\caption{RIXS spectrum at momentum (0.14, 0, 0.45), i.e., at relatively large in-plane momentum. The paramagnon peak (blue arrow) can be clearly distinguished and its line shape and energy are consistent with the RIXS data in Ref.~\cite{dellea17a} measured on a similar SLCO film. The high energy shoulder contains contributions from the plasmon excitation, which cannot be resolved unambiguously at this momentum.}
\label{fig:paramagn}
\end{figure}

The peak observed around 190 meV for $\vert H \vert \geq 0.06$ in
Fig.2(a),(b) in the main text
is assigned to a paramagnon excitation. The presence of a paramagnon in SLCO is in line with Ref.~\cite{dellea17a}, which investigated the paramagnon dispersion in a similar SLCO film, but focused on large in-plane momenta and was hampered by a lower energy resolution. To verify our assignment from
Fig.2(a), (b) in the main text
, we recorded complementary RIXS spectra up to relatively large $H$ values that were also measured in Ref.~\cite{dellea17a}. We find that the center of gravity of the peak in our spectrum disperses up to $\sim$ 400 meV for $H = 0.14$ [Fig.~\ref{fig:paramagn}], which is in excellent agreement with Ref.~\cite{dellea17a} and corroborates our assignment to a paramagnon. 
We note that for momenta smaller than $\vert H \vert = 0.06$, the paramagnon peak cannot be discerned from other spectral features, owing to its decreasing intensity. This is consistent with the notion that the intensity of the paramagnon becomes smallest around the in-plane zone-center \cite{dellea17a}.

\section*{Theoretical scheme}
\label{app:theory}

The layered $t$-$J$-$V$ model is given by the following Hamiltonian 
\begin{align}
H =& -\sum_{i, j,\sigma} t_{i j}\tilde{c}^\dag_{i\sigma}\tilde{c}_{j\sigma} + 
\sum_{\langle i,j \rangle} J_{ij} \left( \vec{S}_i \cdot \vec{S}_j - \frac{1}{4} n_i n_j \right) \nonumber \\
&+\frac{1}{2} \sum_{i,j} V_{ij} n_i n_j \,,
\label{tJV}  
\end{align}
where the sites $i$ and $j$ run over a three-dimensional lattice. 
The hopping $t_{i j}$ takes the value $t$ $(t')$ between the first (second) nearest-neighbor sites on a square lattice. The hopping integral between the layers is scaled by $t_z$. 
$\tilde{c}^\dag_{i\sigma}$ ($\tilde{c}_{i\sigma}$) is 
the creation (annihilation) operator of electrons with spin $\sigma$ $(\sigma=\uparrow, \downarrow)$  
in the Fock space without any double occupancy.  $\langle i,j \rangle$ denotes a nearest-neighbor pair of sites. 
The exchange interaction $J_{i j} \equiv J$ is considered only within the plane, as   
the exchange term between the planes ($J_\perp$) is much smaller \cite{thio88}. In the following we use $J/t=0.3$. 
$n_i=\sum_{\sigma} \tilde{c}^\dag_{i\sigma}\tilde{c}_{i\sigma}$ 
is the electron density operator and $\vec{S}_i$ is the spin operator. 
$V_{ij}$ is the long-range Coulomb interaction in the three-dimensional lattice. 

To deal with the local constraint that the double occupancy of electrons is forbidden at any lattice site in the $t$-$J$-$V$ model, we use a path integral representation for Hubbard operators and developed a non-perturbative technique based on a large-$N$ expansion, where $N$ is the number of electronic degrees of freedom per site and $1/N$ is assumed to be a small parameter \cite{foussats04}. 

In the framework of the large-$N$ expansion the electronic dispersion $\varepsilon_{\vk}$ is given by 
\begin{equation}
\varepsilon_{\vk} = \varepsilon_{\vk}^{\parallel}  + \varepsilon_{\vk}^{\perp} \,,
\label{Ek}
\end{equation}
where the in-plane dispersion $\varepsilon_{\vk}^{\parallel}$ and the out-of-plane dispersion 
$\varepsilon_{\vk}^{\perp}$ are given, respectively, by
\begin{eqnarray}
\varepsilon_{\vk}^{\parallel} =& -2 \left( t \frac{\delta}{2}+\Delta \right) (\cos k_{x}+\cos k_{y})\nonumber\\
&-4t' \frac{\delta}{2} \cos k_{x} \cos k_{y} - \mu  \,, \label{Epara} \\
\varepsilon_{\vk}^{\perp} =& 2 t_{z} \frac{\delta}{2} (\cos k_x-\cos k_y)^2 \cos k_{z}  \,. \label{Eperp}
\end{eqnarray}

The form $(\cos k_x-\cos k_y)^2$ in $\varepsilon_{\vk}^{\perp}$ is frequently invoked for cuprates \cite{andersen95a} including 
the infinite-layer cuprate CaCuO$_2$ \cite{botana20}, which is closely related to Sr$_{0.9}$La$_{0.1}$CuO$_2$. 
Note that the hopping parameters $t$, $t'$, and $t_z$ are renormalized by the doping $\delta$ due to correlation effects. 
The additional term $\Delta$, which is proportional to $J$, in $\varepsilon_{\vk}^{\parallel}$ comes 
from the $J$-term [the second
term in \eq{tJV}] of the $t$-$J$-$V$ model \cite{foussats04}.

For a given doping, 
$\mu$ and $\Delta$ are determined self-consistently by solving
\begin{equation}{\label {Delta-A}}
\Delta = \frac{J}{4N_s N_z} \sum_{\vk} (\cos k_x + \cos k_y) n_F(\varepsilon_\vk) \; , 
\end{equation}
and 
\begin{equation}
(1-\delta)=\frac{2}{N_s N_z} \sum_{\vk} n_F(\varepsilon_\vk)\,,
\end{equation}
where $n_F$ is the Fermi function, $N_s$ is the total 
number of lattice sites on the square lattice, and $N_z$ is the number of layers along the $z$ direction. 
The long-range Coulomb interaction $V_{ij}$ can be written in momentum space \cite{becca96} as  
\begin{equation}
V({\bf q})=\frac{V_c}{A(q_x,q_y) - \cos q_z} \,,
\label{LRC}
\end{equation}
where $V_c= e^2 c(2 \epsilon_{\perp} a^2)^{-1}$ and 
\begin{equation}
A(q_x,q_y)=\alpha (2 - \cos q_x - \cos q_y)+1 \,  
\end{equation}
with $\alpha=\frac{\tilde{\epsilon}}{(a/c)^2}$ and $\tilde{\epsilon}=\epsilon_\parallel/\epsilon_\perp$, where $\epsilon_\parallel$ and $\epsilon_\perp$ are the 
dielectric constants parallel and perpendicular to the planes, respectively, $a$ is the lattice constant on the square
lattice and $c$ the distance between the layers. The electric charge of electrons is denoted as $e$. Thus, the values of the dielectric constants $\epsilon_\parallel$ and $\epsilon_\perp$ are encoded in $V_c$ and $\alpha$.

In the large-$N$ scheme, the charge-charge correlation function can be computed at the leading order as 
\begin{eqnarray}\label{CH}
\chi_{c}(\vq,\mathrm{i}\omega_n)= N \left ( \frac{\delta}{2} \right )^{2} D_{11}(\vq,\mathrm{i}\omega_n)  \,,
\end{eqnarray}
where $D_{11}$ is the element $(1,1)$ of the $6 \times 6$ bosonic propagator defined as 
\begin{equation}
[D_{ab}(\vq,\mathrm{i}\omega_n)]^{-1}
= [D^{(0)}_{ab}(\vq,\mathrm{i}\omega_n)]^{-1} - \Pi_{ab}(\vq,\mathrm{i}\omega_n)\,,
\label{dyson}
\end{equation}
with the bosonic Matsubara frequencies $\omega_n$; $a$ and $b$ run from 1 to 6.  The bare bosonic propagator is given by 
\begin{widetext}
\begin{eqnarray}\label{D0}
\left[ D^{(0)}_{ab}({\bf q},\mathrm{i}\omega_{n}) \right]^{-1} = N \left(
 \begin{array}{cccccc}
\frac{\delta^2}{2} \left[ V(\vq)-J(\vq)\right]
& \delta/2 & 0 & 0 & 0 & 0 \\
   \delta/2 & 0 & 0 & 0 & 0 & 0 \\
   0 & 0 & \frac{4}{J}\Delta^{2} & 0 & 0 & 0 \\
   0 & 0 & 0 & \frac{4}{J}\Delta^{2} & 0 & 0 \\
   0 & 0 & 0 & 0 & \frac{4}{J}\Delta^{2} & 0 \\
   0 & 0 & 0 & 0 & 0 & \frac{4}{J}\Delta^{2} \
 \end{array}
\right),
\end{eqnarray}
\end{widetext}
where $J(\vq) = \frac{J}{2} (\cos q_x + \cos q_y)$. In \eq{dyson}, the $6 \times 6$ bosonic self-energy is computed at leading order as 
\begin{widetext}
\begin{eqnarray}
&& \Pi_{ab}(\vq,\mathrm{i}\omega_n)
            = -\frac{N}{N_s N_z}\sum_{\vk} h_a(\vk,\vq,\varepsilon_\vk-\varepsilon_{\vk-\vq}) 
            \frac{n_F(\varepsilon_{\vk-\vq})-n_F(\varepsilon_\vk)}
                                  {\mathrm{i}\omega_n-\varepsilon_\vk+\varepsilon_{\vk-\vq}} 
            h_b(\vk,\vq,\varepsilon_\vk-\varepsilon_{\vk-\vq}) \nonumber \\
&& \hspace{25mm} - \delta_{a\,1} \delta_{b\,1} \frac{N}{N_s N_z}
                                       \sum_\vk \frac{\varepsilon_\vk-\varepsilon_{\vk-\vq}}{2}n_F(\varepsilon_\vk) \; ,
                                       \label{Pi}
\end{eqnarray}
\end{widetext}
where the six-component interaction vertex is given by 
\begin{widetext}
\begin{align}
 h_a(\vk,\vq,\nu) =& \left\{
                   \frac{2\varepsilon_{\vk-\vq}+\nu+2\mu}{2}+
                   2\Delta \left[ \cos\left(k_x-\frac{q_x}{2}\right)\cos\left(\frac{q_x}{2}\right) +
                                  \cos\left(k_y-\frac{q_y}{2}\right)\cos\left(\frac{q_y}{2}\right) \right];1;
                 \right. \nonumber \\
               & \left. -2\Delta \cos\left(k_x-\frac{q_x}{2}\right); -2\Delta \cos\left(k_y-\frac{q_y}{2}\right);
                         2\Delta \sin\left(k_x-\frac{q_x}{2}\right);  2\Delta \sin\left(k_y-\frac{q_y}{2}\right)
                 \right\} \, .
\label{vertex-h}
\end{align}
\end{widetext}

After performing the analytical continuation 
$\mathrm{i}\omega_n \rightarrow \omega+\mathrm{i} \Gamma$ in \eq{CH}, we obtain  the imaginary part of 
the charge-charge correlation function $\chi''_{c}(\vq,\omega)$, which can be directly compared with the RIXS intensity. 
While $\Gamma (>0)$ is infinitesimally small, we employ a finite broadening $\Gamma$  \cite{greco19a,greco20a}. 
This $\Gamma$ may mimic not only effects of the experimental resolution but also a broadening of the spectrum due to electron correlations  \cite{prelovsek99a}, 
so that we can successfully reproduce the peak width of experimental data.

\section*{Fitting procedure in the $t$-$J$-$V$ model} 
We perform calculations by taking the number of planes $N_z=30$ and setting the temperature to zero. 
We fix the broadening $\Gamma/t$ to be $0.1$. 
Since the values of $t'/t$ are already well-established in cuprates, we choose these parameters as an input according to the materials. The doping $\delta$ is the same as that estimated by charge neutrality of the materials. The computed parameters are $t_z$, $\alpha$, and $V_c$, which we determine by fitting the theoretical curves to 
the experimental data. Note that we obtain a charge excitation spectrum in the whole space of $\omega$ and $\vq$ 
once the parameters are fixed. Hence, we take a full set of available experimental data into account, namely the dispersions along $H$ and $L$, and 
consider the following function, 
\begin{equation}
F_{\rm err} (t_z, \alpha, V_c) = \sum_{i} \left| \omega^{\rm th}(H_i, K_i, L_i) -   \omega^{\rm ex}(H_i, K_i, L_i) \right|^2 
\label{error}
\end{equation}
where $i$ specifies momenta at which experimental data were taken and 
$\omega^{\rm th (ex)}$ is the plasmon energy obtained in the $t$-$J$-$V$ model (experiments). We then minimize 
\eq{error} with respect to $t_z$, $\alpha$, and $V_c$.
We remark that the plasmon intensity is not included as a parameter in the fitting procedure for the $t$-$J$ model, because a direct comparison of the absolute RIXS intensity and the calculated intensity is beyond the scope of this study. For instance, absorption-edge-specific resonance effects or non-Fermi-liquid-producing interactions \cite{hepting18a} can influence the RIXS intensity and its evolution as a function of momentum transfer, which are not included in the present $t$-$J$-$V$ model calculations.

In the large-$N$ formalism in the $t$-$J$ model, the charge excitations [\eq{CH}] are described by the $6 \times 6$ bosonic propagator 
$D_{ab}$ [\eq{dyson}]. The $2 \times 2$ sector with $a=b=1,2$ controls on-site charge fluctuation including plasmons and the $4 \times 4$ sector with $a=b=3-6$ controls bond-charge fluctuations. Reference~\onlinecite{bejas17} showed that these two sectors are essentially decoupled from each other and plasmons are already described even if we neglect the contribution from the $4 \times 4$ sector. We found that \eq{error} evaluated in terms of the full $6 \times 6$ matrix tends to become nearly three times larger than that in terms of the $2 \times 2$ matrix for SLCO, while they are comparable for the cases of LCCO and LSCO. Hence, we optimize \eq{error} by focusing on the $2 \times 2$ sector in \eq{dyson}, which also makes it easier to perform extensive calculations to minimize the error \eq{error}.    

For SLCO, we use $t'/t=0.2$ \cite{botana20a}. By minimizing \eq{error}, we obtain $t_z/t=0.055$, $\alpha=4.1$, and $V_c/t=18.5$. This results in $\epsilon_\parallel/\epsilon_0=5.89$ and $\epsilon_\perp/\epsilon_0=1.06$ for $a=3.960$ {\AA} and $c=3.405$ {\AA}. Notably, these values of the dielectric constants are similar to $\epsilon_\parallel/\epsilon_0 =4.3$ and $\epsilon_\perp/\epsilon_0=1$ reported for the closely related material SrCuO$_2$, using an \textit{ab-initio} rigid-ion model (RIM) for the local (rigid) charge response and taking charge fluctuations into account for the nonlocal part of the charge response \cite{klenner94a}. Furthermore, the obtained value of $t_z/t$ is in agreement with the value reported in Ref.~\onlinecite{botana20a} for the infinite-layer cuprate
CaCuO$_2$. 
As a measure of the absolute value of energy, we adopt $t/2=0.5$ eV \cite{hybertsen90,botana20a}; 
the factor $1/2$ here comes from a large-$N$ formalism, 
where $t$ is scaled by $1/N$ and we take $N=2$ when making a comparison with experiments.

We find that the minimized error does not change substantially for a wide range of $\alpha$ values. In fact, the energy of the acoustic-like plasmon branches is mainly governed by $t_z$, and $\alpha$ has only a marginal effect. Thus, the accurate determination of $\alpha$ is not critical for the present study. On the other hand, the situation becomes vice versa for the optical plasmon branch and $\alpha$ has a dominant effect. We obtain the optical plasma frequency $\omega_{\rm pl}^{\rm op}=0.78$ eV. If we allow the fitting error to increase by $20\%$ by changing $\alpha$ and $V_c$ while keeping $t_z/t=0.055$, \eq{error} yields $\alpha=1.58$ and $V_c/t=16.8$. In consequence, $\omega_{\rm pl}^{\rm op}$ becomes as high as $1.2$ eV, whereas the acoustic-like plasmon energies increase by only $4\%$. In principle, experimental input for the optical plasmon energy $\omega_{\rm pl}^{\rm op}$ could narrow down the range of $\alpha$. Unfortunately, to the best of our knowledge, $\omega_{\rm pl}^{\rm op}$ of SLCO has not been reported yet.

We also consider the hypothetical case of $t_z = 0$ for SLCO, i.e. the absence of interlayer hopping. In this case the gap would be only due to broadening, which for $\Gamma/t=0.1$ yields a gap of only 58 meV. This value is significantly smaller than the observed gap of 120 meV, which invalidates such a scenario. 

\section*{Variation of the broadening $\Gamma$}
\begin{figure}
 \begin{centering}
\includegraphics[width=1.\columnwidth]{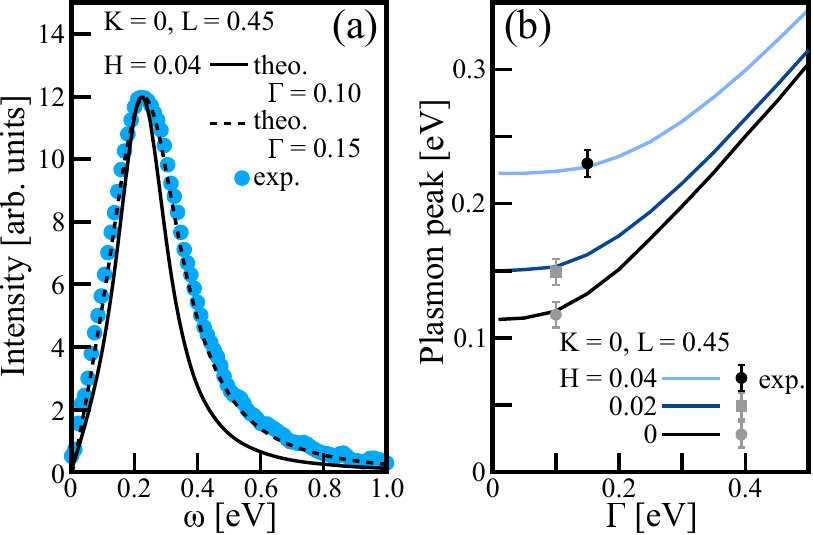}
\par\end{centering}
\caption{(a) Plasmon peak observed in RIXS for SLCO at momentum $(0.04, 0, 0.45)$ together with the theoretical curves for $\Gamma/t=0.1$ (solid black line) and 0.15 (dashed black line). The intensity of the calculated spectrum is scaled such that it matches the maximum of the experimental data. (b) Plasmon energy as a function of $\Gamma$ at several choices of momenta.  Experimental data (symbols) are also plotted by finding the value of $\Gamma$ to reproduce the line shape of the plasmon peak. 
}
\label{fig:vsGamma}
\end{figure}

We fixed $\Gamma/t=0.1$ in the analysis of the RIXS data presented in the main text. This choice of $\Gamma$ reproduces the line shape of the plasmon peak at momentum $(0.02, 0 ,0.45)$ almost perfectly
[Fig.3(a) in the main text].
We note, however, that the experimental plasmon peak broadens more rapidly at large $H$ than the calculated spectrum with $\Gamma/t=0.1$. For instance, the line shape of the plasmon peak at $(0.04, 0, 0.45)$ shows a smaller difference on the high-energy side between the RIXS experiment and the large-$N$ theory of the $t$-$J$-$V$ model for $\Gamma/t=0.15$ [Fig.~\ref{fig:vsGamma}(a)]. Hence, in principle, a $\vq$-dependence of the broadening should be incorporated in the fitting. Nevertheless, the peak position in Fig.~\ref{fig:vsGamma}(a) is almost identical for the
two
choices of $\Gamma$. In fact, especially in the parameter space relevant to the experiments, the plasmon energy depends only marginally on the value of $\Gamma$. 

Figure~\ref{fig:vsGamma}(b) presents the calculated plasmon energy as a function of $\Gamma$ at momenta ($H$, 0, 0.45) with $H = 0, 0.02,$ and 0.04. We also plot the experimental plasmon energies by choosing the appropriate value of $\Gamma$ to reproduce the line shape of the plasmon peak at each momentum. We find that the experimental data, namely the plasmon peak energies, nicely sit on the theoretical curve. Moreover, for small broadenings ($\Gamma \lesssim 0.1$), the theoretical plasmon energy does not depend on $\Gamma$ markedly. Thus, we employ the same value of $\Gamma/t=0.1$ to describe the full plasmon dispersion observed in the RIXS experiments
[Fig.3(b)-(d) in the main text].

Figure~\ref{fig:vsGamma}(b) also implies that the gap of $\sim 120$ meV at $H = 0$ is obtained even for infinitesimally small $\Gamma=+0$ and cannot originate from a broadening of the plasmon peak. Recalling that the charge gap vanishes for $t_z=0$ and $\Gamma=+0$ \cite{grecu73a,grecu75a,greco16a}, this confirms that the  gap observed in SLCO originates from the presence of the finite interlayer hopping $t_z/t=0.055$.

\section*{Possible plasmon gap in LCCO and LSCO} 
We have established the presence of the plasmon gap at $H=K=0$ for a finite $L$ in SLCO. Here we explore a possible gap for LCCO and LSCO by performing a similar analysis of available experimental data. 

\begin{figure*}
 \begin{centering}
\includegraphics[width=14.3cm]{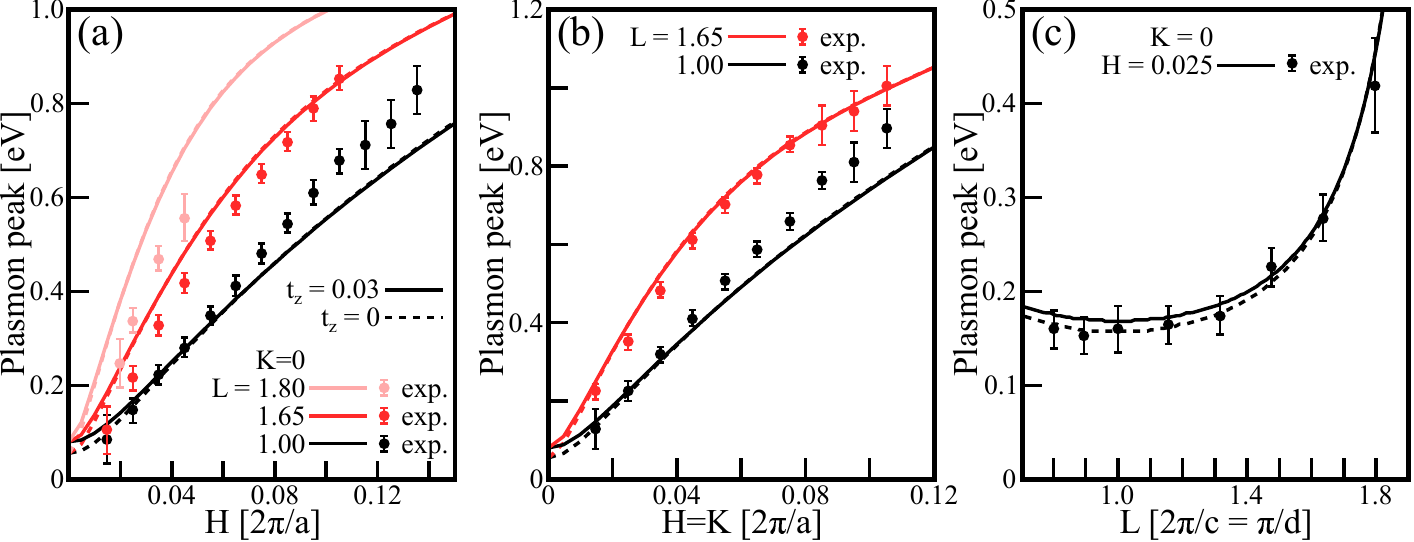}
\par\end{centering}
\caption{Plasmon dispersion in LCCO along different directions reported in Ref.~\onlinecite{hepting18a} (symbols) and the corresponding theoretical results obtained in the $t$-$J$-$V$ model for $t_z/t = 0.03$ (solid lines) and 0 (dashed lines). (a) Momentum transfer along the $H$ direction for $K=0$ and $L = 1, 1.65$, and 1.8. (b) Momentum transfer along the $(H,H)$ direction for $L = 1$ and 1.65. (c) Momentum transfer along the $L$ direction for $H=0.025$ and $K=0$. Note that the momentum transfer is given in units of $2\pi/c$ as well as $\pi/d$, with $d$ being the spacing between nearest-neighbor CuO$_2$ planes. Thus, $d=c/2$ in LSCO and LCCO, whereas $d=c$ in SLCO.}
\label{fig:LCCO}
\end{figure*}

Figure~\ref{fig:LCCO} shows RIXS data of LCCO ($x = 0.175$) reported in Ref.~[\onlinecite{hepting18a}], together with the results of the present $t$-$J$-$V$ model fitting for $t'/t=0.3$, $t/2=0.5$ eV, and $\Gamma/t=0.1$. We obtain $t_z/t=0.03$ $(\alpha=2.9, V_c/t=18$, which corresponds to $\epsilon_\parallel/\epsilon_0 = 2.35$ and $\epsilon_\perp/\epsilon_0 = 1.94$) as the upper bound and $t_z=0$ $(\alpha=3, V_c/t=18.5$, which corresponds to $\epsilon_\parallel/\epsilon_0 = 2.37$ and $\epsilon_\perp/\epsilon_0 = 1.89$) as the lower bound. In the former case, we estimate the gap at $H=K=0$ to be $\sim 82$ meV, while it is $\sim 58$ meV in the latter case. Note that the charge gap for $t_z=0$ arises from the broadening ($\Gamma/t=0.1$). We also checked that both parameters lead to an optical plasmon energy $\omega_{\rm pl}^{\rm op} \sim 1.1$ eV, which is in excellent agreement with $\omega_{\rm pl}^{\rm op}$ measured by optics for the closely similar material NCCO \cite{singley01}. Recalling that we have only three fitting parameters in \eq{error} for the whole spectra, we consider the agreement between theory and experiment as reasonably good for both values of $t_z$ in Fig.~\ref{fig:LCCO}. For momentum transfer along the $L$ direction [Fig.~\ref{fig:LCCO}(c)], $t_z = 0$ matches the experimental data slightly better. Nevertheless, the total quadratic error [\eq{error}] is almost flat as a function of $t_z$. It is only about $2\%$ higher for $t_z=0.03$ than for $t_z=0$, i.e. too subtle to make definitive statements about the value of $t_z$. For momentum transfer along the $H$ direction [Fig.~\ref{fig:LCCO}(a),(b)], the calculated curves for $t_z/t=0.03$ and 0 deviate from each other only for very small momenta. Hence, future high-resolution RIXS data at smaller momenta are highly desirable for a more accurate determination of $t_z$ in LCCO.

\begin{figure*}
 \begin{centering}
\includegraphics[width=9.6cm]{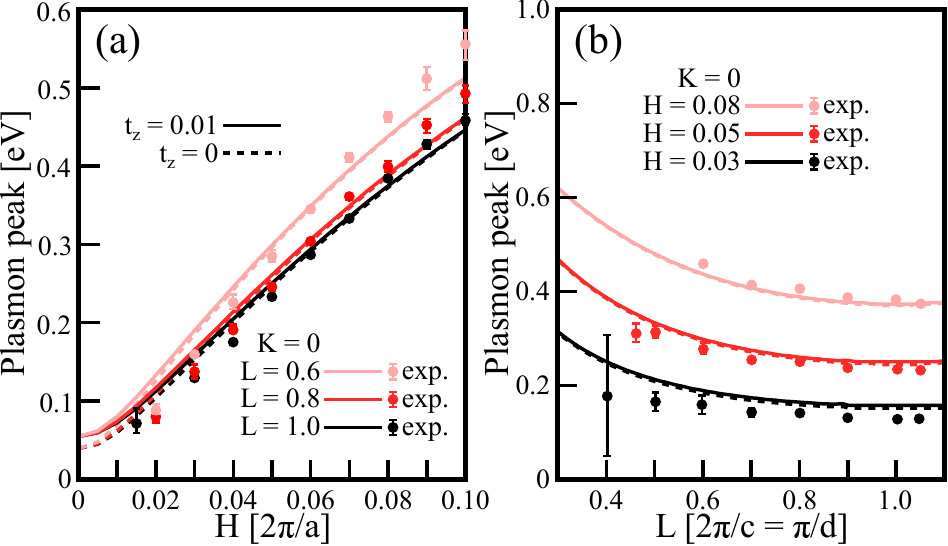}
\par\end{centering}
\caption{Plasmon dispersion in LSCO along different directions reported in Ref.~\onlinecite{nag20a} (symbols) and the corresponding theoretical results obtained in the $t$-$J$-$V$ model for $t_z/t = 0.01$ (solid lines) and 0 (dashed lines). (a) Momentum transfer along the $H$ direction for $K=0$ and $L = 0.6, 0.8$, and 1. (b) Momentum transfer along the $L$ direction for $K=0$ and $H = 0.03, 0.05$, and 0.08.}
\label{fig:LSCO}
\end{figure*}

Figure~\ref{fig:LSCO} displays RIXS data of LSCO ($x = 0.16$) reported in Ref.~[\onlinecite{nag20a}] and also results from the present $t$-$J$-$V$ model fitting. We use $t'/t=-0.2$ \cite{pavarini01a}, $t/2=0.35$ eV \cite{nag20a}, and $\Gamma/t=0.1$ and impose as an additional constraint that the optical plasmon energy should be $\omega_{\rm pl}^{\rm op} \approx 0.85$ eV, as it is already known from optical measurements \cite{suzuki89,uchida91}. As in the case of LCCO, the total fitting error depends weakly on the value of $t_z$ and we cannot determine $t_z$ uniquely. Instead we obtain 
$\alpha=3.5$, $V_c/t=31$ ($\epsilon_\parallel/\epsilon_0 = 2.23$, $\epsilon_\perp/\epsilon_0 = 1.92$), and $t_z/t = 0.01$ as the upper bound. The corresponding gap at $H=K=0$ is $\sim 55$ meV. As the lower bound, we may choose $t_z/t=0$, yielding a gap of $\sim 41$ meV.  This gap is due to the broadening $\Gamma=0.1t$, which was introduced to reproduce the plasmon peak width observed in RIXS. We note that $t_z$ and the upper limit of the gap broadly agree with those discussed in Ref.~\onlinecite{nag20a}, where the present error minimization scheme was not applied. 
Furthermore, we note that for metallic La$_2$CuO$_4$, Ref.~\onlinecite{bauer09a} predicted a plasmon-phonon gap of 60 meV, which is remarkably similar to the upper bound obtained for LSCO by the present $t$-$J$-$V$ model fitting.  

The gap and the interlayer hopping $t_z$ in LCCO and LSCO are approximately a factor of 2 smaller than those in SLCO. This can be rationalized by considering that the interlayer distance in LCCO and LSCO is larger than in SLCO, thus a smaller $t_z/t$ is expected. Hence, in contrast to SLCO, the gap due to $t_z/t$ could be masked by a gap due to the broadening $\Gamma$, and high-resolution RIXS will be required to determine the plasmon gap in LCCO and LSCO.

\section*{Intensity of the plasmon peak}

\begin{figure}
 \begin{centering}
\includegraphics[width=.75\columnwidth]{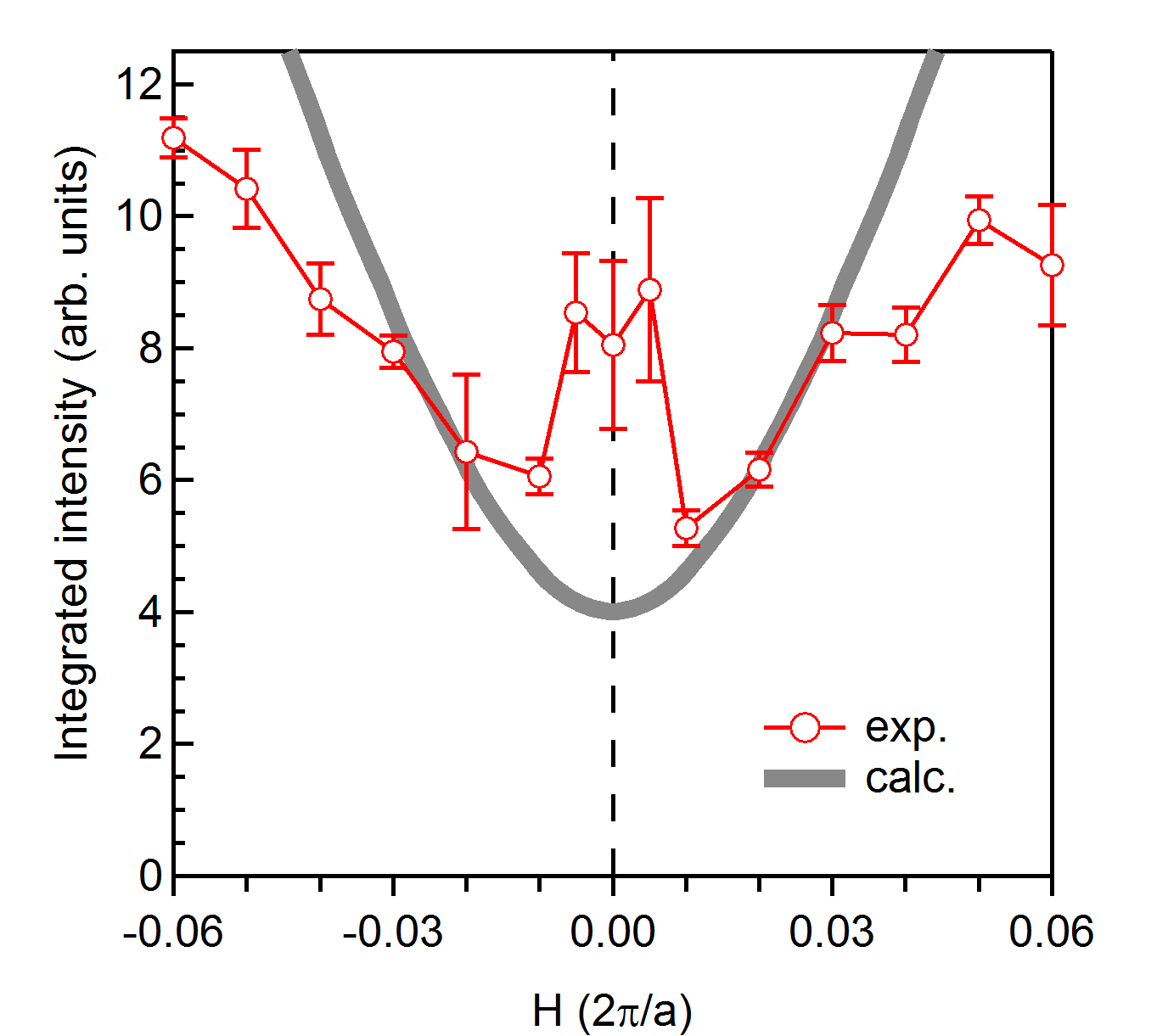}
\par\end{centering}
\caption{Integrated intensity of the plasmon peak of SLCO along the $(H, 0, 0.45)$ direction. Red symbols correspond to the fit results of the expermimental data and the solid gray line is from the $t$-$J$-$V$ model calculations. Note that for $\vert H \vert \leq 0.005$  the  HE  phonon  and  the  plasmon  were  combined  in  the  fits  and  treated  as  a  single peak. Also note that as explained in Fig. 3(a) of the main text, the calculated intensity and broadening $\Gamma$ are scaled to match the experimental data at $H = 0.02$.  }
\label{fig:area}
\end{figure}

Figure~\ref{fig:area} shows the integrated intensity of the plasmon peak of SLCO. The red symbols correspond to the integrated intensity extracted from the fits. 
The gray line is the integrated intensity of the computed imaginary part of the charge susceptibility $\chi_c''({\bf q},\omega)$. As explained in Fig.~3(a) of the main text, the computed intensity and broadening $\Gamma$ are scaled to match the experimental data at $H = 0.02$. For larger $H$, the integrated RIXS intensity and the computed intensity follow a similar trend. The deviation, which increases with increasing $H$, could be an artifact from the fitting of the RIXS spectra, for instance due to the overlap with the increasing paramagnon peak, leading to an underappreciation of the spectral weight of the plasmon peak. In the scope of the present work, however, especially the region around the 2D BZ center (small $H$) is of interest. For very small momenta ($\vert H \vert \leq 0.005$) the experimental intensity clearly increases above the calculated intensity [Fig.~\ref{fig:area}]. Yet, for these small momenta the HE phonon and the plasmon were combined in the fits and treated as a single peak, because the two components could not be resolved separately. Thus, it can not be evaluated whether the enhanced intensity is merely due to a superposition between the plasmon and the HE phonon, or additional effects. For instance, we observe that the elastic line becomes very intense for $\vert H \vert \leq 0.005$ and deviates from a pure Gaussian lineshape. The occurrence of the deviation from a Gaussian lineshape in close proximity to the specular scattering condition at (0, 0, 0.45) can be clearly discerned on the energy gain side of the RIXS spectra in Figs.~\ref{fig:fits}(a),(b), but might also exist on the energy loss side of the RIXS spectra, although the overlap with the other spectral components prevents us from clearly resolving this additional spectral weight as a separate component. Nevertheless, it is possible that such an additional spectral weight accounts for the enhanced intensity in Fig.~\ref{fig:area} (if there is any excess intensity on top of the superposition of the plasmon and the HE phonon). 
Note that in our analysis with the $t$-$J$-$V$ model, RIXS data for $\vert H \vert \leq 0.005$ were excluded.

\section*{Plasmon-phonon coupling}

This section elaborates why we consider plasmon-phonon coupling in SLCO as unlikely, whereas it might be realized in LSCO. In general, such coupling can occur in materials where phonons couple according to their symmetry to charge fluctuations and thus to plasmons. In the case of cuprates, it was predicted for La$_2$CuO$_4$, where a $c$-axis polarized phonon mode involving the apical oxygen ions couples to charge fluctuations \cite{falter94a,bauer09a,falter98,pintschovius05}. The energy of the free plasmon in (metallic) La$_2$CuO$_4$ is small enough to cross the energy of the aforementioned phonon in close proximity to the $Z$-point. This leads to an anticrossing of the two modes with a predicted gap of the upper mixed plasmon-phonon mode of $\sim 60$ meV \cite{bauer09a}. In SLCO, however, the dispersions of all plasmon branches (minimum energy $\sim$ 120 meV) are located above the regime of the phonon energies. Thus, it is implausible that the gap in SLCO is predominantly due to plasmon-phonon coupling and the corresponding anticrossing scenario. In this context, we also remark that a difference in the plasmon-phonon coupling properties of SLCO and LSCO is rooted in the different magnitudes of $t_z$ in the two materials ($t_z^{\text{SLCO}}$ $\gg$ $t_z^{\text{LSCO}}$), or, more generally, in the different strengths of the $k_z$ dispersions. In fact, when assuming an excessively strong $k_z$ dispersion in La$_2$CuO$_4$, the plasmon also lies above the phonon regime at the $Z$-point and in turn substantial plasmon-phonon coupling would be prevented also in this material \cite{falter98}.

Furthermore, in contrast to LSCO, the infinite-layer crystal structure of SLCO lacks the apical oxygen ions responsible for the $c$-axis polarized phonon mode that can couple to plasmons \cite{falter94a,bauer09a,falter98,pintschovius05} and the high-energy phonons in SLCO are not suitable for a coupling to plasmons. Specifically, as calculated for SrCuO$_2$ \cite{klenner94a}, the 91.4 meV phonon at the $Z$-point, which likely corresponds to the experimentally observed 95 meV phonon of SLCO, is an $A_{2u}$ buckling mode of the Cu and O ions in the CuO$_2$ planes, which does not couple to charge fluctuations mediated by the nonlocal part of the charge response. The absence of the coupling to charge fluctuations is also the reason why this phonon exhibits the same (high) frequency as in a rigid ion model (RIM) \cite{klenner94a}, which defines only the local part of the charge response.
The only phonon in SrCuO$_2$ at the $Z$-point that could couple according to its symmetry to plasmons, is the $A_{1g}$ mode with 31.5 meV, which is clearly too far away from the plasmon energy for a plausible coupling scenario.

From the present theoretical perspective, our $t$-$J$-$V$ model calculations of the plasmon energy, which do not include plasmon-phonon interactions, smoothly capture the experimental data for $\vert H \vert \leq 0.005$
[Fig.3(d) in the main text]
, implying a weak plasmon-phonon coupling, if any.
We also note that even in the case of an anticrossing due to plasmon-phonon coupling, the intensity of the upper branch is rather expected to decrease in the range of strong coupling \cite{brar14}, which is opposite to the observed intensity for $\vert H \vert \leq 0.005$ in Fig.~\ref{fig:area}.

Taking all of the above considerations into account, we conclude that plasmon-phonon coupling in SLCO is unlikely. We remark, however, that plasmon-phonon coupling is a common phenomenon in other materials \cite{koch10, bezares17}. Therefore, it will be interesting to use high-resolution RIXS to probe possible plasmon-phonon couplings in LSCO, where it has been predicted by quantitative calculations \cite{bauer09a}, and in other cuprates, where the plasmon dispersion crosses the energy of phonon modes that couple to charge fluctuations.


\begin{thebibliography}{82}%
\makeatletter
\providecommand \@ifxundefined [1]{%
 \@ifx{#1\undefined}
}%
\providecommand \@ifnum [1]{%
 \ifnum #1\expandafter \@firstoftwo
 \else \expandafter \@secondoftwo
 \fi
}%
\providecommand \@ifx [1]{%
 \ifx #1\expandafter \@firstoftwo
 \else \expandafter \@secondoftwo
 \fi
}%
\providecommand \natexlab [1]{#1}%
\providecommand \enquote  [1]{``#1''}%
\providecommand \bibnamefont  [1]{#1}%
\providecommand \bibfnamefont [1]{#1}%
\providecommand \citenamefont [1]{#1}%
\providecommand \href@noop [0]{\@secondoftwo}%
\providecommand \href [0]{\begingroup \@sanitize@url \@href}%
\providecommand \@href[1]{\@@startlink{#1}\@@href}%
\providecommand \@@href[1]{\endgroup#1\@@endlink}%
\providecommand \@sanitize@url [0]{\catcode `\\12\catcode `\$12\catcode
  `\&12\catcode `\#12\catcode `\^12\catcode `\_12\catcode `\%12\relax}%
\providecommand \@@startlink[1]{}%
\providecommand \@@endlink[0]{}%
\providecommand \url  [0]{\begingroup\@sanitize@url \@url }%
\providecommand \@url [1]{\endgroup\@href {#1}{\urlprefix }}%
\providecommand \urlprefix  [0]{URL }%
\providecommand \Eprint [0]{\href }%
\providecommand \doibase [0]{https://doi.org/}%
\providecommand \selectlanguage [0]{\@gobble}%
\providecommand \bibinfo  [0]{\@secondoftwo}%
\providecommand \bibfield  [0]{\@secondoftwo}%
\providecommand \translation [1]{[#1]}%
\providecommand \BibitemOpen [0]{}%
\providecommand \bibitemStop [0]{}%
\providecommand \bibitemNoStop [0]{.\EOS\space}%
\providecommand \EOS [0]{\spacefactor3000\relax}%
\providecommand \BibitemShut  [1]{\csname bibitem#1\endcsname}%
\let\auto@bib@innerbib\@empty
\bibitem [{\citenamefont {Keimer}\ \emph {et~al.}(2015)\citenamefont {Keimer},
  \citenamefont {Kivelson}, \citenamefont {Norman}, \citenamefont {Uchida},\
  and\ \citenamefont {Zaanen}}]{keimer15}%
  \BibitemOpen
  \bibfield  {author} {\bibinfo {author} {\bibfnamefont {B.}~\bibnamefont
  {Keimer}}, \bibinfo {author} {\bibfnamefont {S.~A.}\ \bibnamefont
  {Kivelson}}, \bibinfo {author} {\bibfnamefont {M.~R.}\ \bibnamefont
  {Norman}}, \bibinfo {author} {\bibfnamefont {S.}~\bibnamefont {Uchida}},\
  and\ \bibinfo {author} {\bibfnamefont {J.}~\bibnamefont {Zaanen}},\
  }\bibfield  {title} {\bibinfo {title} {From quantum matter to
  high-temperature superconductivity in copper oxides},\ }\href
  {http://dx.doi.org/10.1038/nature14165} {\bibfield  {journal} {\bibinfo
  {journal} {Nature}\ }\textbf {\bibinfo {volume} {518}},\ \bibinfo {pages}
  {179} (\bibinfo {year} {2015})}\BibitemShut {NoStop}%
\bibitem [{\citenamefont {Armitage}\ \emph {et~al.}(2010)\citenamefont
  {Armitage}, \citenamefont {Fournier},\ and\ \citenamefont
  {Greene}}]{armitage10}%
  \BibitemOpen
  \bibfield  {author} {\bibinfo {author} {\bibfnamefont {N.~P.}\ \bibnamefont
  {Armitage}}, \bibinfo {author} {\bibfnamefont {P.}~\bibnamefont {Fournier}},\
  and\ \bibinfo {author} {\bibfnamefont {R.~L.}\ \bibnamefont {Greene}},\
  }\bibfield  {title} {\bibinfo {title} {Progress and perspectives on
  electron-doped cuprates},\ }\href
  {https://link.aps.org/doi/10.1103/RevModPhys.82.2421} {\bibfield  {journal}
  {\bibinfo  {journal} {Rev. Mod. Phys.}\ }\textbf {\bibinfo {volume} {82}},\
  \bibinfo {pages} {2421} (\bibinfo {year} {2010})}\BibitemShut {NoStop}%
\bibitem [{\citenamefont {Lee}\ \emph {et~al.}(2006)\citenamefont {Lee},
  \citenamefont {Nagaosa},\ and\ \citenamefont {Wen}}]{lee06}%
  \BibitemOpen
  \bibfield  {author} {\bibinfo {author} {\bibfnamefont {P.~A.}\ \bibnamefont
  {Lee}}, \bibinfo {author} {\bibfnamefont {N.}~\bibnamefont {Nagaosa}},\ and\
  \bibinfo {author} {\bibfnamefont {X.-G.}\ \bibnamefont {Wen}},\ }\bibfield
  {title} {\bibinfo {title} {Doping a mott insulator: Physics of
  high-temperature superconductivity},\ }\href
  {https://doi.org/10.1103/RevModPhys.78.17} {\bibfield  {journal} {\bibinfo
  {journal} {Rev. Mod. Phys.}\ }\textbf {\bibinfo {volume} {78}},\ \bibinfo
  {pages} {17} (\bibinfo {year} {2006})}\BibitemShut {NoStop}%
\bibitem [{\citenamefont {Scalapino}(2012)}]{scalapino12}%
  \BibitemOpen
  \bibfield  {author} {\bibinfo {author} {\bibfnamefont {D.~J.}\ \bibnamefont
  {Scalapino}},\ }\bibfield  {title} {\bibinfo {title} {A common thread: The
  pairing interaction for unconventional superconductors},\ }\href
  {https://doi.org/10.1103/RevModPhys.84.1383} {\bibfield  {journal} {\bibinfo
  {journal} {Rev. Mod. Phys.}\ }\textbf {\bibinfo {volume} {84}},\ \bibinfo
  {pages} {1383} (\bibinfo {year} {2012})}\BibitemShut {NoStop}%
\bibitem [{\citenamefont {Dai}\ \emph {et~al.}(2001)\citenamefont {Dai},
  \citenamefont {Mook}, \citenamefont {Hunt},\ and\ \citenamefont
  {Do\ifmmode~\breve{g}\else \u{g}\fi{}an}}]{dai01}%
  \BibitemOpen
  \bibfield  {author} {\bibinfo {author} {\bibfnamefont {P.}~\bibnamefont
  {Dai}}, \bibinfo {author} {\bibfnamefont {H.~A.}\ \bibnamefont {Mook}},
  \bibinfo {author} {\bibfnamefont {R.~D.}\ \bibnamefont {Hunt}},\ and\
  \bibinfo {author} {\bibfnamefont {F.}~\bibnamefont {Do\ifmmode~\breve{g}\else
  \u{g}\fi{}an}},\ }\bibfield  {title} {\bibinfo {title} {Evolution of the
  resonance and incommensurate spin fluctuations in superconducting
  YBa$_{2}$Cu$_{3}$O$_{6+x}$},\ }\href
  {https://doi.org/10.1103/PhysRevB.63.054525} {\bibfield  {journal} {\bibinfo
  {journal} {Phys. Rev. B}\ }\textbf {\bibinfo {volume} {63}},\ \bibinfo
  {pages} {054525} (\bibinfo {year} {2001})}\BibitemShut {NoStop}%
\bibitem [{\citenamefont {Le~Tacon}\ \emph {et~al.}(2011)\citenamefont
  {Le~Tacon}, \citenamefont {Ghiringhelli}, \citenamefont {Chaloupka},
  \citenamefont {Sala}, \citenamefont {Hinkov}, \citenamefont {Haverkort},
  \citenamefont {Minola}, \citenamefont {Bakr}, \citenamefont {Zhou},
  \citenamefont {Blanco-Canosa}, \citenamefont {Monney}, \citenamefont {Song},
  \citenamefont {Sun}, \citenamefont {Lin}, \citenamefont {De~Luca},
  \citenamefont {Salluzzo}, \citenamefont {Khaliullin}, \citenamefont
  {Schmitt}, \citenamefont {Braicovich},\ and\ \citenamefont
  {Keimer}}]{letacon11}%
  \BibitemOpen
  \bibfield  {author} {\bibinfo {author} {\bibfnamefont {M.}~\bibnamefont
  {Le~Tacon}}, \bibinfo {author} {\bibfnamefont {G.}~\bibnamefont
  {Ghiringhelli}}, \bibinfo {author} {\bibfnamefont {J.}~\bibnamefont
  {Chaloupka}}, \bibinfo {author} {\bibfnamefont {M.~M.}\ \bibnamefont {Sala}},
  \bibinfo {author} {\bibfnamefont {V.}~\bibnamefont {Hinkov}}, \bibinfo
  {author} {\bibfnamefont {M.~W.}\ \bibnamefont {Haverkort}}, \bibinfo {author}
  {\bibfnamefont {M.}~\bibnamefont {Minola}}, \bibinfo {author} {\bibfnamefont
  {M.}~\bibnamefont {Bakr}}, \bibinfo {author} {\bibfnamefont {K.~J.}\
  \bibnamefont {Zhou}}, \bibinfo {author} {\bibfnamefont {S.}~\bibnamefont
  {Blanco-Canosa}}, \bibinfo {author} {\bibfnamefont {C.}~\bibnamefont
  {Monney}}, \bibinfo {author} {\bibfnamefont {Y.~T.}\ \bibnamefont {Song}},
  \bibinfo {author} {\bibfnamefont {G.~L.}\ \bibnamefont {Sun}}, \bibinfo
  {author} {\bibfnamefont {C.~T.}\ \bibnamefont {Lin}}, \bibinfo {author}
  {\bibfnamefont {G.~M.}\ \bibnamefont {De~Luca}}, \bibinfo {author}
  {\bibfnamefont {M.}~\bibnamefont {Salluzzo}}, \bibinfo {author}
  {\bibfnamefont {G.}~\bibnamefont {Khaliullin}}, \bibinfo {author}
  {\bibfnamefont {T.}~\bibnamefont {Schmitt}}, \bibinfo {author} {\bibfnamefont
  {L.}~\bibnamefont {Braicovich}},\ and\ \bibinfo {author} {\bibfnamefont
  {B.}~\bibnamefont {Keimer}},\ }\bibfield  {title} {\bibinfo {title} {Intense
  paramagnon excitations in a large family of high-temperature
  superconductors},\ }\href {https://doi.org/10.1038/nphys2041} {\bibfield
  {journal} {\bibinfo  {journal} {Nat. Phys.}\ }\textbf {\bibinfo {volume}
  {7}},\ \bibinfo {pages} {725} (\bibinfo {year} {2011})}\BibitemShut {NoStop}%
\bibitem [{\citenamefont {Dean}\ \emph {et~al.}(2013)\citenamefont {Dean},
  \citenamefont {Dellea}, \citenamefont {Springell}, \citenamefont
  {Yakhou-Harris}, \citenamefont {Kummer}, \citenamefont {Brookes},
  \citenamefont {Liu}, \citenamefont {Sun}, \citenamefont {Strle},
  \citenamefont {Schmitt}, \citenamefont {Braicovich}, \citenamefont
  {Ghiringhelli}, \citenamefont {Bo{\v{z}}ovi{\'{c}}},\ and\ \citenamefont
  {Hill}}]{dean13}%
  \BibitemOpen
  \bibfield  {author} {\bibinfo {author} {\bibfnamefont {M.~P.~M.}\
  \bibnamefont {Dean}}, \bibinfo {author} {\bibfnamefont {G.}~\bibnamefont
  {Dellea}}, \bibinfo {author} {\bibfnamefont {R.~S.}\ \bibnamefont
  {Springell}}, \bibinfo {author} {\bibfnamefont {F.}~\bibnamefont
  {Yakhou-Harris}}, \bibinfo {author} {\bibfnamefont {K.}~\bibnamefont
  {Kummer}}, \bibinfo {author} {\bibfnamefont {N.~B.}\ \bibnamefont {Brookes}},
  \bibinfo {author} {\bibfnamefont {X.}~\bibnamefont {Liu}}, \bibinfo {author}
  {\bibfnamefont {Y.-J.}\ \bibnamefont {Sun}}, \bibinfo {author} {\bibfnamefont
  {J.}~\bibnamefont {Strle}}, \bibinfo {author} {\bibfnamefont
  {T.}~\bibnamefont {Schmitt}}, \bibinfo {author} {\bibfnamefont
  {L.}~\bibnamefont {Braicovich}}, \bibinfo {author} {\bibfnamefont
  {G.}~\bibnamefont {Ghiringhelli}}, \bibinfo {author} {\bibfnamefont
  {I.}~\bibnamefont {Bo{\v{z}}ovi{\'{c}}}},\ and\ \bibinfo {author}
  {\bibfnamefont {J.~P.}\ \bibnamefont {Hill}},\ }\bibfield  {title} {\bibinfo
  {title} {Persistence of magnetic excitations in La$_{2-x}$Sr$_x$CuO$_4$ from the
  undoped insulator to the heavily overdoped non-superconducting metal},\
  }\href {https://doi.org/10.1038/nmat3723} {\bibfield  {journal} {\bibinfo
  {journal} {Nat. Mater.}\ }\textbf {\bibinfo {volume} {12}},\ \bibinfo
  {pages} {1019} (\bibinfo {year} {2013})}\BibitemShut {NoStop}%
\bibitem [{\citenamefont {Vilardi}\ \emph {et~al.}(2019)\citenamefont
  {Vilardi}, \citenamefont {Taranto},\ and\ \citenamefont
  {Metzner}}]{vilardi19}%
  \BibitemOpen
  \bibfield  {author} {\bibinfo {author} {\bibfnamefont {D.}~\bibnamefont
  {Vilardi}}, \bibinfo {author} {\bibfnamefont {C.}~\bibnamefont {Taranto}},\
  and\ \bibinfo {author} {\bibfnamefont {W.}~\bibnamefont {Metzner}},\
  }\bibfield  {title} {\bibinfo {title} {Antiferromagnetic and $d$-wave pairing
  correlations in the strongly interacting two-dimensional hubbard model from
  the functional renormalization group},\ }\href
  {https://doi.org/10.1103/PhysRevB.99.104501} {\bibfield  {journal} {\bibinfo
  {journal} {Phys. Rev. B}\ }\textbf {\bibinfo {volume} {99}},\ \bibinfo
  {pages} {104501} (\bibinfo {year} {2019})}\BibitemShut {NoStop}%
\bibitem [{\citenamefont {Franck}(1994)}]{franck94}%
  \BibitemOpen
  \bibfield  {author} {\bibinfo {author} {\bibfnamefont {J.~P.}\ \bibnamefont
  {Franck}},\ }\href@noop {} {\emph {\bibinfo {title} {Experimental studies of
  the isotope effect in high temperature superconductors Physical Properties of
  High Temperature Superconductors IV ed Ginsberg D M (World Scientific)}}}\
  (\bibinfo {year} {1994})\BibitemShut {NoStop}%
\bibitem [{\citenamefont {Cuk}\ \emph {et~al.}(2004)\citenamefont {Cuk},
  \citenamefont {Baumberger}, \citenamefont {Lu}, \citenamefont {Ingle},
  \citenamefont {Zhou}, \citenamefont {Eisaki}, \citenamefont {Kaneko},
  \citenamefont {Hussain}, \citenamefont {Devereaux}, \citenamefont {Nagaosa},\
  and\ \citenamefont {Shen}}]{cuk04}%
  \BibitemOpen
  \bibfield  {author} {\bibinfo {author} {\bibfnamefont {T.}~\bibnamefont
  {Cuk}}, \bibinfo {author} {\bibfnamefont {F.}~\bibnamefont {Baumberger}},
  \bibinfo {author} {\bibfnamefont {D.~H.}\ \bibnamefont {Lu}}, \bibinfo
  {author} {\bibfnamefont {N.}~\bibnamefont {Ingle}}, \bibinfo {author}
  {\bibfnamefont {X.~J.}\ \bibnamefont {Zhou}}, \bibinfo {author}
  {\bibfnamefont {H.}~\bibnamefont {Eisaki}}, \bibinfo {author} {\bibfnamefont
  {N.}~\bibnamefont {Kaneko}}, \bibinfo {author} {\bibfnamefont
  {Z.}~\bibnamefont {Hussain}}, \bibinfo {author} {\bibfnamefont {T.~P.}\
  \bibnamefont {Devereaux}}, \bibinfo {author} {\bibfnamefont {N.}~\bibnamefont
  {Nagaosa}},\ and\ \bibinfo {author} {\bibfnamefont {Z.-X.}\ \bibnamefont
  {Shen}},\ }\bibfield  {title} {\bibinfo {title} {Coupling of the ${B}_{1g}$
  phonon to the antinodal electronic states of
  Bi$_{2}$Sr$_{2}$Ca$_{0.92}$Y$_{0.08}$Cu$_{2}$O$_{8+\delta}$},\
  }\href {https://doi.org/10.1103/PhysRevLett.93.117003} {\bibfield  {journal}
  {\bibinfo  {journal} {Phys. Rev. Lett.}\ }\textbf {\bibinfo {volume} {93}},\
  \bibinfo {pages} {117003} (\bibinfo {year} {2004})}\BibitemShut {NoStop}%
\bibitem [{\citenamefont {Heid}\ \emph {et~al.}(2009)\citenamefont {Heid},
  \citenamefont {Zeyher}, \citenamefont {Manske},\ and\ \citenamefont
  {Bohnen}}]{heid09}%
  \BibitemOpen
  \bibfield  {author} {\bibinfo {author} {\bibfnamefont {R.}~\bibnamefont
  {Heid}}, \bibinfo {author} {\bibfnamefont {R.}~\bibnamefont {Zeyher}},
  \bibinfo {author} {\bibfnamefont {D.}~\bibnamefont {Manske}},\ and\ \bibinfo
  {author} {\bibfnamefont {K.-P.}\ \bibnamefont {Bohnen}},\ }\bibfield  {title}
  {\bibinfo {title} {Phonon-induced pairing interaction in
  YBa$_{2}$Cu$_{3}$O$_{7}$ within the local-density
  approximation},\ }\href {https://doi.org/10.1103/PhysRevB.80.024507}
  {\bibfield  {journal} {\bibinfo  {journal} {Phys. Rev. B}\ }\textbf {\bibinfo
  {volume} {80}},\ \bibinfo {pages} {024507} (\bibinfo {year}
  {2009})}\BibitemShut {NoStop}%
\bibitem [{\citenamefont {Giustino}\ \emph {et~al.}(2008)\citenamefont
  {Giustino}, \citenamefont {Cohen},\ and\ \citenamefont {Louie}}]{giustino08}%
  \BibitemOpen
  \bibfield  {author} {\bibinfo {author} {\bibfnamefont {F.}~\bibnamefont
  {Giustino}}, \bibinfo {author} {\bibfnamefont {M.~L.}\ \bibnamefont
  {Cohen}},\ and\ \bibinfo {author} {\bibfnamefont {S.~G.}\ \bibnamefont
  {Louie}},\ }\href{https://doi.org/10.1038/nature06874} {\bibfield  {journal} {\bibinfo  {journal} {Nature}\ }\textbf {\bibinfo {volume} {452}},\ \bibinfo {pages} {975}
  (\bibinfo {year} {2008})}\BibitemShut {NoStop}%
\bibitem [{\citenamefont {Reznik}\ \emph {et~al.}(2008)\citenamefont {Reznik},
  \citenamefont {Sangiovanni}, \citenamefont {Gunnarsson},\ and\ \citenamefont
  {Devereaux}}]{reznik08}%
  \BibitemOpen
  \bibfield  {author} {\bibinfo {author} {\bibfnamefont {D.}~\bibnamefont
  {Reznik}}, \bibinfo {author} {\bibfnamefont {G.}~\bibnamefont {Sangiovanni}},
  \bibinfo {author} {\bibfnamefont {O.}~\bibnamefont {Gunnarsson}},\ and\
  \bibinfo {author} {\bibfnamefont {T.~P.}\ \bibnamefont {Devereaux}},\
  }\href{https://doi.org/10.1038/nature07364} {\bibfield  {journal} {\bibinfo  {journal} {Nature}\
  }\textbf {\bibinfo {volume} {455}},\ \bibinfo {pages} {E6} (\bibinfo {year}
  {2008})}\BibitemShut {NoStop}%
\bibitem [{\citenamefont {Maksimov}\ \emph {et~al.}(2010)\citenamefont
  {Maksimov}, \citenamefont {Kuli\'c},\ and\ \citenamefont
  {Dolgov}}]{maksimov10}%
  \BibitemOpen
  \bibfield  {author} {\bibinfo {author} {\bibfnamefont {E.~G.}\ \bibnamefont
  {Maksimov}}, \bibinfo {author} {\bibfnamefont {M.~L.}\ \bibnamefont
  {Kuli\'c}},\ and\ \bibinfo {author} {\bibfnamefont {O.~V.}\ \bibnamefont
  {Dolgov}},\ }\href{https://doi.org/10.1155/2010/423725} {\bibfield  {journal} {\bibinfo  {journal}
  {Adv. Condens. Matter Phys.}\ }\textbf {\bibinfo {volume} {2010}},\
  \bibinfo {pages} {1} (\bibinfo {year} {2010})}\BibitemShut {NoStop}%
\bibitem [{\citenamefont {Johnston}\ \emph {et~al.}(2010)\citenamefont
  {Johnston}, \citenamefont {Vernay}, \citenamefont {Moritz}, \citenamefont
  {Shen}, \citenamefont {Nagaosa}, \citenamefont {Zaanen},\ and\ \citenamefont
  {Devereaux}}]{johnston10}%
  \BibitemOpen
  \bibfield  {author} {\bibinfo {author} {\bibfnamefont {S.}~\bibnamefont
  {Johnston}}, \bibinfo {author} {\bibfnamefont {F.}~\bibnamefont {Vernay}},
  \bibinfo {author} {\bibfnamefont {B.}~\bibnamefont {Moritz}}, \bibinfo
  {author} {\bibfnamefont {Z.-X.}\ \bibnamefont {Shen}}, \bibinfo {author}
  {\bibfnamefont {N.}~\bibnamefont {Nagaosa}}, \bibinfo {author} {\bibfnamefont
  {J.}~\bibnamefont {Zaanen}},\ and\ \bibinfo {author} {\bibfnamefont {T.~P.}\
  \bibnamefont {Devereaux}},\ }\bibfield  {title} {\bibinfo {title} {Systematic
  study of electron-phonon coupling to oxygen modes across the cuprates},\
  }\href {https://doi.org/10.1103/PhysRevB.82.064513} {\bibfield  {journal}
  {\bibinfo  {journal} {Phys. Rev. B}\ }\textbf {\bibinfo {volume} {82}},\
  \bibinfo {pages} {064513} (\bibinfo {year} {2010})}\BibitemShut {NoStop}%
\bibitem [{\citenamefont {Ruvalds}(1987)}]{ruvalds87}%
  \BibitemOpen
  \bibfield  {author} {\bibinfo {author} {\bibfnamefont {J.}~\bibnamefont
  {Ruvalds}},\ }\bibfield  {title} {\bibinfo {title} {Plasmons and
  high-temperature superconductivity in alloys of copper oxides},\ }\href
  {https://doi.org/10.1103/PhysRevB.35.8869} {\bibfield  {journal} {\bibinfo
  {journal} {Phys. Rev. B}\ }\textbf {\bibinfo {volume} {35}},\ \bibinfo
  {pages} {8869} (\bibinfo {year} {1987})}\BibitemShut {NoStop}%
\bibitem [{\citenamefont {Kresin}\ and\ \citenamefont
  {Morawitz}(1988)}]{kresin88}%
  \BibitemOpen
  \bibfield  {author} {\bibinfo {author} {\bibfnamefont {V.~Z.}\ \bibnamefont
  {Kresin}}\ and\ \bibinfo {author} {\bibfnamefont {H.}~\bibnamefont
  {Morawitz}},\ }\bibfield  {title} {\bibinfo {title} {Layer plasmons and
  high-${T}_{c}$ superconductivity},\ }\href
  {https://link.aps.org/doi/10.1103/PhysRevB.37.7854} {\bibfield  {journal}
  {\bibinfo  {journal} {Phys. Rev. B}\ }\textbf {\bibinfo {volume} {37}},\
  \bibinfo {pages} {7854} (\bibinfo {year} {1988})}\BibitemShut {NoStop}%
\bibitem [{\citenamefont {Cui}\ and\ \citenamefont {Tsai}(1991)}]{cui91}%
  \BibitemOpen
  \bibfield  {author} {\bibinfo {author} {\bibfnamefont {S.-M.}\ \bibnamefont
  {Cui}}\ and\ \bibinfo {author} {\bibfnamefont {C.-H.}\ \bibnamefont {Tsai}},\
  }\bibfield  {title} {\bibinfo {title} {Plasmon theory of
  high-${\mathit{T}}_{\mathit{c}}$ superconductivity},\ }\href
  {https://doi.org/10.1103/PhysRevB.44.12500} {\bibfield  {journal} {\bibinfo
  {journal} {Phys. Rev. B}\ }\textbf {\bibinfo {volume} {44}},\ \bibinfo
  {pages} {12500} (\bibinfo {year} {1991})}\BibitemShut {NoStop}%
\bibitem [{\citenamefont {Ishii}\ and\ \citenamefont
  {Ruvalds}(1993)}]{ishii93}%
  \BibitemOpen
  \bibfield  {author} {\bibinfo {author} {\bibfnamefont {Y.}~\bibnamefont
  {Ishii}}\ and\ \bibinfo {author} {\bibfnamefont {J.}~\bibnamefont
  {Ruvalds}},\ }\bibfield  {title} {\bibinfo {title} {Acoustic plasmons and
  cuprate superconductivity},\ }\href
  {https://doi.org/10.1103/PhysRevB.48.3455} {\bibfield  {journal} {\bibinfo
  {journal} {Phys. Rev. B}\ }\textbf {\bibinfo {volume} {48}},\ \bibinfo
  {pages} {3455} (\bibinfo {year} {1993})}\BibitemShut {NoStop}%
\bibitem [{\citenamefont {Malozovsky}\ \emph {et~al.}(1993)\citenamefont
  {Malozovsky}, \citenamefont {Bose}, \citenamefont {Longe},\ and\
  \citenamefont {Fan}}]{malozovsky93}%
  \BibitemOpen
  \bibfield  {author} {\bibinfo {author} {\bibfnamefont {Y.~M.}\ \bibnamefont
  {Malozovsky}}, \bibinfo {author} {\bibfnamefont {S.~M.}\ \bibnamefont
  {Bose}}, \bibinfo {author} {\bibfnamefont {P.}~\bibnamefont {Longe}},\ and\
  \bibinfo {author} {\bibfnamefont {J.~D.}\ \bibnamefont {Fan}},\ }\bibfield
  {title} {\bibinfo {title} {Eliashberg equations and superconductivity in a
  layered two-dimensional metal},\ }\href
  {https://doi.org/10.1103/PhysRevB.48.10504} {\bibfield  {journal} {\bibinfo
  {journal} {Phys. Rev. B}\ }\textbf {\bibinfo {volume} {48}},\ \bibinfo
  {pages} {10504} (\bibinfo {year} {1993})}\BibitemShut {NoStop}%
\bibitem [{\citenamefont {Varshney}\ and\ \citenamefont
  {Singh}(1995)}]{varshney95}%
  \BibitemOpen
  \bibfield  {author} {\bibinfo {author} {\bibfnamefont {D.}~\bibnamefont
  {Varshney}}\ and\ \bibinfo {author} {\bibfnamefont {R.~K.}\ \bibnamefont
  {Singh}},\ }\bibfield  {title} {\bibinfo {title} {Superconductivity in
  lanthanum cuprates: A layered-electron-gas model},\ }\href
  {https://doi.org/10.1103/PhysRevB.52.7629} {\bibfield  {journal} {\bibinfo
  {journal} {Phys. Rev. B}\ }\textbf {\bibinfo {volume} {52}},\ \bibinfo
  {pages} {7629} (\bibinfo {year} {1995})}\BibitemShut {NoStop}%
\bibitem [{\citenamefont {Pashitskii}\ and\ \citenamefont
  {Pentegov}(2008)}]{pashitskii08}%
  \BibitemOpen
  \bibfield  {author} {\bibinfo {author} {\bibfnamefont {E.}~\bibnamefont
  {Pashitskii}}\ and\ \bibinfo {author} {\bibfnamefont {V.}~\bibnamefont
  {Pentegov}},\ }\href{https://doi.org/10.1063/1.2834256} {\bibfield  {journal} {\bibinfo  {journal} {J.
  Low Temp. Phys.}\ }\textbf {\bibinfo {volume} {34}},\ \bibinfo {pages} {113}
  (\bibinfo {year} {2008})}\BibitemShut {NoStop}%
\bibitem [{\citenamefont {Falter}\ and\ \citenamefont
  {Klenner}(1994)}]{falter94}%
  \BibitemOpen
  \bibfield  {author} {\bibinfo {author} {\bibfnamefont {C.}~\bibnamefont
  {Falter}}\ and\ \bibinfo {author} {\bibfnamefont {M.}~\bibnamefont
  {Klenner}},\ }\bibfield  {title} {\bibinfo {title} {Nonadiabatic and nonlocal
  electron-phonon interaction and phonon-plasmon mixing in the high-temperature
  superconductors},\ }\href {https://doi.org/10.1103/PhysRevB.50.9426}
  {\bibfield  {journal} {\bibinfo  {journal} {Phys. Rev. B}\ }\textbf {\bibinfo
  {volume} {50}},\ \bibinfo {pages} {9426} (\bibinfo {year}
  {1994})}\BibitemShut {NoStop}%
\bibitem [{\citenamefont {Bauer}\ and\ \citenamefont {Falter}(2009)}]{bauer09}%
  \BibitemOpen
  \bibfield  {author} {\bibinfo {author} {\bibfnamefont {T.}~\bibnamefont
  {Bauer}}\ and\ \bibinfo {author} {\bibfnamefont {C.}~\bibnamefont {Falter}},\
  }\bibfield  {title} {\bibinfo {title} {Impact of dynamical screening on the
  phonon dynamics of metallic La$_{2}$CuO$_{4}$},\ }\href
  {https://doi.org/10.1103/PhysRevB.80.094525} {\bibfield  {journal} {\bibinfo
  {journal} {Phys. Rev. B}\ }\textbf {\bibinfo {volume} {80}},\ \bibinfo
  {pages} {094525} (\bibinfo {year} {2009})}\BibitemShut {NoStop}%
\bibitem [{\citenamefont {Bill}\ \emph {et~al.}(2003)\citenamefont {Bill},
  \citenamefont {Morawitz},\ and\ \citenamefont {Kresin}}]{bill03}%
  \BibitemOpen
  \bibfield  {author} {\bibinfo {author} {\bibfnamefont {A.}~\bibnamefont
  {Bill}}, \bibinfo {author} {\bibfnamefont {H.}~\bibnamefont {Morawitz}},\
  and\ \bibinfo {author} {\bibfnamefont {V.~Z.}\ \bibnamefont {Kresin}},\
  }\bibfield  {title} {\bibinfo {title} {Electronic collective modes and
  superconductivity in layered conductors},\ }\href
  {https://link.aps.org/doi/10.1103/PhysRevB.68.144519} {\bibfield  {journal}
  {\bibinfo  {journal} {Phys. Rev. B}\ }\textbf {\bibinfo {volume} {68}},\
  \bibinfo {pages} {144519} (\bibinfo {year} {2003})}\BibitemShut {NoStop}%
\bibitem [{\citenamefont {Anderson}(1992)}]{anderson92}%
  \BibitemOpen
  \bibfield  {author} {\bibinfo {author} {\bibfnamefont {P.~W.}\ \bibnamefont
  {Anderson}},\ }\bibfield  {title} {\bibinfo {title} {Experimental constraints
  on the theory of high-$T_c$ superconductivity},\ }\href
  {http://www.jstor.org/stable/2876898} {\bibfield  {journal} {\bibinfo
  {journal} {Science}\ }\textbf {\bibinfo {volume} {256}},\ \bibinfo {pages}
  {1526} (\bibinfo {year} {1992})}\BibitemShut {NoStop}%
\bibitem [{\citenamefont {Chakravarty}\ \emph {et~al.}(1993)\citenamefont
  {Chakravarty}, \citenamefont {Sudb{\o}}, \citenamefont {Anderson},\ and\
  \citenamefont {Strong}}]{chakravarty93}%
  \BibitemOpen
  \bibfield  {author} {\bibinfo {author} {\bibfnamefont {S.}~\bibnamefont
  {Chakravarty}}, \bibinfo {author} {\bibfnamefont {A.}~\bibnamefont
  {Sudb{\o}}}, \bibinfo {author} {\bibfnamefont {P.~W.}\ \bibnamefont
  {Anderson}},\ and\ \bibinfo {author} {\bibfnamefont {S.}~\bibnamefont
  {Strong}},\ }\bibfield  {title} {\bibinfo {title} {Interlayer tunneling and
  gap anisotropy in high-temperature superconductors},\ }\href
  {https://doi.org/10.1126/science.261.5119.337} {\bibfield  {journal}
  {\bibinfo  {journal} {Science}\ }\textbf {\bibinfo {volume} {261}},\ \bibinfo
  {pages} {337} (\bibinfo {year} {1993})}\BibitemShut {NoStop}%
\bibitem [{\citenamefont {Anderson}(1995)}]{anderson95}%
  \BibitemOpen
  \bibfield  {author} {\bibinfo {author} {\bibfnamefont {P.~W.}\ \bibnamefont
  {Anderson}},\ }\bibfield  {title} {\bibinfo {title} {Interlayer tunneling
  mechanism for high-tc superconductivity: Comparison with c axis infrared
  experiments},\ }\href {https://doi.org/10.1126/science.268.5214.1154}
  {\bibfield  {journal} {\bibinfo  {journal} {Science}\ }\textbf {\bibinfo
  {volume} {268}},\ \bibinfo {pages} {1154} (\bibinfo {year}
  {1995})}\BibitemShut {NoStop}%
\bibitem [{\citenamefont {Leggett}(1996)}]{leggett96}%
  \BibitemOpen
  \bibfield  {author} {\bibinfo {author} {\bibfnamefont {A.~J.}\ \bibnamefont
  {Leggett}},\ }\bibfield  {title} {\bibinfo {title} {Interlayer tunneling
  models of cuprate superconductivity: Implications of a recent experiment},\
  }\href {https://doi.org/10.1126/science.274.5287.587} {\bibfield  {journal}
  {\bibinfo  {journal} {Science}\ }\textbf {\bibinfo {volume} {274}},\ \bibinfo
  {pages} {587} (\bibinfo {year} {1996})}\BibitemShut {NoStop}%
\bibitem [{\citenamefont {Tsvetkov}\ and\ \citenamefont
  {et~al}(1998)}]{tsvetkov98}%
   \BibitemOpen
  \bibfield  {author} {\bibinfo {author} {\bibfnamefont {A.~A.}\ \bibnamefont
  {Tsvetkov}}, \bibinfo {author} {\bibfnamefont {D.}~\bibnamefont {van~der
  Marel}}, \bibinfo {author} {\bibfnamefont {K.~A.}\ \bibnamefont {Moler}},
  \bibinfo {author} {\bibfnamefont {J.~R.}\ \bibnamefont {Kirtley}}, \bibinfo
  {author} {\bibfnamefont {J.~L.}\ \bibnamefont {de~Boer}}, \bibinfo {author}
  {\bibfnamefont {A.}~\bibnamefont {Meetsma}}, \bibinfo {author} {\bibfnamefont
  {Z.~F.}\ \bibnamefont {Ren}}, \bibinfo {author} {\bibfnamefont
  {N.}~\bibnamefont {Koleshnikov}}, \bibinfo {author} {\bibfnamefont
  {D.}~\bibnamefont {Dulic}}, \bibinfo {author} {\bibfnamefont
  {A.}~\bibnamefont {Damascelli}}, \bibinfo {author} {\bibfnamefont
  {M.}~\bibnamefont {Gr{\"u}ninger}}, \bibinfo {author} {\bibfnamefont
  {J.}~\bibnamefont {Sch{\"u}tzmann}}, \bibinfo {author} {\bibfnamefont
  {J.~W.}\ \bibnamefont {van~der Eb}}, \bibinfo {author} {\bibfnamefont
  {H.~S.}\ \bibnamefont {Somal}}, \ and\ \bibinfo {author} {\bibfnamefont
  {J.~H.}\ \bibnamefont {Wang}},\ }
  \bibfield  {title} {\bibinfo
  {title} {Global and local measures of the intrinsic Josephson coupling in Tl$_2$Ba$_2$CuO$_6$ as a test of the interlayer tunnelling model},\ }
  \href {\doibase 10.1038/26439} {\bibfield
  {journal} {\bibinfo  {journal} {Nature}\ }\textbf {\bibinfo {volume} {395}},\
  \bibinfo {pages} {360} (\bibinfo {year} {1998})}\BibitemShut {NoStop}%
\bibitem [{\citenamefont {Moler}\ \emph {et~al.}(1998)\citenamefont {Moler},
  \citenamefont {Kirtley}, \citenamefont {Hinks}, \citenamefont {Li},\ and\
  \citenamefont {Xu}}]{moler98}%
  \BibitemOpen
  \bibfield  {author} {\bibinfo {author} {\bibfnamefont {K.~A.}\ \bibnamefont
  {Moler}}, \bibinfo {author} {\bibfnamefont {J.~R.}\ \bibnamefont {Kirtley}},
  \bibinfo {author} {\bibfnamefont {D.~G.}\ \bibnamefont {Hinks}}, \bibinfo
  {author} {\bibfnamefont {T.~W.}\ \bibnamefont {Li}},\ and\ \bibinfo {author}
  {\bibfnamefont {M.}~\bibnamefont {Xu}},\ }\bibfield  {title} {\bibinfo
  {title} {Images of interlayer josephson vortices in Tl$_2$Ba$_2$CuO$_{6+\delta}$},\ }\href
  {https://doi.org/10.1126/science.279.5354.1193} {\bibfield  {journal}
  {\bibinfo  {journal} {Science}\ }\textbf {\bibinfo {volume} {279}},\ \bibinfo
  {pages} {1193} (\bibinfo {year} {1998})}\BibitemShut {NoStop}%
\bibitem [{\citenamefont {Anderson}(1998)}]{anderson98}%
  \BibitemOpen
  \bibfield  {author} {\bibinfo {author} {\bibfnamefont {P.~W.}\ \bibnamefont
  {Anderson}},\ }\bibfield  {title} {\bibinfo {title} {c-axis electrodynamics
  as evidence for the interlayer theory of high-temperature
  superconductivity},\ }\href {https://doi.org/10.1126/science.279.5354.1196}
  {\bibfield  {journal} {\bibinfo  {journal} {Science}\ }\textbf {\bibinfo
  {volume} {279}},\ \bibinfo {pages} {1196} (\bibinfo {year}
  {1998})}\BibitemShut {NoStop}%
\bibitem [{\citenamefont {Kirtley}\ \emph {et~al.}(1998)\citenamefont
  {Kirtley}, \citenamefont {Moler}, \citenamefont {Villard},\ and\
  \citenamefont {Maignan}}]{kirtley98}%
  \BibitemOpen
  \bibfield  {author} {\bibinfo {author} {\bibfnamefont {J.~R.}\ \bibnamefont
  {Kirtley}}, \bibinfo {author} {\bibfnamefont {K.~A.}\ \bibnamefont {Moler}},
  \bibinfo {author} {\bibfnamefont {G.}~\bibnamefont {Villard}},\ and\ \bibinfo
  {author} {\bibfnamefont {A.}~\bibnamefont {Maignan}},\ }\bibfield  {title}
  {\bibinfo {title} {$\mathit{c}$-axis penetration depth of Hg-1201 single
  crystals},\ }\href {https://doi.org/10.1103/PhysRevLett.81.2140} {\bibfield
  {journal} {\bibinfo  {journal} {Phys. Rev. Lett.}\ }\textbf {\bibinfo
  {volume} {81}},\ \bibinfo {pages} {2140} (\bibinfo {year}
  {1998})}\BibitemShut {NoStop}%
\bibitem [{\citenamefont {Chakravarty}\ \emph {et~al.}(1999)\citenamefont
  {Chakravarty}, \citenamefont {Kee},\ and\ \citenamefont
  {Abrahams}}]{chakravary99}%
  \BibitemOpen
  \bibfield  {author} {\bibinfo {author} {\bibfnamefont {S.}~\bibnamefont
  {Chakravarty}}, \bibinfo {author} {\bibfnamefont {H.-Y.}\ \bibnamefont
  {Kee}},\ and\ \bibinfo {author} {\bibfnamefont {E.}~\bibnamefont
  {Abrahams}},\ }\bibfield  {title} {\bibinfo {title} {Frustrated kinetic
  energy, the optical sum rule, and the mechanism of superconductivity},\
  }\href {https://doi.org/10.1103/PhysRevLett.82.2366} {\bibfield  {journal}
  {\bibinfo  {journal} {Phys. Rev. Lett.}\ }\textbf {\bibinfo {volume} {82}},\
  \bibinfo {pages} {2366} (\bibinfo {year} {1999})}\BibitemShut {NoStop}%
\bibitem [{\citenamefont {Basov}\ \emph {et~al.}(1999)\citenamefont {Basov},
  \citenamefont {Woods}, \citenamefont {Katz}, \citenamefont {Singley},
  \citenamefont {Dynes}, \citenamefont {Xu}, \citenamefont {Hinks},
  \citenamefont {Homes},\ and\ \citenamefont {Strongin}}]{basov99}%
  \BibitemOpen
  \bibfield  {author} {\bibinfo {author} {\bibfnamefont {D.~N.}\ \bibnamefont
  {Basov}}, \bibinfo {author} {\bibfnamefont {S.~I.}\ \bibnamefont {Woods}},
  \bibinfo {author} {\bibfnamefont {A.~S.}\ \bibnamefont {Katz}}, \bibinfo
  {author} {\bibfnamefont {E.~J.}\ \bibnamefont {Singley}}, \bibinfo {author}
  {\bibfnamefont {R.~C.}\ \bibnamefont {Dynes}}, \bibinfo {author}
  {\bibfnamefont {M.}~\bibnamefont {Xu}}, \bibinfo {author} {\bibfnamefont
  {D.~G.}\ \bibnamefont {Hinks}}, \bibinfo {author} {\bibfnamefont {C.~C.}\
  \bibnamefont {Homes}},\ and\ \bibinfo {author} {\bibfnamefont
  {M.}~\bibnamefont {Strongin}},\ }\bibfield  {title} {\bibinfo {title} {Sum
  rules and interlayer conductivity of high-Tc cuprates},\ }\href
  {https://doi.org/10.1126/science.283.5398.49} {\bibfield  {journal} {\bibinfo
   {journal} {Science}\ }\textbf {\bibinfo {volume} {283}},\ \bibinfo {pages}
  {49} (\bibinfo {year} {1999})}\BibitemShut {NoStop}%
\bibitem [{\citenamefont {Uchida}\ \emph {et~al.}(1996)\citenamefont {Uchida},
  \citenamefont {Tamasaku},\ and\ \citenamefont {Tajima}}]{uchida96}%
  \BibitemOpen
  \bibfield  {author} {\bibinfo {author} {\bibfnamefont {S.}~\bibnamefont
  {Uchida}}, \bibinfo {author} {\bibfnamefont {K.}~\bibnamefont {Tamasaku}},\
  and\ \bibinfo {author} {\bibfnamefont {S.}~\bibnamefont {Tajima}},\
  }\bibfield  {title} {\bibinfo {title} {$c$-axis optical spectra and charge
  dynamics in
  La$_{2-x}$Sr$_{x}$CuO$_{4}$},\
  }\href {https://doi.org/10.1103/PhysRevB.53.14558} {\bibfield  {journal}
  {\bibinfo  {journal} {Phys. Rev. B}\ }\textbf {\bibinfo {volume} {53}},\
  \bibinfo {pages} {14558} (\bibinfo {year} {1996})}\BibitemShut {NoStop}%
\bibitem [{\citenamefont {Hussey}\ \emph {et~al.}(2003)\citenamefont {Hussey},
  \citenamefont {Abdel-Jawad}, \citenamefont {Carrington}, \citenamefont
  {Mackenzie},\ and\ \citenamefont {Balicas}}]{hussey03}%
  \BibitemOpen
  \bibfield  {author} {\bibinfo {author} {\bibfnamefont {N.~E.}\ \bibnamefont
  {Hussey}}, \bibinfo {author} {\bibfnamefont {M.}~\bibnamefont {Abdel-Jawad}},
  \bibinfo {author} {\bibfnamefont {A.}~\bibnamefont {Carrington}}, \bibinfo
  {author} {\bibfnamefont {A.~P.}\ \bibnamefont {Mackenzie}},\ and\ \bibinfo
  {author} {\bibfnamefont {L.~A.}\ \bibnamefont {Balicas}},\ }\href {https://doi.org/10.1038/nature01981}
  {\bibfield  {journal} {\bibinfo  {journal} {Nature}\ }\textbf {\bibinfo
  {volume} {425}},\ \bibinfo {pages} {814} (\bibinfo {year}
  {2003})}\BibitemShut {NoStop}%
\bibitem [{\citenamefont {Homes}\ \emph {et~al.}(2005)\citenamefont {Homes},
  \citenamefont {Dordevic}, \citenamefont {Bonn}, \citenamefont {Liang},
  \citenamefont {Hardy},\ and\ \citenamefont {Timusk}}]{homes05}%
  \BibitemOpen
  \bibfield  {author} {\bibinfo {author} {\bibfnamefont {C.~C.}\ \bibnamefont
  {Homes}}, \bibinfo {author} {\bibfnamefont {S.~V.}\ \bibnamefont {Dordevic}},
  \bibinfo {author} {\bibfnamefont {D.~A.}\ \bibnamefont {Bonn}}, \bibinfo
  {author} {\bibfnamefont {R.}~\bibnamefont {Liang}}, \bibinfo {author}
  {\bibfnamefont {W.~N.}\ \bibnamefont {Hardy}},\ and\ \bibinfo {author}
  {\bibfnamefont {T.}~\bibnamefont {Timusk}},\ }\bibfield  {title} {\bibinfo
  {title} {Coherence, incoherence, and scaling along the $c$ axis of
  YBa$_{2}$Cu$_{3}$O$_{6+x}$},\ }\href
  {https://doi.org/10.1103/PhysRevB.71.184515} {\bibfield  {journal} {\bibinfo
  {journal} {Phys. Rev. B}\ }\textbf {\bibinfo {volume} {71}},\ \bibinfo
  {pages} {184515} (\bibinfo {year} {2005})}\BibitemShut {NoStop}%
\bibitem [{\citenamefont {Yamase}\ \emph {et~al.}(2021)\citenamefont {Yamase},
  \citenamefont {Sakurai}, \citenamefont {Fujita}, \citenamefont {Wakimoto},\
  and\ \citenamefont {Yamada}}]{yamase21b}%
  \BibitemOpen
  \bibfield  {author} {\bibinfo {author} {\bibfnamefont {H.}~\bibnamefont
  {Yamase}}, \bibinfo {author} {\bibfnamefont {Y.}~\bibnamefont {Sakurai}},
  \bibinfo {author} {\bibfnamefont {M.}~\bibnamefont {Fujita}}, \bibinfo
  {author} {\bibfnamefont {S.}~\bibnamefont {Wakimoto}},\ and\ \bibinfo
  {author} {\bibfnamefont {K.}~\bibnamefont {Yamada}},\ }\href {https://doi.org/10.1038/s41467-021-22229-6}
  {\bibfield  {journal} {\bibinfo  {journal} {Nat. Commun.}\ }\textbf {\bibinfo
  {volume} {12}},\ \bibinfo {pages} {2223} (\bibinfo {year}
  {2021})}\BibitemShut {NoStop}%
\bibitem [{\citenamefont {Takeuchi}\ \emph {et~al.}(2005)\citenamefont
  {Takeuchi}, \citenamefont {Kondo}, \citenamefont {Kitao}, \citenamefont
  {Kaga}, \citenamefont {Yang}, \citenamefont {Ding}, \citenamefont
  {Kaminski},\ and\ \citenamefont {Campuzano}}]{takeuchi05}%
  \BibitemOpen
  \bibfield  {author} {\bibinfo {author} {\bibfnamefont {T.}~\bibnamefont
  {Takeuchi}}, \bibinfo {author} {\bibfnamefont {T.}~\bibnamefont {Kondo}},
  \bibinfo {author} {\bibfnamefont {T.}~\bibnamefont {Kitao}}, \bibinfo
  {author} {\bibfnamefont {H.}~\bibnamefont {Kaga}}, \bibinfo {author}
  {\bibfnamefont {H.}~\bibnamefont {Yang}}, \bibinfo {author} {\bibfnamefont
  {H.}~\bibnamefont {Ding}}, \bibinfo {author} {\bibfnamefont {A.}~\bibnamefont
  {Kaminski}},\ and\ \bibinfo {author} {\bibfnamefont {J.~C.}\ \bibnamefont
  {Campuzano}},\ }\bibfield  {title} {\bibinfo {title} {Two- to
  three-dimensional crossover in the electronic structure of
  $(\mathrm{Bi},\mathrm{Pb}{)}_{2}(\mathrm{Sr},\mathrm{La}{)}_{2}{\mathrm{CuO}}_{6+\delta}$
  from angle-resolved photoemission spectroscopy},\ }\href
  {https://doi.org/10.1103/PhysRevLett.95.227004} {\bibfield  {journal}
  {\bibinfo  {journal} {Phys. Rev. Lett.}\ }\textbf {\bibinfo {volume} {95}},\
  \bibinfo {pages} {227004} (\bibinfo {year} {2005})}\BibitemShut {NoStop}%
\bibitem [{\citenamefont {Horio}\ \emph {et~al.}(2018)\citenamefont {Horio},
  \citenamefont {Hauser}, \citenamefont {Sassa}, \citenamefont {Mingazheva},
  \citenamefont {Sutter}, \citenamefont {Kramer}, \citenamefont {Cook},
  \citenamefont {Nocerino}, \citenamefont {Forslund}, \citenamefont
  {Tjernberg}, \citenamefont {Kobayashi}, \citenamefont {Chikina},
  \citenamefont {Schr\"oter}, \citenamefont {Krieger}, \citenamefont {Schmitt},
  \citenamefont {Strocov}, \citenamefont {Pyon}, \citenamefont {Takayama},
  \citenamefont {Takagi}, \citenamefont {Lipscombe}, \citenamefont {Hayden},
  \citenamefont {Ishikado}, \citenamefont {Eisaki}, \citenamefont {Neupert},
  \citenamefont {M\aa{}nsson}, \citenamefont {Matt},\ and\ \citenamefont
  {Chang}}]{horio18}%
  \BibitemOpen
  \bibfield  {author} {\bibinfo {author} {\bibfnamefont {M.}~\bibnamefont
  {Horio}}, \bibinfo {author} {\bibfnamefont {K.}~\bibnamefont {Hauser}},
  \bibinfo {author} {\bibfnamefont {Y.}~\bibnamefont {Sassa}}, \bibinfo
  {author} {\bibfnamefont {Z.}~\bibnamefont {Mingazheva}}, \bibinfo {author}
  {\bibfnamefont {D.}~\bibnamefont {Sutter}}, \bibinfo {author} {\bibfnamefont
  {K.}~\bibnamefont {Kramer}}, \bibinfo {author} {\bibfnamefont
  {A.}~\bibnamefont {Cook}}, \bibinfo {author} {\bibfnamefont {E.}~\bibnamefont
  {Nocerino}}, \bibinfo {author} {\bibfnamefont {O.~K.}\ \bibnamefont
  {Forslund}}, \bibinfo {author} {\bibfnamefont {O.}~\bibnamefont {Tjernberg}},
  \bibinfo {author} {\bibfnamefont {M.}~\bibnamefont {Kobayashi}}, \bibinfo
  {author} {\bibfnamefont {A.}~\bibnamefont {Chikina}}, \bibinfo {author}
  {\bibfnamefont {N.~B.~M.}\ \bibnamefont {Schr\"oter}}, \bibinfo {author}
  {\bibfnamefont {J.~A.}\ \bibnamefont {Krieger}}, \bibinfo {author}
  {\bibfnamefont {T.}~\bibnamefont {Schmitt}}, \bibinfo {author} {\bibfnamefont
  {V.~N.}\ \bibnamefont {Strocov}}, \bibinfo {author} {\bibfnamefont
  {S.}~\bibnamefont {Pyon}}, \bibinfo {author} {\bibfnamefont {T.}~\bibnamefont
  {Takayama}}, \bibinfo {author} {\bibfnamefont {H.}~\bibnamefont {Takagi}},
  \bibinfo {author} {\bibfnamefont {O.~J.}\ \bibnamefont {Lipscombe}}, \bibinfo
  {author} {\bibfnamefont {S.~M.}\ \bibnamefont {Hayden}}, \bibinfo {author}
  {\bibfnamefont {M.}~\bibnamefont {Ishikado}}, \bibinfo {author}
  {\bibfnamefont {H.}~\bibnamefont {Eisaki}}, \bibinfo {author} {\bibfnamefont
  {T.}~\bibnamefont {Neupert}}, \bibinfo {author} {\bibfnamefont
  {M.}~\bibnamefont {M\aa{}nsson}}, \bibinfo {author} {\bibfnamefont {C.~E.}\
  \bibnamefont {Matt}},\ and\ \bibinfo {author} {\bibfnamefont
  {J.}~\bibnamefont {Chang}},\ }\bibfield  {title} {\bibinfo {title}
  {Three-dimensional fermi surface of overdoped La-based cuprates},\ }\href
  {https://doi.org/10.1103/PhysRevLett.121.077004} {\bibfield  {journal}
  {\bibinfo  {journal} {Phys. Rev. Lett.}\ }\textbf {\bibinfo {volume} {121}},\
  \bibinfo {pages} {077004} (\bibinfo {year} {2018})}\BibitemShut {NoStop}%
\bibitem [{\citenamefont {Matt}\ \emph {et~al.}(2018)\citenamefont {Matt},
  \citenamefont {Sutter}, \citenamefont {Cook}, \citenamefont {Sassa},
  \citenamefont {M{\aa}nsson}, \citenamefont {Tjernberg}, \citenamefont {Das},
  \citenamefont {Horio}, \citenamefont {Destraz}, \citenamefont {Fatuzzo},
  \citenamefont {Hauser}, \citenamefont {Shi}, \citenamefont {Kobayashi},
  \citenamefont {Strocov}, \citenamefont {Schmitt}, \citenamefont {Dudin},
  \citenamefont {Hoesch}, \citenamefont {Pyon}, \citenamefont {Takayama},
  \citenamefont {Takagi}, \citenamefont {Lipscombe}, \citenamefont {Hayden},
  \citenamefont {Kurosawa}, \citenamefont {Momono}, \citenamefont {Oda},
  \citenamefont {Neupert},\ and\ \citenamefont {Chang}}]{matt18}%
  \BibitemOpen
  \bibfield  {author} {\bibinfo {author} {\bibfnamefont {C.~E.}\ \bibnamefont
  {Matt}}, \bibinfo {author} {\bibfnamefont {D.}~\bibnamefont {Sutter}},
  \bibinfo {author} {\bibfnamefont {A.~M.}\ \bibnamefont {Cook}}, \bibinfo
  {author} {\bibfnamefont {Y.}~\bibnamefont {Sassa}}, \bibinfo {author}
  {\bibfnamefont {M.}~\bibnamefont {M{\aa}nsson}}, \bibinfo {author}
  {\bibfnamefont {O.}~\bibnamefont {Tjernberg}}, \bibinfo {author}
  {\bibfnamefont {L.}~\bibnamefont {Das}}, \bibinfo {author} {\bibfnamefont
  {M.}~\bibnamefont {Horio}}, \bibinfo {author} {\bibfnamefont
  {D.}~\bibnamefont {Destraz}}, \bibinfo {author} {\bibfnamefont {C.~G.}\
  \bibnamefont {Fatuzzo}}, \bibinfo {author} {\bibfnamefont {K.}~\bibnamefont
  {Hauser}}, \bibinfo {author} {\bibfnamefont {M.}~\bibnamefont {Shi}},
  \bibinfo {author} {\bibfnamefont {M.}~\bibnamefont {Kobayashi}}, \bibinfo
  {author} {\bibfnamefont {V.~N.}\ \bibnamefont {Strocov}}, \bibinfo {author}
  {\bibfnamefont {T.}~\bibnamefont {Schmitt}}, \bibinfo {author} {\bibfnamefont
  {P.}~\bibnamefont {Dudin}}, \bibinfo {author} {\bibfnamefont
  {M.}~\bibnamefont {Hoesch}}, \bibinfo {author} {\bibfnamefont
  {S.}~\bibnamefont {Pyon}}, \bibinfo {author} {\bibfnamefont {T.}~\bibnamefont
  {Takayama}}, \bibinfo {author} {\bibfnamefont {H.}~\bibnamefont {Takagi}},
  \bibinfo {author} {\bibfnamefont {O.~J.}\ \bibnamefont {Lipscombe}}, \bibinfo
  {author} {\bibfnamefont {S.~M.}\ \bibnamefont {Hayden}}, \bibinfo {author}
  {\bibfnamefont {T.}~\bibnamefont {Kurosawa}}, \bibinfo {author}
  {\bibfnamefont {N.}~\bibnamefont {Momono}}, \bibinfo {author} {\bibfnamefont
  {M.}~\bibnamefont {Oda}}, \bibinfo {author} {\bibfnamefont {T.}~\bibnamefont
  {Neupert}},\ and\ \bibinfo {author} {\bibfnamefont {J.}~\bibnamefont
  {Chang}},\ }\bibfield  {title} {\bibinfo {title} {Direct observation of
  orbital hybridisation in a cuprate superconductor},\ }\href
  {https://doi.org/10.1038/s41467-018-03266-0} {\bibfield  {journal} {\bibinfo
  {journal} {Nat. Commun.}\ }\textbf {\bibinfo {volume} {9}},\ \bibinfo {pages}
  {972} (\bibinfo {year} {2018})}\BibitemShut {NoStop}%
\bibitem [{\citenamefont {Zha}\ \emph {et~al.}(1996)\citenamefont {Zha},
  \citenamefont {Cooper},\ and\ \citenamefont {Pines}}]{zha96}%
  \BibitemOpen
  \bibfield  {author} {\bibinfo {author} {\bibfnamefont {Y.}~\bibnamefont
  {Zha}}, \bibinfo {author} {\bibfnamefont {S.~L.}\ \bibnamefont {Cooper}},\
  and\ \bibinfo {author} {\bibfnamefont {D.}~\bibnamefont {Pines}},\ }\bibfield
   {title} {\bibinfo {title} {Model of c-axis resistivity of
  high-T$_c$ cuprates},\ }\href
  {https://doi.org/10.1103/PhysRevB.53.8253} {\bibfield  {journal} {\bibinfo
  {journal} {Phys. Rev. B}\ }\textbf {\bibinfo {volume} {53}},\ \bibinfo
  {pages} {8253} (\bibinfo {year} {1996})}\BibitemShut {NoStop}%
\bibitem [{\citenamefont {Grissonnanche}\ \emph {et~al.}(2021)\citenamefont
  {Grissonnanche}, \citenamefont {Fang}, \citenamefont {Legros}, \citenamefont
  {Verret}, \citenamefont {Lalibert{\'{e}}}, \citenamefont {Collignon},
  \citenamefont {Zhou}, \citenamefont {Graf}, \citenamefont {Goddard},
  \citenamefont {Taillefer},\ and\ \citenamefont {Ramshaw}}]{grissonnanche21}%
  \BibitemOpen
  \bibfield  {author} {\bibinfo {author} {\bibfnamefont {G.}~\bibnamefont
  {Grissonnanche}}, \bibinfo {author} {\bibfnamefont {Y.}~\bibnamefont {Fang}},
  \bibinfo {author} {\bibfnamefont {A.}~\bibnamefont {Legros}}, \bibinfo
  {author} {\bibfnamefont {S.}~\bibnamefont {Verret}}, \bibinfo {author}
  {\bibfnamefont {F.}~\bibnamefont {Lalibert{\'{e}}}}, \bibinfo {author}
  {\bibfnamefont {C.}~\bibnamefont {Collignon}}, \bibinfo {author}
  {\bibfnamefont {J.}~\bibnamefont {Zhou}}, \bibinfo {author} {\bibfnamefont
  {D.}~\bibnamefont {Graf}}, \bibinfo {author} {\bibfnamefont {P.~A.}\
  \bibnamefont {Goddard}}, \bibinfo {author} {\bibfnamefont {L.}~\bibnamefont
  {Taillefer}},\ and\ \bibinfo {author} {\bibfnamefont {B.~J.}\ \bibnamefont
  {Ramshaw}},\ }\bibfield  {title} {\bibinfo {title} {{Linear-in temperature
  resistivity from an isotropic Planckian scattering rate}},\ }\href
  {https://doi.org/10.1038/s41586-021-03697-8} {\bibfield  {journal} {\bibinfo
  {journal} {Nature}\ }\textbf {\bibinfo {volume} {595}},\ \bibinfo {pages}
  {667} (\bibinfo {year} {2021})}\BibitemShut {NoStop}%
\bibitem [{\citenamefont {Andersen}\ \emph {et~al.}(1995)\citenamefont
  {Andersen}, \citenamefont {Liechtenstein}, \citenamefont {Jepsen},\ and\
  \citenamefont {Paulsen}}]{andersen95}%
  \BibitemOpen
  \bibfield  {author} {\bibinfo {author} {\bibfnamefont {O.}~\bibnamefont
  {Andersen}}, \bibinfo {author} {\bibfnamefont {A.}~\bibnamefont
  {Liechtenstein}}, \bibinfo {author} {\bibfnamefont {O.}~\bibnamefont
  {Jepsen}},\ and\ \bibinfo {author} {\bibfnamefont {F.}~\bibnamefont
  {Paulsen}},\ }\bibfield  {title} {\bibinfo {title} {Lda energy bands,
  low-energy hamiltonians, t', t'', t$_\perp$(k), and J$_\perp$},\ }\href
  {https://www.sciencedirect.com/science/article/pii/0022369795002693}
  {\bibfield  {journal} {\bibinfo  {journal} {J. Phys. Chem. Solids}\ }\textbf
  {\bibinfo {volume} {56}},\ \bibinfo {pages} {1573} (\bibinfo {year}
  {1995})}\BibitemShut {NoStop}%
\bibitem [{\citenamefont {Markiewicz}\ \emph {et~al.}(2005)\citenamefont
  {Markiewicz}, \citenamefont {Sahrakorpi}, \citenamefont {Lindroos},
  \citenamefont {Lin},\ and\ \citenamefont {Bansil}}]{markiewicz05}%
  \BibitemOpen
  \bibfield  {author} {\bibinfo {author} {\bibfnamefont {R.~S.}\ \bibnamefont
  {Markiewicz}}, \bibinfo {author} {\bibfnamefont {S.}~\bibnamefont
  {Sahrakorpi}}, \bibinfo {author} {\bibfnamefont {M.}~\bibnamefont
  {Lindroos}}, \bibinfo {author} {\bibfnamefont {H.}~\bibnamefont {Lin}},\ and\
  \bibinfo {author} {\bibfnamefont {A.}~\bibnamefont {Bansil}},\ }\bibfield
  {title} {\bibinfo {title} {One-band tight-binding model parametrization of
  the high-${T}_{c}$ cuprates including the effect of ${k}_{z}$ dispersion},\
  }\href {https://link.aps.org/doi/10.1103/PhysRevB.72.054519} {\bibfield
  {journal} {\bibinfo  {journal} {Phys. Rev. B}\ }\textbf {\bibinfo {volume}
  {72}},\ \bibinfo {pages} {054519} (\bibinfo {year} {2005})}\BibitemShut
  {NoStop}%
\bibitem [{\citenamefont {Greco}\ \emph {et~al.}(2016)\citenamefont {Greco},
  \citenamefont {Yamase},\ and\ \citenamefont {Bejas}}]{greco16}%
  \BibitemOpen
  \bibfield  {author} {\bibinfo {author} {\bibfnamefont {A.}~\bibnamefont
  {Greco}}, \bibinfo {author} {\bibfnamefont {H.}~\bibnamefont {Yamase}},\ and\
  \bibinfo {author} {\bibfnamefont {M.}~\bibnamefont {Bejas}},\ }\bibfield
  {title} {\bibinfo {title} {{Plasmon excitations in layered high-${T}_{c}$
  cuprates}},\ }\href {https://doi.org/10.1103/PhysRevB.94.075139} {\bibfield
  {journal} {\bibinfo  {journal} {Phys. Rev. B}\ }\textbf {\bibinfo {volume}
  {94}},\ \bibinfo {pages} {075139} (\bibinfo {year} {2016})}\BibitemShut
  {NoStop}%
\bibitem [{\citenamefont {Grecu}(1973)}]{grecu73}%
  \BibitemOpen
  \bibfield  {author} {\bibinfo {author} {\bibfnamefont {D.}~\bibnamefont
  {Grecu}},\ }\bibfield  {title} {\bibinfo {title} {Plasma frequency of the
  electron gas in layered structures},\ }\href
  {https://link.aps.org/doi/10.1103/PhysRevB.8.1958} {\bibfield  {journal}
  {\bibinfo  {journal} {Phys. Rev. B}\ }\textbf {\bibinfo {volume} {8}},\
  \bibinfo {pages} {1958} (\bibinfo {year} {1973})}\BibitemShut {NoStop}%
\bibitem [{\citenamefont {Fetter}(1974)}]{fetter74}%
  \BibitemOpen
  \bibfield  {author} {\bibinfo {author} {\bibfnamefont {A.~L.}\ \bibnamefont
  {Fetter}},\ }\bibfield  {title} {\bibinfo {title} {Electrodynamics of a
  layered electron gas. ii. periodic array},\ }\href
  {http://www.sciencedirect.com/science/article/pii/0003491674903972}
  {\bibfield  {journal} {\bibinfo  {journal} {Ann. Phys.}\ }\textbf
  {\bibinfo {volume} {88}},\ \bibinfo {pages} {1 } (\bibinfo {year}
  {1974})}\BibitemShut {NoStop}%
\bibitem [{\citenamefont {Grecu}(1975)}]{grecu75}%
  \BibitemOpen
  \bibfield  {author} {\bibinfo {author} {\bibfnamefont {D.}~\bibnamefont
  {Grecu}},\ }\bibfield  {title} {\bibinfo {title} {Self-consistent field
  approximation for the plasma frequencies of an electron gas in a layered thin
  film},\ }\href {https://doi.org/10.1088/0022-3719/8/16/014} {\bibfield
  {journal} {\bibinfo  {journal} {J. Phys. C: Solid State Phys.}\ }\textbf {\bibinfo
  {volume} {8}},\ \bibinfo {pages} {2627} (\bibinfo {year} {1975})}\BibitemShut
  {NoStop}%
\bibitem [{\citenamefont {Hepting}\ \emph {et~al.}(2018)\citenamefont
  {Hepting}, \citenamefont {Chaix}, \citenamefont {Huang}, \citenamefont
  {Fumagalli}, \citenamefont {Peng}, \citenamefont {Moritz}, \citenamefont
  {Kummer}, \citenamefont {Brookes}, \citenamefont {Lee}, \citenamefont
  {Hashimoto}, \citenamefont {Sarkar}, \citenamefont {He}, \citenamefont
  {Rotundu}, \citenamefont {Lee}, \citenamefont {Greene}, \citenamefont
  {Braicovich}, \citenamefont {Ghiringhelli}, \citenamefont {Shen},
  \citenamefont {Devereaux},\ and\ \citenamefont {Lee}}]{hepting18}%
  \BibitemOpen
  \bibfield  {author} {\bibinfo {author} {\bibfnamefont {M.}~\bibnamefont
  {Hepting}}, \bibinfo {author} {\bibfnamefont {L.}~\bibnamefont {Chaix}},
  \bibinfo {author} {\bibfnamefont {E.~W.}\ \bibnamefont {Huang}}, \bibinfo
  {author} {\bibfnamefont {R.}~\bibnamefont {Fumagalli}}, \bibinfo {author}
  {\bibfnamefont {Y.~Y.}\ \bibnamefont {Peng}}, \bibinfo {author}
  {\bibfnamefont {B.}~\bibnamefont {Moritz}}, \bibinfo {author} {\bibfnamefont
  {K.}~\bibnamefont {Kummer}}, \bibinfo {author} {\bibfnamefont {N.~B.}\
  \bibnamefont {Brookes}}, \bibinfo {author} {\bibfnamefont {W.~C.}\
  \bibnamefont {Lee}}, \bibinfo {author} {\bibfnamefont {M.}~\bibnamefont
  {Hashimoto}}, \bibinfo {author} {\bibfnamefont {T.}~\bibnamefont {Sarkar}},
  \bibinfo {author} {\bibfnamefont {J.-F.}\ \bibnamefont {He}}, \bibinfo
  {author} {\bibfnamefont {C.~R.}\ \bibnamefont {Rotundu}}, \bibinfo {author}
  {\bibfnamefont {Y.~S.}\ \bibnamefont {Lee}}, \bibinfo {author} {\bibfnamefont
  {R.~L.}\ \bibnamefont {Greene}}, \bibinfo {author} {\bibfnamefont
  {L.}~\bibnamefont {Braicovich}}, \bibinfo {author} {\bibfnamefont
  {G.}~\bibnamefont {Ghiringhelli}}, \bibinfo {author} {\bibfnamefont {Z.~X.}\
  \bibnamefont {Shen}}, \bibinfo {author} {\bibfnamefont {T.~P.}\ \bibnamefont
  {Devereaux}},\ and\ \bibinfo {author} {\bibfnamefont {W.~S.}\ \bibnamefont
  {Lee}},\ }\bibfield  {title} {\bibinfo {title} {Three-dimensional collective
  charge excitations in electron-doped copper oxide superconductors},\ }\href
  {https://doi.org/10.1038/s41586-018-0648-3} {\bibfield  {journal} {\bibinfo
  {journal} {Nature}\ }\textbf {\bibinfo {volume} {563}},\ \bibinfo {pages}
  {374} (\bibinfo {year} {2018})}\BibitemShut {NoStop}%
\bibitem [{\citenamefont {Lin}\ \emph {et~al.}(2020)\citenamefont {Lin},
  \citenamefont {Yuan}, \citenamefont {Jin}, \citenamefont {Yin}, \citenamefont
  {Li}, \citenamefont {Zhou}, \citenamefont {Lu}, \citenamefont {Dantz},
  \citenamefont {Schmitt}, \citenamefont {Ding}, \citenamefont {Guo},
  \citenamefont {Dean},\ and\ \citenamefont {Liu}}]{jlin20}%
  \BibitemOpen
  \bibfield  {author} {\bibinfo {author} {\bibfnamefont {J.}~\bibnamefont
  {Lin}}, \bibinfo {author} {\bibfnamefont {J.}~\bibnamefont {Yuan}}, \bibinfo
  {author} {\bibfnamefont {K.}~\bibnamefont {Jin}}, \bibinfo {author}
  {\bibfnamefont {Z.}~\bibnamefont {Yin}}, \bibinfo {author} {\bibfnamefont
  {G.}~\bibnamefont {Li}}, \bibinfo {author} {\bibfnamefont {K.-J.}\
  \bibnamefont {Zhou}}, \bibinfo {author} {\bibfnamefont {X.}~\bibnamefont
  {Lu}}, \bibinfo {author} {\bibfnamefont {M.}~\bibnamefont {Dantz}}, \bibinfo
  {author} {\bibfnamefont {T.}~\bibnamefont {Schmitt}}, \bibinfo {author}
  {\bibfnamefont {H.}~\bibnamefont {Ding}}, \bibinfo {author} {\bibfnamefont
  {H.}~\bibnamefont {Guo}}, \bibinfo {author} {\bibfnamefont {M.~P.~M.}\
  \bibnamefont {Dean}},\ and\ \bibinfo {author} {\bibfnamefont
  {X.}~\bibnamefont {Liu}},\ }\bibfield  {title} {\bibinfo {title} {{Doping
  evolution of the charge excitations and electron correlations in
  electron-doped superconducting
  La$_{\mathrm{2-x}}$Ce$_{\mathrm{x}}$CuO$_4$}},\ }\href
  {https://doi.org/10.1038/s41535-019-0205-9} {\bibfield  {journal} {\bibinfo
  {journal} {npj Quantum Mater.}\ }\textbf {\bibinfo {volume} {5}},\ \bibinfo
  {pages} {4} (\bibinfo {year} {2020})}\BibitemShut {NoStop}%
\bibitem [{\citenamefont {Nag}\ \emph {et~al.}(2020)\citenamefont {Nag},
  \citenamefont {Zhu}, \citenamefont {Bejas}, \citenamefont {Li}, \citenamefont
  {Robarts}, \citenamefont {Yamase}, \citenamefont {Petsch}, \citenamefont
  {Song}, \citenamefont {Eisaki}, \citenamefont {Walters}, \citenamefont
  {Garc\'{\i}a-Fern\'andez}, \citenamefont {Greco}, \citenamefont {Hayden},\
  and\ \citenamefont {Zhou}}]{nag20}%
  \BibitemOpen
  \bibfield  {author} {\bibinfo {author} {\bibfnamefont {A.}~\bibnamefont
  {Nag}}, \bibinfo {author} {\bibfnamefont {M.}~\bibnamefont {Zhu}}, \bibinfo
  {author} {\bibfnamefont {M.}~\bibnamefont {Bejas}}, \bibinfo {author}
  {\bibfnamefont {J.}~\bibnamefont {Li}}, \bibinfo {author} {\bibfnamefont
  {H.~C.}\ \bibnamefont {Robarts}}, \bibinfo {author} {\bibfnamefont
  {H.}~\bibnamefont {Yamase}}, \bibinfo {author} {\bibfnamefont {A.~N.}\
  \bibnamefont {Petsch}}, \bibinfo {author} {\bibfnamefont {D.}~\bibnamefont
  {Song}}, \bibinfo {author} {\bibfnamefont {H.}~\bibnamefont {Eisaki}},
  \bibinfo {author} {\bibfnamefont {A.~C.}\ \bibnamefont {Walters}}, \bibinfo
  {author} {\bibfnamefont {M.}~\bibnamefont {Garc\'{\i}a-Fern\'andez}},
  \bibinfo {author} {\bibfnamefont {A.}~\bibnamefont {Greco}}, \bibinfo
  {author} {\bibfnamefont {S.~M.}\ \bibnamefont {Hayden}},\ and\ \bibinfo
  {author} {\bibfnamefont {K.-J.}\ \bibnamefont {Zhou}},\ }\bibfield  {title}
  {\bibinfo {title} {Detection of acoustic plasmons in hole-doped lanthanum and
  bismuth cuprate superconductors using resonant inelastic x-ray scattering},\
  }\href {https://link.aps.org/doi/10.1103/PhysRevLett.125.257002} {\bibfield
  {journal} {\bibinfo  {journal} {Phys. Rev. Lett.}\ }\textbf {\bibinfo
  {volume} {125}},\ \bibinfo {pages} {257002} (\bibinfo {year}
  {2020})}\BibitemShut {NoStop}%
\bibitem [{\citenamefont {Singh}\ \emph {et~al.}(2022)\citenamefont {Singh},
  \citenamefont {Huang}, \citenamefont {Lane}, \citenamefont {Li},
  \citenamefont {Okamoto}, \citenamefont {Komiya}, \citenamefont {Markiewicz},
  \citenamefont {Bansil}, \citenamefont {Fujimori}, \citenamefont {Chen},\ and\
  \citenamefont {Huang}}]{singh21}%
  \BibitemOpen
  \bibfield  {author} {\bibinfo {author} {\bibfnamefont {A.}~\bibnamefont
  {Singh}}, \bibinfo {author} {\bibfnamefont {H.~Y.}\ \bibnamefont {Huang}},
  \bibinfo {author} {\bibfnamefont {C.}~\bibnamefont {Lane}}, \bibinfo {author}
  {\bibfnamefont {J.~H.}\ \bibnamefont {Li}}, \bibinfo {author} {\bibfnamefont
  {J.}~\bibnamefont {Okamoto}}, \bibinfo {author} {\bibfnamefont
  {S.}~\bibnamefont {Komiya}}, \bibinfo {author} {\bibfnamefont {R.~S.}\
  \bibnamefont {Markiewicz}}, \bibinfo {author} {\bibfnamefont
  {A.}~\bibnamefont {Bansil}}, \bibinfo {author} {\bibfnamefont
  {A.}~\bibnamefont {Fujimori}}, \bibinfo {author} {\bibfnamefont {C.~T.}\
  \bibnamefont {Chen}},\ and\ \bibinfo {author} {\bibfnamefont {D.~J.}\
  \bibnamefont {Huang}},\ } {\bibinfo {title} {Acoustic plasmons and conducting carriers in hole-doped cuprate superconductors,\ }}\href {https://doi.org/10.1103/PhysRevB.105.235105} {\bibfield  {journal} {\bibinfo
  {journal} {Phys. Rev. B}\ }\textbf {\bibinfo {volume} {105}},\ \bibinfo
  {pages} {235105} (\bibinfo {year} {2022})}\BibitemShut {NoStop}%
\bibitem [{\citenamefont {Galdi}\ \emph {et~al.}(2018)\citenamefont {Galdi},
  \citenamefont {Orgiani}, \citenamefont {Sacco}, \citenamefont {Gobaut},
  \citenamefont {Torelli}, \citenamefont {Aruta}, \citenamefont {Brookes},
  \citenamefont {Minola}, \citenamefont {Harter}, \citenamefont {Shen},
  \citenamefont {Schlom},\ and\ \citenamefont {Maritato}}]{galdi18}%
  \BibitemOpen
  \bibfield  {author} {\bibinfo {author} {\bibfnamefont {A.}~\bibnamefont
  {Galdi}}, \bibinfo {author} {\bibfnamefont {P.}~\bibnamefont {Orgiani}},
  \bibinfo {author} {\bibfnamefont {C.}~\bibnamefont {Sacco}}, \bibinfo
  {author} {\bibfnamefont {B.}~\bibnamefont {Gobaut}}, \bibinfo {author}
  {\bibfnamefont {P.}~\bibnamefont {Torelli}}, \bibinfo {author} {\bibfnamefont
  {C.}~\bibnamefont {Aruta}}, \bibinfo {author} {\bibfnamefont {N.~B.}\
  \bibnamefont {Brookes}}, \bibinfo {author} {\bibfnamefont {M.}~\bibnamefont
  {Minola}}, \bibinfo {author} {\bibfnamefont {J.~W.}\ \bibnamefont {Harter}},
  \bibinfo {author} {\bibfnamefont {K.~M.}\ \bibnamefont {Shen}}, \bibinfo
  {author} {\bibfnamefont {D.~G.}\ \bibnamefont {Schlom}},\ and\ \bibinfo
  {author} {\bibfnamefont {L.}~\bibnamefont {Maritato}},\ }\bibfield  {title}
  {\bibinfo {title} {X-ray absorption spectroscopy study of annealing process
  on Sr$_{1-x}$La$_x$CuO$_2$ electron-doped cuprate thin films},\ }\href
  {https://doi.org/10.1063/1.5021559} {\bibfield  {journal} {\bibinfo
  {journal} {J. Appl. Phys.}\ }\textbf {\bibinfo {volume} {123}},\ \bibinfo
  {pages} {123901} (\bibinfo {year} {2018})}\BibitemShut {NoStop}%
\bibitem [{\citenamefont {Fournier}(2015)}]{fournier15}%
  \BibitemOpen
  \bibfield  {author} {\bibinfo {author} {\bibfnamefont {P.}~\bibnamefont
  {Fournier}},\ }\bibfield  {title} {\bibinfo {title} {T' and infinite-layer
  electron-doped cuprates},\ }\href
  {https://doi.org/https://doi.org/10.1016/j.physc.2015.02.036} {\bibfield
  {journal} {\bibinfo  {journal} {Physica C}\ }\textbf {\bibinfo {volume}
  {514}},\ \bibinfo {pages} {314} (\bibinfo {year} {2015})}\BibitemShut
  {NoStop}%
\bibitem [{\citenamefont {Tomaschko}\ \emph {et~al.}(2012)\citenamefont
  {Tomaschko}, \citenamefont {Scharinger}, \citenamefont {Leca}, \citenamefont
  {Nagel}, \citenamefont {Kemmler}, \citenamefont {Selistrovski}, \citenamefont
  {Koelle},\ and\ \citenamefont {Kleiner}}]{tomaschko12}%
  \BibitemOpen
  \bibfield  {author} {\bibinfo {author} {\bibfnamefont {J.}~\bibnamefont
  {Tomaschko}}, \bibinfo {author} {\bibfnamefont {S.}~\bibnamefont
  {Scharinger}}, \bibinfo {author} {\bibfnamefont {V.}~\bibnamefont {Leca}},
  \bibinfo {author} {\bibfnamefont {J.}~\bibnamefont {Nagel}}, \bibinfo
  {author} {\bibfnamefont {M.}~\bibnamefont {Kemmler}}, \bibinfo {author}
  {\bibfnamefont {T.}~\bibnamefont {Selistrovski}}, \bibinfo {author}
  {\bibfnamefont {D.}~\bibnamefont {Koelle}},\ and\ \bibinfo {author}
  {\bibfnamefont {R.}~\bibnamefont {Kleiner}},\ }\bibfield  {title} {\bibinfo
  {title} {Phase-sensitive evidence for
  ${d}_{{x}^{2}\ensuremath{-}{y}^{2}}$-pairing symmetry in the parent-structure
  high-${T}_{c}$ cuprate superconductor
  Sr${}_{1\ensuremath{-}x}$La${}_{x}$CuO${}_{2}$},\ }\href
  {https://doi.org/10.1103/PhysRevB.86.094509} {\bibfield  {journal} {\bibinfo
  {journal} {Phys. Rev. B}\ }\textbf {\bibinfo {volume} {86}},\ \bibinfo
  {pages} {094509} (\bibinfo {year} {2012})}\BibitemShut {NoStop}%
\bibitem [{\citenamefont {Zhou}\ \emph {et~al.}(2022)\citenamefont {Zhou},
  \citenamefont {Walters}, \citenamefont {Garcia-Fernandez}, \citenamefont
  {Rice}, \citenamefont {Hand}, \citenamefont {Nag}, \citenamefont {Li},
  \citenamefont {Agrestini}, \citenamefont {Garland}, \citenamefont {Wang},
  \citenamefont {Alcock}, \citenamefont {Nistea}, \citenamefont {Nutter},
  \citenamefont {Rubies}, \citenamefont {Knap}, \citenamefont {Gaughran},
  \citenamefont {Yuan}, \citenamefont {Chang}, \citenamefont {Emmins},\ and\
  \citenamefont {Howell}}]{zhou22}%
  \BibitemOpen
  \bibfield  {author} {\bibinfo {author} {\bibfnamefont {K.-J.}\ \bibnamefont
  {Zhou}}, \bibinfo {author} {\bibfnamefont {A.}~\bibnamefont {Walters}},
  \bibinfo {author} {\bibfnamefont {M.}~\bibnamefont {Garcia-Fernandez}},
  \bibinfo {author} {\bibfnamefont {T.}~\bibnamefont {Rice}}, \bibinfo {author}
  {\bibfnamefont {M.}~\bibnamefont {Hand}}, \bibinfo {author} {\bibfnamefont
  {A.}~\bibnamefont {Nag}}, \bibinfo {author} {\bibfnamefont {J.}~\bibnamefont
  {Li}}, \bibinfo {author} {\bibfnamefont {S.}~\bibnamefont {Agrestini}},
  \bibinfo {author} {\bibfnamefont {P.}~\bibnamefont {Garland}}, \bibinfo
  {author} {\bibfnamefont {H.}~\bibnamefont {Wang}}, \bibinfo {author}
  {\bibfnamefont {S.}~\bibnamefont {Alcock}}, \bibinfo {author} {\bibfnamefont
  {I.}~\bibnamefont {Nistea}}, \bibinfo {author} {\bibfnamefont
  {B.}~\bibnamefont {Nutter}}, \bibinfo {author} {\bibfnamefont
  {N.}~\bibnamefont {Rubies}}, \bibinfo {author} {\bibfnamefont
  {G.}~\bibnamefont {Knap}}, \bibinfo {author} {\bibfnamefont {M.}~\bibnamefont
  {Gaughran}}, \bibinfo {author} {\bibfnamefont {F.}~\bibnamefont {Yuan}},
  \bibinfo {author} {\bibfnamefont {P.}~\bibnamefont {Chang}}, \bibinfo
  {author} {\bibfnamefont {J.}~\bibnamefont {Emmins}},\ and\ \bibinfo {author}
  {\bibfnamefont {G.}~\bibnamefont {Howell}},\ }\bibfield  {title} {\bibinfo
  {title} {I21: an advanced high-resolution resonant inelastic x-ray scattering
  beamline at diamond light source},\ }\href
  {https://doi.org/10.1107/S1600577522000601} {\bibfield  {journal} {\bibinfo
  {journal} {J. Synchrotron Rad.}\ }\textbf {\bibinfo {volume} {29}},\ \bibinfo
  {pages} {563} (\bibinfo {year} {2022})}\BibitemShut {NoStop}%
  \bibitem [{sup()}]{suppmat}%
  \BibitemOpen
  \href@noop {} {}\bibinfo {note} {See Supplemental Material at [URL will be
  inserted by publisher] for details of the RIXS raw data and fits, theoretical
  scheme, fitting procedure in the $t$-$J$-$V$ model, broadening $\Gamma$, possible plasmon gap in LCCO and LSCO, intensity of the plasmon, and plasmon-phonon coupling}\BibitemShut
  {NoStop}%

\bibitem [{\citenamefont {Gunnarsson}\ and\ \citenamefont
  {R\"osch}(2008)}]{gunnarsson08}%
  \BibitemOpen
  \bibfield  {author} {\bibinfo {author} {\bibfnamefont {O.}~\bibnamefont
  {Gunnarsson}}\ and\ \bibinfo {author} {\bibfnamefont {O.}~\bibnamefont
  {R\"osch}},\ }\bibfield  {title} {\bibinfo {title} {Interplay between
  electron{\textendash}phonon and coulomb interactions in cuprates},\ }\href
  {https://doi.org/10.1088/0953-8984/20/04/043201} {\bibfield  {journal}
  {\bibinfo  {journal} {J. Phys.: Condens. Matter}\ }\textbf {\bibinfo {volume}
  {20}},\ \bibinfo {pages} {043201} (\bibinfo {year} {2008})}\BibitemShut
  {NoStop}%
\bibitem [{\citenamefont {Koval}\ and\ \citenamefont {Migoni}(1996)}]{koval96}%
  \BibitemOpen
  \bibfield  {author} {\bibinfo {author} {\bibfnamefont {S.}~\bibnamefont
  {Koval}}\ and\ \bibinfo {author} {\bibfnamefont {R.}~\bibnamefont {Migoni}},\
  }\bibfield  {title} {\bibinfo {title} {Lattice dynamics of the
  infinite-layered compounds},\ }\href
  {https://doi.org/https://doi.org/10.1016/0921-4534(95)00688-5} {\bibfield
  {journal} {\bibinfo  {journal} {Physica C}\ }\textbf {\bibinfo {volume}
  {257}},\ \bibinfo {pages} {255} (\bibinfo {year} {1996})}\BibitemShut
  {NoStop}%
\bibitem [{\citenamefont {Mun}\ \emph {et~al.}(2001)\citenamefont {Mun},
  \citenamefont {Roh}, \citenamefont {Kim}, \citenamefont {Jung}, \citenamefont
  {Kim},\ and\ \citenamefont {Lee}}]{mun01}%
  \BibitemOpen
  \bibfield  {author} {\bibinfo {author} {\bibfnamefont {M.-O.}\ \bibnamefont
  {Mun}}, \bibinfo {author} {\bibfnamefont {Y.~S.}\ \bibnamefont {Roh}},
  \bibinfo {author} {\bibfnamefont {J.~H.}\ \bibnamefont {Kim}}, \bibinfo
  {author} {\bibfnamefont {C.}~\bibnamefont {Jung}}, \bibinfo {author}
  {\bibfnamefont {J.}~\bibnamefont {Kim}},\ and\ \bibinfo {author}
  {\bibfnamefont {S.-I.}\ \bibnamefont {Lee}},\ }\bibfield  {title} {\bibinfo
  {title} {Optical phonons of superconducting infinite-layer compounds
  Sr$_{0.9}$Ln$_{0.1}$CuO$_2$ (Ln=La and Sm)},\ }\href
  {https://doi.org/https://doi.org/10.1016/S0921-4534(01)00867-X} {\bibfield
  {journal} {\bibinfo  {journal} {Physica C}\ }\textbf {\bibinfo {volume}
  {364-365}},\ \bibinfo {pages} {629} (\bibinfo {year} {2001})}\BibitemShut
  {NoStop}%
\bibitem [{\citenamefont {Klenner}\ \emph {et~al.}(1994)\citenamefont
  {Klenner}, \citenamefont {Falter},\ and\ \citenamefont {Chen}}]{klenner94}%
  \BibitemOpen
  \bibfield  {author} {\bibinfo {author} {\bibfnamefont {M.}~\bibnamefont
  {Klenner}}, \bibinfo {author} {\bibfnamefont {C.}~\bibnamefont {Falter}},\
  and\ \bibinfo {author} {\bibfnamefont {Q.}~\bibnamefont {Chen}},\ }\bibfield
  {title} {\bibinfo {title} {Calculated phonon dispersion of infinite-layer
  compounds and the effects of charge fluctuations},\ }\href
  {https://doi.org/10.1007/BF01313348} {\bibfield  {journal} {\bibinfo
  {journal} {Z. Phys. B}\ }\textbf {\bibinfo {volume} {95}},\ \bibinfo {pages}
  {417} (\bibinfo {year} {1994})}\BibitemShut {NoStop}%
\bibitem [{\citenamefont {Dellea}\ \emph {et~al.}(2017)\citenamefont {Dellea},
  \citenamefont {Minola}, \citenamefont {Galdi}, \citenamefont {Di~Castro},
  \citenamefont {Aruta}, \citenamefont {Brookes}, \citenamefont {Jia},
  \citenamefont {Mazzoli}, \citenamefont {Moretti~Sala}, \citenamefont
  {Moritz}, \citenamefont {Orgiani}, \citenamefont {Schlom}, \citenamefont
  {Tebano}, \citenamefont {Balestrino}, \citenamefont {Braicovich},
  \citenamefont {Devereaux}, \citenamefont {Maritato},\ and\ \citenamefont
  {Ghiringhelli}}]{dellea17}%
  \BibitemOpen
  \bibfield  {author} {\bibinfo {author} {\bibfnamefont {G.}~\bibnamefont
  {Dellea}}, \bibinfo {author} {\bibfnamefont {M.}~\bibnamefont {Minola}},
  \bibinfo {author} {\bibfnamefont {A.}~\bibnamefont {Galdi}}, \bibinfo
  {author} {\bibfnamefont {D.}~\bibnamefont {Di~Castro}}, \bibinfo {author}
  {\bibfnamefont {C.}~\bibnamefont {Aruta}}, \bibinfo {author} {\bibfnamefont
  {N.~B.}\ \bibnamefont {Brookes}}, \bibinfo {author} {\bibfnamefont {C.~J.}\
  \bibnamefont {Jia}}, \bibinfo {author} {\bibfnamefont {C.}~\bibnamefont
  {Mazzoli}}, \bibinfo {author} {\bibfnamefont {M.}~\bibnamefont
  {Moretti~Sala}}, \bibinfo {author} {\bibfnamefont {B.}~\bibnamefont
  {Moritz}}, \bibinfo {author} {\bibfnamefont {P.}~\bibnamefont {Orgiani}},
  \bibinfo {author} {\bibfnamefont {D.~G.}\ \bibnamefont {Schlom}}, \bibinfo
  {author} {\bibfnamefont {A.}~\bibnamefont {Tebano}}, \bibinfo {author}
  {\bibfnamefont {G.}~\bibnamefont {Balestrino}}, \bibinfo {author}
  {\bibfnamefont {L.}~\bibnamefont {Braicovich}}, \bibinfo {author}
  {\bibfnamefont {T.~P.}\ \bibnamefont {Devereaux}}, \bibinfo {author}
  {\bibfnamefont {L.}~\bibnamefont {Maritato}},\ and\ \bibinfo {author}
  {\bibfnamefont {G.}~\bibnamefont {Ghiringhelli}},\ }\bibfield  {title}
  {\bibinfo {title} {Spin and charge excitations in artificial hole- and
  electron-doped infinite layer cuprate superconductors},\ }\href
  {https://doi.org/10.1103/PhysRevB.96.115117} {\bibfield  {journal} {\bibinfo
  {journal} {Phys. Rev. B}\ }\textbf {\bibinfo {volume} {96}},\ \bibinfo
  {pages} {115117} (\bibinfo {year} {2017})}\BibitemShut {NoStop}%
\bibitem [{\citenamefont {Markiewicz}\ \emph {et~al.}(2008)\citenamefont
  {Markiewicz}, \citenamefont {Hasan},\ and\ \citenamefont
  {Bansil}}]{markiewicz08}%
  \BibitemOpen
  \bibfield  {author} {\bibinfo {author} {\bibfnamefont {R.~S.}\ \bibnamefont
  {Markiewicz}}, \bibinfo {author} {\bibfnamefont {M.~Z.}\ \bibnamefont
  {Hasan}},\ and\ \bibinfo {author} {\bibfnamefont {A.}~\bibnamefont
  {Bansil}},\ }\bibfield  {title} {\bibinfo {title} {Acoustic plasmons and
  doping evolution of mott physics in resonant inelastic x-ray scattering from
  cuprate superconductors},\ }\href
  {https://doi.org/10.1103/PhysRevB.77.094518} {\bibfield  {journal} {\bibinfo
  {journal} {Phys. Rev. B}\ }\textbf {\bibinfo {volume} {77}},\ \bibinfo
  {pages} {094518} (\bibinfo {year} {2008})}\BibitemShut {NoStop}%
\bibitem [{\citenamefont {Fidrysiak}\ and\ \citenamefont
  {Spa\l{}ek}(2021)}]{fidrysiak21}%
  \BibitemOpen
  \bibfield  {author} {\bibinfo {author} {\bibfnamefont {M.}~\bibnamefont
  {Fidrysiak}}\ and\ \bibinfo {author} {\bibfnamefont {J.}~\bibnamefont
  {Spa\l{}ek}},\ }\bibfield  {title} {\bibinfo {title} {Unified theory of spin
  and charge excitations in high-${T}_{c}$ cuprate superconductors: A
  quantitative comparison with experiment and interpretation},\ }\href
  {https://doi.org/10.1103/PhysRevB.104.L020510} {\bibfield  {journal}
  {\bibinfo  {journal} {Phys. Rev. B}\ }\textbf {\bibinfo {volume} {104}},\
  \bibinfo {pages} {L020510} (\bibinfo {year} {2021})}\BibitemShut {NoStop}%
\bibitem [{\citenamefont {Greco}\ \emph {et~al.}(2019)\citenamefont {Greco},
  \citenamefont {Yamase},\ and\ \citenamefont {Bejas}}]{greco19}%
  \BibitemOpen
  \bibfield  {author} {\bibinfo {author} {\bibfnamefont {A.}~\bibnamefont
  {Greco}}, \bibinfo {author} {\bibfnamefont {H.}~\bibnamefont {Yamase}},\ and\
  \bibinfo {author} {\bibfnamefont {M.}~\bibnamefont {Bejas}},\ }\bibfield
  {title} {\bibinfo {title} {Origin of high-energy charge excitations observed
  by resonant inelastic x-ray scattering in cuprate superconductors},\ }\href
  {https://doi.org/10.1038/s42005-018-0099-z} {\bibfield  {journal} {\bibinfo
  {journal} {Commun. Phys.}\ }\textbf {\bibinfo {volume} {2}},\ \bibinfo
  {pages} {3} (\bibinfo {year} {2019})}\BibitemShut {NoStop}%
\bibitem [{\citenamefont {Greco}\ \emph {et~al.}(2020)\citenamefont {Greco},
  \citenamefont {Yamase},\ and\ \citenamefont {Bejas}}]{greco20}%
  \BibitemOpen
  \bibfield  {author} {\bibinfo {author} {\bibfnamefont {A.}~\bibnamefont
  {Greco}}, \bibinfo {author} {\bibfnamefont {H.}~\bibnamefont {Yamase}},\ and\
  \bibinfo {author} {\bibfnamefont {M.}~\bibnamefont {Bejas}},\ }\bibfield
  {title} {\bibinfo {title} {Close inspection of plasmon excitations in cuprate
  superconductors},\ }\href
  {https://link.aps.org/doi/10.1103/PhysRevB.102.024509} {\bibfield  {journal}
  {\bibinfo  {journal} {Phys. Rev. B}\ }\textbf {\bibinfo {volume} {102}},\
  \bibinfo {pages} {024509} (\bibinfo {year} {2020})}\BibitemShut {NoStop}%
\bibitem [{\citenamefont {Zhang}\ and\ \citenamefont {Rice}(1988)}]{fczhang88}%
  \BibitemOpen
  \bibfield  {author} {\bibinfo {author} {\bibfnamefont {F.~C.}\ \bibnamefont
  {Zhang}}\ and\ \bibinfo {author} {\bibfnamefont {T.~M.}\ \bibnamefont
  {Rice}},\ }\bibfield  {title} {\bibinfo {title} {{Effective Hamiltonian for
  the superconducting {Cu} oxides}},\ }\href
  {https://doi.org/10.1103/PhysRevB.37.3759} {\bibfield  {journal} {\bibinfo
  {journal} {Phys. Rev. B}\ }\textbf {\bibinfo {volume} {37}},\ \bibinfo
  {pages} {3759} (\bibinfo {year} {1988})}\BibitemShut {NoStop}%
\bibitem [{\citenamefont {Prelov\ifmmode~\check{s}\else \v{s}\fi{}ek}\ and\
  \citenamefont {Horsch}(1999)}]{prelovsek99}%
  \BibitemOpen
  \bibfield  {author} {\bibinfo {author} {\bibfnamefont {P.}~\bibnamefont
  {Prelov\ifmmode~\check{s}\else \v{s}\fi{}ek}}\ and\ \bibinfo {author}
  {\bibfnamefont {P.}~\bibnamefont {Horsch}},\ }\bibfield  {title} {\bibinfo
  {title} {Electron-energy loss spectra and plasmon resonance in cuprates},\
  }\href {https://doi.org/10.1103/PhysRevB.60.R3735} {\bibfield  {journal}
  {\bibinfo  {journal} {Phys. Rev. B}\ }\textbf {\bibinfo {volume} {60}},\
  \bibinfo {pages} {R3735} (\bibinfo {year} {1999})}\BibitemShut {NoStop}%
\bibitem [{\citenamefont {Botana}\ and\ \citenamefont
  {Norman}(2020)}]{botana20}%
  \BibitemOpen
  \bibfield  {author} {\bibinfo {author} {\bibfnamefont {A.~S.}\ \bibnamefont
  {Botana}}\ and\ \bibinfo {author} {\bibfnamefont {M.~R.}\ \bibnamefont
  {Norman}},\ }\bibfield  {title} {\bibinfo {title} {Similarities and
  differences between LaNiO$_{2}$ and CaCuO$_{2}$ and
  implications for superconductivity},\ }\href
  {https://doi.org/10.1103/PhysRevX.10.011024} {\bibfield  {journal} {\bibinfo
  {journal} {Phys. Rev. X}\ }\textbf {\bibinfo {volume} {10}},\ \bibinfo
  {pages} {011024} (\bibinfo {year} {2020})}\BibitemShut {NoStop}%
\bibitem [{\citenamefont {Bozovic}(1990)}]{bozovic90}%
  \BibitemOpen
  \bibfield  {author} {\bibinfo {author} {\bibfnamefont {I.}~\bibnamefont
  {Bozovic}},\ }\bibfield  {title} {\bibinfo {title} {Plasmons in cuprate
  superconductors},\ }\href {https://doi.org/10.1103/PhysRevB.42.1969}
  {\bibfield  {journal} {\bibinfo  {journal} {Phys. Rev. B}\ }\textbf {\bibinfo
  {volume} {42}},\ \bibinfo {pages} {1969} (\bibinfo {year}
  {1990})}\BibitemShut {NoStop}%
\bibitem [{\citenamefont {Fink}\ \emph {et~al.}(2001)\citenamefont {Fink},
  \citenamefont {Knupfer}, \citenamefont {Atzkern},\ and\ \citenamefont
  {Golden}}]{fink01}%
  \BibitemOpen
  \bibfield  {author} {\bibinfo {author} {\bibfnamefont {J.}~\bibnamefont
  {Fink}}, \bibinfo {author} {\bibfnamefont {M.}~\bibnamefont {Knupfer}},
  \bibinfo {author} {\bibfnamefont {S.}~\bibnamefont {Atzkern}},\ and\ \bibinfo
  {author} {\bibfnamefont {M.}~\bibnamefont {Golden}},\ }\bibfield  {title}
  {\bibinfo {title} {Electronic correlations in solids, studied using electron
  energy-loss spectroscopy},\ }\href
  {https://doi.org/https://doi.org/10.1016/S0368-2048(01)00254-7} {\bibfield
  {journal} {\bibinfo  {journal} {J. Electron Spectrosc. Relat. Phenom.}\ }\textbf {\bibinfo
  {volume} {117-118}},\ \bibinfo {pages} {287} (\bibinfo {year}
  {2001})}\BibitemShut {NoStop}%
\bibitem [{\citenamefont {Pavarini}\ \emph {et~al.}(2001)\citenamefont
  {Pavarini}, \citenamefont {Dasgupta}, \citenamefont {Saha-Dasgupta},
  \citenamefont {Jepsen},\ and\ \citenamefont {Andersen}}]{pavarini01}%
  \BibitemOpen
  \bibfield  {author} {\bibinfo {author} {\bibfnamefont {E.}~\bibnamefont
  {Pavarini}}, \bibinfo {author} {\bibfnamefont {I.}~\bibnamefont {Dasgupta}},
  \bibinfo {author} {\bibfnamefont {T.}~\bibnamefont {Saha-Dasgupta}}, \bibinfo
  {author} {\bibfnamefont {O.}~\bibnamefont {Jepsen}},\ and\ \bibinfo {author}
  {\bibfnamefont {O.~K.}\ \bibnamefont {Andersen}},\ }\bibfield  {title}
  {\bibinfo {title} {Band-structure trend in hole-doped cuprates and
  correlation with $T_{\mathit{c}\mathrm{max}}$},\ }\href
  {https://link.aps.org/doi/10.1103/PhysRevLett.87.047003} {\bibfield
  {journal} {\bibinfo  {journal} {Phys. Rev. Lett.}\ }\textbf {\bibinfo
  {volume} {87}},\ \bibinfo {pages} {047003} (\bibinfo {year}
  {2001})}\BibitemShut {NoStop}%
\bibitem [{\citenamefont {Iyo}\ \emph {et~al.}(2007)\citenamefont {Iyo},
  \citenamefont {Tanaka}, \citenamefont {Kito}, \citenamefont {Kodama},
  \citenamefont {M.~Shirage}, \citenamefont {D.~Shivagan}, \citenamefont
  {Matsuhata}, \citenamefont {Tokiwa},\ and\ \citenamefont {Watanabe}}]{iyo07}%
  \BibitemOpen
  \bibfield  {author} {\bibinfo {author} {\bibfnamefont {A.}~\bibnamefont
  {Iyo}}, \bibinfo {author} {\bibfnamefont {Y.}~\bibnamefont {Tanaka}},
  \bibinfo {author} {\bibfnamefont {H.}~\bibnamefont {Kito}}, \bibinfo {author}
  {\bibfnamefont {Y.}~\bibnamefont {Kodama}}, \bibinfo {author} {\bibfnamefont
  {P.}~\bibnamefont {M.~Shirage}}, \bibinfo {author} {\bibfnamefont
  {D.}~\bibnamefont {D.~Shivagan}}, \bibinfo {author} {\bibfnamefont
  {H.}~\bibnamefont {Matsuhata}}, \bibinfo {author} {\bibfnamefont
  {K.}~\bibnamefont {Tokiwa}},\ and\ \bibinfo {author} {\bibfnamefont
  {T.}~\bibnamefont {Watanabe}},\ }\bibfield  {title} {\bibinfo {title} {Tc vs
  n relationship for multilayered high-tc superconductors},\ }\href
  {https://doi.org/10.1143/JPSJ.76.094711} {\bibfield  {journal} {\bibinfo
  {journal} {J. Phys. Soc. Jpn.}\ }\textbf {\bibinfo {volume} {76}},\ \bibinfo
  {pages} {094711} (\bibinfo {year} {2007})}\BibitemShut {NoStop}%
\bibitem [{\citenamefont {Vincini}\ \emph {et~al.}(2019)\citenamefont
  {Vincini}, \citenamefont {Tajima}, \citenamefont {Miyasaka},\ and\
  \citenamefont {Tanaka}}]{vincini19}%
  \BibitemOpen
  \bibfield  {author} {\bibinfo {author} {\bibfnamefont {G.}~\bibnamefont
  {Vincini}}, \bibinfo {author} {\bibfnamefont {S.}~\bibnamefont {Tajima}},
  \bibinfo {author} {\bibfnamefont {S.}~\bibnamefont {Miyasaka}},\ and\
  \bibinfo {author} {\bibfnamefont {K.}~\bibnamefont {Tanaka}},\ }\bibfield
  {title} {\bibinfo {title} {{Multilayer effects in Bi$_2$Sr$_2$Ca$_2$Cu$_3$O$_{10+z}$
  superconductors}},\ }\href {https://doi.org/10.1088/1361-6668/ab4246}
  {\bibfield  {journal} {\bibinfo  {journal} {Supercond. Sci. Technol.}\
  }\textbf {\bibinfo {volume} {32}},\ \bibinfo {pages} {113001} (\bibinfo
  {year} {2019})}\BibitemShut {NoStop}%
\bibitem [{\citenamefont {Mahan}(1990)}]{mahan}%
  \BibitemOpen
  \bibfield  {author} {\bibinfo {author} {\bibfnamefont {G.~D.}\ \bibnamefont
  {Mahan}},\ }\href@noop {} {\emph {\bibinfo {title} {Many-Particle
  Physics}}},\ \bibinfo {edition} {2nd}\ ed.\ (\bibinfo  {publisher} {Plunum
  Press},\ \bibinfo {year} {1990})\BibitemShut {NoStop}%
\bibitem [{\citenamefont {Cudazzo}\ \emph {et~al.}(2012)\citenamefont
  {Cudazzo}, \citenamefont {Gatti},\ and\ \citenamefont {Rubio}}]{cudazzo12}%
  \BibitemOpen
  \bibfield  {author} {\bibinfo {author} {\bibfnamefont {P.}~\bibnamefont
  {Cudazzo}}, \bibinfo {author} {\bibfnamefont {M.}~\bibnamefont {Gatti}},\
  and\ \bibinfo {author} {\bibfnamefont {A.}~\bibnamefont {Rubio}},\ }\bibfield
   {title} {\bibinfo {title} {Plasmon dispersion in layered transition-metal
  dichalcogenides},\ }\href {https://doi.org/10.1103/PhysRevB.86.075121}
  {\bibfield  {journal} {\bibinfo  {journal} {Phys. Rev. B}\ }\textbf {\bibinfo
  {volume} {86}},\ \bibinfo {pages} {075121} (\bibinfo {year}
  {2012})}\BibitemShut {NoStop}%
\bibitem [{\citenamefont {Groenewald}\ \emph {et~al.}(2016)\citenamefont
  {Groenewald}, \citenamefont {R\"osner}, \citenamefont {Sch\"onhoff},
  \citenamefont {Haas},\ and\ \citenamefont {Wehling}}]{groenewald16}%
  \BibitemOpen
  \bibfield  {author} {\bibinfo {author} {\bibfnamefont {R.~E.}\ \bibnamefont
  {Groenewald}}, \bibinfo {author} {\bibfnamefont {M.}~\bibnamefont
  {R\"osner}}, \bibinfo {author} {\bibfnamefont {G.}~\bibnamefont
  {Sch\"onhoff}}, \bibinfo {author} {\bibfnamefont {S.}~\bibnamefont {Haas}},\
  and\ \bibinfo {author} {\bibfnamefont {T.~O.}\ \bibnamefont {Wehling}},\
  }\bibfield  {title} {\bibinfo {title} {Valley plasmonics in transition metal
  dichalcogenides},\ }\href {https://doi.org/10.1103/PhysRevB.93.205145}
  {\bibfield  {journal} {\bibinfo  {journal} {Phys. Rev. B}\ }\textbf {\bibinfo
  {volume} {93}},\ \bibinfo {pages} {205145} (\bibinfo {year}
  {2016})}\BibitemShut {NoStop}%
\bibitem [{\citenamefont {Thygesen}(2017)}]{thygesen17}%
  \BibitemOpen
  \bibfield  {author} {\bibinfo {author} {\bibfnamefont {K.~S.}\ \bibnamefont
  {Thygesen}},\ }\bibfield  {title} {\bibinfo {title} {Calculating excitons,
  plasmons, and quasiparticles in 2D materials and van der waals
  heterostructures},\ }\href {https://doi.org/10.1088/2053-1583/aa6432}
  {\bibfield  {journal} {\bibinfo  {journal} {2D Materials}\ }\textbf {\bibinfo
  {volume} {4}},\ \bibinfo {pages} {022004} (\bibinfo {year}
  {2017})}\BibitemShut {NoStop}%
\bibitem [{\citenamefont {Li}\ \emph {et~al.}(2019)\citenamefont {Li},
  \citenamefont {Lee}, \citenamefont {Wang}, \citenamefont {Osada},
  \citenamefont {Crossley}, \citenamefont {Lee}, \citenamefont {Cui},
  \citenamefont {Hikita},\ and\ \citenamefont {Hwang}}]{Li19}%
  \BibitemOpen
  \bibfield  {author} {\bibinfo {author} {\bibfnamefont {D.}~\bibnamefont
  {Li}}, \bibinfo {author} {\bibfnamefont {K.}~\bibnamefont {Lee}}, \bibinfo
  {author} {\bibfnamefont {B.}~\bibnamefont {Wang}}, \bibinfo {author}
  {\bibfnamefont {M.}~\bibnamefont {Osada}}, \bibinfo {author} {\bibfnamefont
  {S.}~\bibnamefont {Crossley}}, \bibinfo {author} {\bibfnamefont
  {H.}~\bibnamefont {Lee}}, \bibinfo {author} {\bibfnamefont {Y.}~\bibnamefont
  {Cui}}, \bibinfo {author} {\bibfnamefont {Y.}~\bibnamefont {Hikita}},\ and\
  \bibinfo {author} {\bibfnamefont {H.}~\bibnamefont {Hwang}},\ }\bibfield
  {title} {\bibinfo {title} {Superconductivity in an infinite-layer
  nickelate},\ }\href {https://doi.org/10.1038/s41586-019-1496-5} {\bibfield
  {journal} {\bibinfo  {journal} {Nature}\ }\textbf {\bibinfo {volume} {572}},\
  \bibinfo {pages} {624} (\bibinfo {year} {2019})}\BibitemShut {NoStop}%
\bibitem [{\citenamefont {Hepting}\ \emph {et~al.}(2020)\citenamefont
  {Hepting}, \citenamefont {Li}, \citenamefont {Jia}, \citenamefont {Lu},
  \citenamefont {Paris}, \citenamefont {Tseng}, \citenamefont {Feng},
  \citenamefont {Osada}, \citenamefont {Been}, \citenamefont {Hikita},
  \citenamefont {Chuang}, \citenamefont {Hussain}, \citenamefont {Zhou},
  \citenamefont {Nag}, \citenamefont {Garcia-Fernandez}, \citenamefont {Rossi},
  \citenamefont {Huang}, \citenamefont {Huang}, \citenamefont {Shen},
  \citenamefont {Schmitt}, \citenamefont {Hwang}, \citenamefont {Moritz},
  \citenamefont {Zaanen}, \citenamefont {Devereaux},\ and\ \citenamefont
  {Lee}}]{hepting20}%
  \BibitemOpen
  \bibfield  {author} {\bibinfo {author} {\bibfnamefont {M.}~\bibnamefont
  {Hepting}}, \bibinfo {author} {\bibfnamefont {D.}~\bibnamefont {Li}},
  \bibinfo {author} {\bibfnamefont {C.~J.}\ \bibnamefont {Jia}}, \bibinfo
  {author} {\bibfnamefont {H.}~\bibnamefont {Lu}}, \bibinfo {author}
  {\bibfnamefont {E.}~\bibnamefont {Paris}}, \bibinfo {author} {\bibfnamefont
  {Y.}~\bibnamefont {Tseng}}, \bibinfo {author} {\bibfnamefont
  {X.}~\bibnamefont {Feng}}, \bibinfo {author} {\bibfnamefont {M.}~\bibnamefont
  {Osada}}, \bibinfo {author} {\bibfnamefont {E.}~\bibnamefont {Been}},
  \bibinfo {author} {\bibfnamefont {Y.}~\bibnamefont {Hikita}}, \bibinfo
  {author} {\bibfnamefont {Y.-D.}\ \bibnamefont {Chuang}}, \bibinfo {author}
  {\bibfnamefont {Z.}~\bibnamefont {Hussain}}, \bibinfo {author} {\bibfnamefont
  {K.~J.}\ \bibnamefont {Zhou}}, \bibinfo {author} {\bibfnamefont
  {A.}~\bibnamefont {Nag}}, \bibinfo {author} {\bibfnamefont {M.}~\bibnamefont
  {Garcia-Fernandez}}, \bibinfo {author} {\bibfnamefont {M.}~\bibnamefont
  {Rossi}}, \bibinfo {author} {\bibfnamefont {H.~Y.}\ \bibnamefont {Huang}},
  \bibinfo {author} {\bibfnamefont {D.~J.}\ \bibnamefont {Huang}}, \bibinfo
  {author} {\bibfnamefont {Z.~X.}\ \bibnamefont {Shen}}, \bibinfo {author}
  {\bibfnamefont {T.}~\bibnamefont {Schmitt}}, \bibinfo {author} {\bibfnamefont
  {H.~Y.}\ \bibnamefont {Hwang}}, \bibinfo {author} {\bibfnamefont
  {B.}~\bibnamefont {Moritz}}, \bibinfo {author} {\bibfnamefont
  {J.}~\bibnamefont {Zaanen}}, \bibinfo {author} {\bibfnamefont {T.~P.}\
  \bibnamefont {Devereaux}},\ and\ \bibinfo {author} {\bibfnamefont {W.~S.}\
  \bibnamefont {Lee}},\ }\bibfield  {title} {\bibinfo {title} {Electronic
  structure of the parent compound of superconducting infinite-layer
  nickelates},\ }\href {https://doi.org/10.1038/s41563-019-0585-z} {\bibfield
  {journal} {\bibinfo  {journal} {Nat. Mater.}\ }\textbf {\bibinfo
  {volume} {19}},\ \bibinfo {pages} {381} (\bibinfo {year} {2020})}\BibitemShut
  {NoStop}%
\bibitem [{\citenamefont {Zeng}\ \emph {et~al.}(2022)\citenamefont {Zeng},
  \citenamefont {Li}, \citenamefont {Chow}, \citenamefont {Cao}, \citenamefont
  {Zhang}, \citenamefont {Tang}, \citenamefont {Yin}, \citenamefont {Lim},
  \citenamefont {Hu}, \citenamefont {Yang},\ and\ \citenamefont
  {Ariando}}]{zeng22}%
  \BibitemOpen
  \bibfield  {author} {\bibinfo {author} {\bibfnamefont {S.}~\bibnamefont
  {Zeng}}, \bibinfo {author} {\bibfnamefont {C.}~\bibnamefont {Li}}, \bibinfo
  {author} {\bibfnamefont {L.~E.}\ \bibnamefont {Chow}}, \bibinfo {author}
  {\bibfnamefont {Y.}~\bibnamefont {Cao}}, \bibinfo {author} {\bibfnamefont
  {Z.}~\bibnamefont {Zhang}}, \bibinfo {author} {\bibfnamefont {C.~S.}\
  \bibnamefont {Tang}}, \bibinfo {author} {\bibfnamefont {X.}~\bibnamefont
  {Yin}}, \bibinfo {author} {\bibfnamefont {Z.~S.}\ \bibnamefont {Lim}},
  \bibinfo {author} {\bibfnamefont {J.}~\bibnamefont {Hu}}, \bibinfo {author}
  {\bibfnamefont {P.}~\bibnamefont {Yang}}, \ and\ \bibinfo {author}
  {\bibfnamefont {A.}~\bibnamefont {Ariando}},\ }\href {\doibase
  10.1126/sciadv.abl9927} {\bibfield  {journal} {\bibinfo  {journal} {Sci.
  Adv.}\ }\textbf {\bibinfo {volume} {8}},\ \bibinfo {pages} {eabl9927}
  (\bibinfo {year} {2022})} \BibitemShut
  {NoStop}%
\end{thebibliography}

\begin{thebibliography}{30}%
\makeatletter
\providecommand \@ifxundefined [1]{%
 \@ifx{#1\undefined}
}%
\providecommand \@ifnum [1]{%
 \ifnum #1\expandafter \@firstoftwo
 \else \expandafter \@secondoftwo
 \fi
}%
\providecommand \@ifx [1]{%
 \ifx #1\expandafter \@firstoftwo
 \else \expandafter \@secondoftwo
 \fi
}%
\providecommand \natexlab [1]{#1}%
\providecommand \enquote  [1]{``#1''}%
\providecommand \bibnamefont  [1]{#1}%
\providecommand \bibfnamefont [1]{#1}%
\providecommand \citenamefont [1]{#1}%
\providecommand \href@noop [0]{\@secondoftwo}%
\providecommand \href [0]{\begingroup \@sanitize@url \@href}%
\providecommand \@href[1]{\@@startlink{#1}\@@href}%
\providecommand \@@href[1]{\endgroup#1\@@endlink}%
\providecommand \@sanitize@url [0]{\catcode `\\12\catcode `\$12\catcode
  `\&12\catcode `\#12\catcode `\^12\catcode `\_12\catcode `\%12\relax}%
\providecommand \@@startlink[1]{}%
\providecommand \@@endlink[0]{}%
\providecommand \url  [0]{\begingroup\@sanitize@url \@url }%
\providecommand \@url [1]{\endgroup\@href {#1}{\urlprefix }}%
\providecommand \urlprefix  [0]{URL }%
\providecommand \Eprint [0]{\href }%
\providecommand \doibase [0]{https://doi.org/}%
\providecommand \selectlanguage [0]{\@gobble}%
\providecommand \bibinfo  [0]{\@secondoftwo}%
\providecommand \bibfield  [0]{\@secondoftwo}%
\providecommand \translation [1]{[#1]}%
\providecommand \BibitemOpen [0]{}%
\providecommand \bibitemStop [0]{}%
\providecommand \bibitemNoStop [0]{.\EOS\space}%
\providecommand \EOS [0]{\spacefactor3000\relax}%
\providecommand \BibitemShut  [1]{\csname bibitem#1\endcsname}%
\let\auto@bib@innerbib\@empty
\bibitem [{\citenamefont {Hepting}\ \emph {et~al.}(2018)\citenamefont
  {Hepting}, \citenamefont {Chaix}, \citenamefont {Huang}, \citenamefont
  {Fumagalli}, \citenamefont {Peng}, \citenamefont {Moritz}, \citenamefont
  {Kummer}, \citenamefont {Brookes}, \citenamefont {Lee}, \citenamefont
  {Hashimoto}, \citenamefont {Sarkar}, \citenamefont {He}, \citenamefont
  {Rotundu}, \citenamefont {Lee}, \citenamefont {Greene}, \citenamefont
  {Braicovich}, \citenamefont {Ghiringhelli}, \citenamefont {Shen},
  \citenamefont {Devereaux},\ and\ \citenamefont {Lee}}]{hepting18a}%
  \BibitemOpen
  \bibfield  {author} {\bibinfo {author} {\bibfnamefont {M.}~\bibnamefont
  {Hepting}}, \bibinfo {author} {\bibfnamefont {L.}~\bibnamefont {Chaix}},
  \bibinfo {author} {\bibfnamefont {E.~W.}\ \bibnamefont {Huang}}, \bibinfo
  {author} {\bibfnamefont {R.}~\bibnamefont {Fumagalli}}, \bibinfo {author}
  {\bibfnamefont {Y.~Y.}\ \bibnamefont {Peng}}, \bibinfo {author}
  {\bibfnamefont {B.}~\bibnamefont {Moritz}}, \bibinfo {author} {\bibfnamefont
  {K.}~\bibnamefont {Kummer}}, \bibinfo {author} {\bibfnamefont {N.~B.}\
  \bibnamefont {Brookes}}, \bibinfo {author} {\bibfnamefont {W.~C.}\
  \bibnamefont {Lee}}, \bibinfo {author} {\bibfnamefont {M.}~\bibnamefont
  {Hashimoto}}, \bibinfo {author} {\bibfnamefont {T.}~\bibnamefont {Sarkar}},
  \bibinfo {author} {\bibfnamefont {J.-F.}\ \bibnamefont {He}}, \bibinfo
  {author} {\bibfnamefont {C.~R.}\ \bibnamefont {Rotundu}}, \bibinfo {author}
  {\bibfnamefont {Y.~S.}\ \bibnamefont {Lee}}, \bibinfo {author} {\bibfnamefont
  {R.~L.}\ \bibnamefont {Greene}}, \bibinfo {author} {\bibfnamefont
  {L.}~\bibnamefont {Braicovich}}, \bibinfo {author} {\bibfnamefont
  {G.}~\bibnamefont {Ghiringhelli}}, \bibinfo {author} {\bibfnamefont {Z.~X.}\
  \bibnamefont {Shen}}, \bibinfo {author} {\bibfnamefont {T.~P.}\ \bibnamefont
  {Devereaux}},\ and\ \bibinfo {author} {\bibfnamefont {W.~S.}\ \bibnamefont
  {Lee}},\ }\bibfield  {title} {\bibinfo {title} {Three-dimensional collective
  charge excitations in electron-doped copper oxide superconductors},\ }\href
  {https://doi.org/10.1038/s41586-018-0648-3} {\bibfield  {journal} {\bibinfo
  {journal} {Nature}\ }\textbf {\bibinfo {volume} {563}},\ \bibinfo {pages}
  {374} (\bibinfo {year} {2018})}\BibitemShut {NoStop}%
\bibitem [{\citenamefont {Nag}\ \emph {et~al.}(2020)\citenamefont {Nag},
  \citenamefont {Zhu}, \citenamefont {Bejas}, \citenamefont {Li}, \citenamefont
  {Robarts}, \citenamefont {Yamase}, \citenamefont {Petsch}, \citenamefont
  {Song}, \citenamefont {Eisaki}, \citenamefont {Walters}, \citenamefont
  {Garc\'{\i}a-Fern\'andez}, \citenamefont {Greco}, \citenamefont {Hayden},\
  and\ \citenamefont {Zhou}}]{nag20a}%
  \BibitemOpen
  \bibfield  {author} {\bibinfo {author} {\bibfnamefont {A.}~\bibnamefont
  {Nag}}, \bibinfo {author} {\bibfnamefont {M.}~\bibnamefont {Zhu}}, \bibinfo
  {author} {\bibfnamefont {M.}~\bibnamefont {Bejas}}, \bibinfo {author}
  {\bibfnamefont {J.}~\bibnamefont {Li}}, \bibinfo {author} {\bibfnamefont
  {H.~C.}\ \bibnamefont {Robarts}}, \bibinfo {author} {\bibfnamefont
  {H.}~\bibnamefont {Yamase}}, \bibinfo {author} {\bibfnamefont {A.~N.}\
  \bibnamefont {Petsch}}, \bibinfo {author} {\bibfnamefont {D.}~\bibnamefont
  {Song}}, \bibinfo {author} {\bibfnamefont {H.}~\bibnamefont {Eisaki}},
  \bibinfo {author} {\bibfnamefont {A.~C.}\ \bibnamefont {Walters}}, \bibinfo
  {author} {\bibfnamefont {M.}~\bibnamefont {Garc\'{\i}a-Fern\'andez}},
  \bibinfo {author} {\bibfnamefont {A.}~\bibnamefont {Greco}}, \bibinfo
  {author} {\bibfnamefont {S.~M.}\ \bibnamefont {Hayden}},\ and\ \bibinfo
  {author} {\bibfnamefont {K.-J.}\ \bibnamefont {Zhou}},\ }\bibfield  {title}
  {\bibinfo {title} {Detection of acoustic plasmons in hole-doped lanthanum and
  bismuth cuprate superconductors using resonant inelastic x-ray scattering},\
  }\href {https://link.aps.org/doi/10.1103/PhysRevLett.125.257002} {\bibfield
  {journal} {\bibinfo  {journal} {Phys. Rev. Lett.}\ }\textbf {\bibinfo
  {volume} {125}},\ \bibinfo {pages} {257002} (\bibinfo {year}
  {2020})}\BibitemShut {NoStop}%
\bibitem [{\citenamefont {Fumagalli}\ \emph {et~al.}(2019)\citenamefont
  {Fumagalli}, \citenamefont {Braicovich}, \citenamefont {Minola},
  \citenamefont {Peng}, \citenamefont {Kummer}, \citenamefont {Betto},
  \citenamefont {Rossi}, \citenamefont {Lefran\ifmmode~\mbox{\c{c}}\else
  \c{c}\fi{}ois}, \citenamefont {Morawe}, \citenamefont {Salluzzo},
  \citenamefont {Suzuki}, \citenamefont {Yakhou}, \citenamefont {Le~Tacon},
  \citenamefont {Keimer}, \citenamefont {Brookes}, \citenamefont {Sala},\ and\
  \citenamefont {Ghiringhelli}}]{fumagalli19}%
  \BibitemOpen
  \bibfield  {author} {\bibinfo {author} {\bibfnamefont {R.}~\bibnamefont
  {Fumagalli}}, \bibinfo {author} {\bibfnamefont {L.}~\bibnamefont
  {Braicovich}}, \bibinfo {author} {\bibfnamefont {M.}~\bibnamefont {Minola}},
  \bibinfo {author} {\bibfnamefont {Y.~Y.}\ \bibnamefont {Peng}}, \bibinfo
  {author} {\bibfnamefont {K.}~\bibnamefont {Kummer}}, \bibinfo {author}
  {\bibfnamefont {D.}~\bibnamefont {Betto}}, \bibinfo {author} {\bibfnamefont
  {M.}~\bibnamefont {Rossi}}, \bibinfo {author} {\bibfnamefont
  {E.}~\bibnamefont {Lefran\ifmmode~\mbox{\c{c}}\else \c{c}\fi{}ois}}, \bibinfo
  {author} {\bibfnamefont {C.}~\bibnamefont {Morawe}}, \bibinfo {author}
  {\bibfnamefont {M.}~\bibnamefont {Salluzzo}}, \bibinfo {author}
  {\bibfnamefont {H.}~\bibnamefont {Suzuki}}, \bibinfo {author} {\bibfnamefont
  {F.}~\bibnamefont {Yakhou}}, \bibinfo {author} {\bibfnamefont
  {M.}~\bibnamefont {Le~Tacon}}, \bibinfo {author} {\bibfnamefont
  {B.}~\bibnamefont {Keimer}}, \bibinfo {author} {\bibfnamefont {N.~B.}\
  \bibnamefont {Brookes}}, \bibinfo {author} {\bibfnamefont {M.~M.}\
  \bibnamefont {Sala}},\ and\ \bibinfo {author} {\bibfnamefont
  {G.}~\bibnamefont {Ghiringhelli}},\ }\bibfield  {title} {\bibinfo {title}
  {Polarization-resolved Cu ${L}_{3}$-edge resonant inelastic x-ray scattering
  of orbital and spin excitations in
  NdBa$_{2}$Cu$_{3}$O$_{7\ensuremath{-}\ensuremath{\delta}}$},\
  }\href {https://doi.org/10.1103/PhysRevB.99.134517} {\bibfield  {journal}
  {\bibinfo  {journal} {Phys. Rev. B}\ }\textbf {\bibinfo {volume} {99}},\
  \bibinfo {pages} {134517} (\bibinfo {year} {2019})}\BibitemShut {NoStop}%
\bibitem [{\citenamefont {Minola}\ \emph {et~al.}(2015)\citenamefont {Minola},
  \citenamefont {Dellea}, \citenamefont {Gretarsson}, \citenamefont {Peng},
  \citenamefont {Lu}, \citenamefont {Porras}, \citenamefont {Loew},
  \citenamefont {Yakhou}, \citenamefont {Brookes}, \citenamefont {Huang},
  \citenamefont {Pelliciari}, \citenamefont {Schmitt}, \citenamefont
  {Ghiringhelli}, \citenamefont {Keimer}, \citenamefont {Braicovich},\ and\
  \citenamefont {Le~Tacon}}]{minola15}%
  \BibitemOpen
  \bibfield  {author} {\bibinfo {author} {\bibfnamefont {M.}~\bibnamefont
  {Minola}}, \bibinfo {author} {\bibfnamefont {G.}~\bibnamefont {Dellea}},
  \bibinfo {author} {\bibfnamefont {H.}~\bibnamefont {Gretarsson}}, \bibinfo
  {author} {\bibfnamefont {Y.~Y.}\ \bibnamefont {Peng}}, \bibinfo {author}
  {\bibfnamefont {Y.}~\bibnamefont {Lu}}, \bibinfo {author} {\bibfnamefont
  {J.}~\bibnamefont {Porras}}, \bibinfo {author} {\bibfnamefont
  {T.}~\bibnamefont {Loew}}, \bibinfo {author} {\bibfnamefont {F.}~\bibnamefont
  {Yakhou}}, \bibinfo {author} {\bibfnamefont {N.~B.}\ \bibnamefont {Brookes}},
  \bibinfo {author} {\bibfnamefont {Y.~B.}\ \bibnamefont {Huang}}, \bibinfo
  {author} {\bibfnamefont {J.}~\bibnamefont {Pelliciari}}, \bibinfo {author}
  {\bibfnamefont {T.}~\bibnamefont {Schmitt}}, \bibinfo {author} {\bibfnamefont
  {G.}~\bibnamefont {Ghiringhelli}}, \bibinfo {author} {\bibfnamefont
  {B.}~\bibnamefont {Keimer}}, \bibinfo {author} {\bibfnamefont
  {L.}~\bibnamefont {Braicovich}},\ and\ \bibinfo {author} {\bibfnamefont
  {M.}~\bibnamefont {Le~Tacon}},\ }\bibfield  {title} {\bibinfo {title}
  {Collective nature of spin excitations in superconducting cuprates probed by
  resonant inelastic x-ray scattering},\ }\href
  {https://doi.org/10.1103/PhysRevLett.114.217003} {\bibfield  {journal}
  {\bibinfo  {journal} {Phys. Rev. Lett.}\ }\textbf {\bibinfo {volume} {114}},\
  \bibinfo {pages} {217003} (\bibinfo {year} {2015})}\BibitemShut {NoStop}%
\bibitem [{\citenamefont {Dellea}\ \emph {et~al.}(2017)\citenamefont {Dellea},
  \citenamefont {Minola}, \citenamefont {Galdi}, \citenamefont {Di~Castro},
  \citenamefont {Aruta}, \citenamefont {Brookes}, \citenamefont {Jia},
  \citenamefont {Mazzoli}, \citenamefont {Moretti~Sala}, \citenamefont
  {Moritz}, \citenamefont {Orgiani}, \citenamefont {Schlom}, \citenamefont
  {Tebano}, \citenamefont {Balestrino}, \citenamefont {Braicovich},
  \citenamefont {Devereaux}, \citenamefont {Maritato},\ and\ \citenamefont
  {Ghiringhelli}}]{dellea17a}%
  \BibitemOpen
  \bibfield  {author} {\bibinfo {author} {\bibfnamefont {G.}~\bibnamefont
  {Dellea}}, \bibinfo {author} {\bibfnamefont {M.}~\bibnamefont {Minola}},
  \bibinfo {author} {\bibfnamefont {A.}~\bibnamefont {Galdi}}, \bibinfo
  {author} {\bibfnamefont {D.}~\bibnamefont {Di~Castro}}, \bibinfo {author}
  {\bibfnamefont {C.}~\bibnamefont {Aruta}}, \bibinfo {author} {\bibfnamefont
  {N.~B.}\ \bibnamefont {Brookes}}, \bibinfo {author} {\bibfnamefont {C.~J.}\
  \bibnamefont {Jia}}, \bibinfo {author} {\bibfnamefont {C.}~\bibnamefont
  {Mazzoli}}, \bibinfo {author} {\bibfnamefont {M.}~\bibnamefont
  {Moretti~Sala}}, \bibinfo {author} {\bibfnamefont {B.}~\bibnamefont
  {Moritz}}, \bibinfo {author} {\bibfnamefont {P.}~\bibnamefont {Orgiani}},
  \bibinfo {author} {\bibfnamefont {D.~G.}\ \bibnamefont {Schlom}}, \bibinfo
  {author} {\bibfnamefont {A.}~\bibnamefont {Tebano}}, \bibinfo {author}
  {\bibfnamefont {G.}~\bibnamefont {Balestrino}}, \bibinfo {author}
  {\bibfnamefont {L.}~\bibnamefont {Braicovich}}, \bibinfo {author}
  {\bibfnamefont {T.~P.}\ \bibnamefont {Devereaux}}, \bibinfo {author}
  {\bibfnamefont {L.}~\bibnamefont {Maritato}},\ and\ \bibinfo {author}
  {\bibfnamefont {G.}~\bibnamefont {Ghiringhelli}},\ }\bibfield  {title}
  {\bibinfo {title} {Spin and charge excitations in artificial hole- and
  electron-doped infinite layer cuprate superconductors},\ }\href
  {https://doi.org/10.1103/PhysRevB.96.115117} {\bibfield  {journal} {\bibinfo
  {journal} {Phys. Rev. B}\ }\textbf {\bibinfo {volume} {96}},\ \bibinfo
  {pages} {115117} (\bibinfo {year} {2017})}\BibitemShut {NoStop}%
\bibitem [{\citenamefont {Thio}\ \emph {et~al.}(1988)\citenamefont {Thio},
  \citenamefont {Thurston}, \citenamefont {Preyer}, \citenamefont {Picone},
  \citenamefont {Kastner}, \citenamefont {Jenssen}, \citenamefont {Gabbe},
  \citenamefont {Chen}, \citenamefont {Birgeneau},\ and\ \citenamefont
  {Aharony}}]{thio88}%
  \BibitemOpen
  \bibfield  {author} {\bibinfo {author} {\bibfnamefont {T.}~\bibnamefont
  {Thio}}, \bibinfo {author} {\bibfnamefont {T.~R.}\ \bibnamefont {Thurston}},
  \bibinfo {author} {\bibfnamefont {N.~W.}\ \bibnamefont {Preyer}}, \bibinfo
  {author} {\bibfnamefont {P.~J.}\ \bibnamefont {Picone}}, \bibinfo {author}
  {\bibfnamefont {M.~A.}\ \bibnamefont {Kastner}}, \bibinfo {author}
  {\bibfnamefont {H.~P.}\ \bibnamefont {Jenssen}}, \bibinfo {author}
  {\bibfnamefont {D.~R.}\ \bibnamefont {Gabbe}}, \bibinfo {author}
  {\bibfnamefont {C.~Y.}\ \bibnamefont {Chen}}, \bibinfo {author}
  {\bibfnamefont {R.~J.}\ \bibnamefont {Birgeneau}},\ and\ \bibinfo {author}
  {\bibfnamefont {A.}~\bibnamefont {Aharony}},\ }\bibfield  {title} {\bibinfo
  {title} {{Antisymmetric exchange and its influence on the magnetic structure
  and conductivity of ${\mathrm{La}}_{2}$Cu${\mathrm{O}}_{4}$}},\ }\href
  {https://doi.org/10.1103/PhysRevB.38.905} {\bibfield  {journal} {\bibinfo
  {journal} {Phys. Rev. B}\ }\textbf {\bibinfo {volume} {38}},\ \bibinfo
  {pages} {905} (\bibinfo {year} {1988})}\BibitemShut {NoStop}%
\bibitem [{\citenamefont {Foussats}\ and\ \citenamefont
  {Greco}(2004)}]{foussats04}%
  \BibitemOpen
  \bibfield  {author} {\bibinfo {author} {\bibfnamefont {A.}~\bibnamefont
  {Foussats}}\ and\ \bibinfo {author} {\bibfnamefont {A.}~\bibnamefont
  {Greco}},\ }\bibfield  {title} {\bibinfo {title} {{Large-$N$ expansion based
  on the Hubbard operator path integral representation and its application to
  the $t\text{\ensuremath{-}}J$ model. II. The case for finite $J$}},\ }\href
  {https://link.aps.org/doi/10.1103/PhysRevB.70.205123} {\bibfield  {journal}
  {\bibinfo  {journal} {Phys. Rev. B}\ }\textbf {\bibinfo {volume} {70}},\
  \bibinfo {pages} {205123} (\bibinfo {year} {2004})}\BibitemShut {NoStop}%
\bibitem [{\citenamefont {Andersen}\ \emph {et~al.}(1995)\citenamefont
  {Andersen}, \citenamefont {Liechtenstein}, \citenamefont {Jepsen},\ and\
  \citenamefont {Paulsen}}]{andersen95a}%
  \BibitemOpen
  \bibfield  {author} {\bibinfo {author} {\bibfnamefont {O.}~\bibnamefont
  {Andersen}}, \bibinfo {author} {\bibfnamefont {A.}~\bibnamefont
  {Liechtenstein}}, \bibinfo {author} {\bibfnamefont {O.}~\bibnamefont
  {Jepsen}},\ and\ \bibinfo {author} {\bibfnamefont {F.}~\bibnamefont
  {Paulsen}},\ }\bibfield  {title} {\bibinfo {title} {Lda energy bands,
  low-energy hamiltonians, t', t'', t$_\perp$(k), and J$_\perp$},\ }\href
  {https://www.sciencedirect.com/science/article/pii/0022369795002693}
  {\bibfield  {journal} {\bibinfo  {journal} {J. Phys. Chem. Solids}\ }\textbf
  {\bibinfo {volume} {56}},\ \bibinfo {pages} {1573} (\bibinfo {year}
  {1995})}\BibitemShut {NoStop}%
\bibitem [{\citenamefont {Botana}\ and\ \citenamefont
  {Norman}(2020)}]{botana20a}%
  \BibitemOpen
  \bibfield  {author} {\bibinfo {author} {\bibfnamefont {A.~S.}\ \bibnamefont
  {Botana}}\ and\ \bibinfo {author} {\bibfnamefont {M.~R.}\ \bibnamefont
  {Norman}},\ }\bibfield  {title} {\bibinfo {title} {Similarities and
  differences between LaNiO$_{2}$ and CaCuO$_{2}$ and
  implications for superconductivity},\ }\href
  {https://doi.org/10.1103/PhysRevX.10.011024} {\bibfield  {journal} {\bibinfo
  {journal} {Phys. Rev. X}\ }\textbf {\bibinfo {volume} {10}},\ \bibinfo
  {pages} {011024} (\bibinfo {year} {2020})}\BibitemShut {NoStop}%
\bibitem [{\citenamefont {Becca}\ \emph {et~al.}(1996)\citenamefont {Becca},
  \citenamefont {Tarquini}, \citenamefont {Grilli},\ and\ \citenamefont
  {Di~Castro}}]{becca96}%
  \BibitemOpen
  \bibfield  {author} {\bibinfo {author} {\bibfnamefont {F.}~\bibnamefont
  {Becca}}, \bibinfo {author} {\bibfnamefont {M.}~\bibnamefont {Tarquini}},
  \bibinfo {author} {\bibfnamefont {M.}~\bibnamefont {Grilli}},\ and\ \bibinfo
  {author} {\bibfnamefont {C.}~\bibnamefont {Di~Castro}},\ }\bibfield  {title}
  {\bibinfo {title} {{Charge-density waves and superconductivity as an
  alternative to phase separation in the infinite-$U$ Hubbard-Holstein
  model}},\ }\href {https://doi.org/10.1103/PhysRevB.54.12443} {\bibfield
  {journal} {\bibinfo  {journal} {Phys. Rev. B}\ }\textbf {\bibinfo {volume}
  {54}},\ \bibinfo {pages} {12443} (\bibinfo {year} {1996})}\BibitemShut
  {NoStop}%
\bibitem [{\citenamefont {Greco}\ \emph {et~al.}(2019)\citenamefont {Greco},
  \citenamefont {Yamase},\ and\ \citenamefont {Bejas}}]{greco19a}%
  \BibitemOpen
  \bibfield  {author} {\bibinfo {author} {\bibfnamefont {A.}~\bibnamefont
  {Greco}}, \bibinfo {author} {\bibfnamefont {H.}~\bibnamefont {Yamase}},\ and\
  \bibinfo {author} {\bibfnamefont {M.}~\bibnamefont {Bejas}},\ }\bibfield
  {title} {\bibinfo {title} {Origin of high-energy charge excitations observed
  by resonant inelastic x-ray scattering in cuprate superconductors},\ }\href
  {https://doi.org/10.1038/s42005-018-0099-z} {\bibfield  {journal} {\bibinfo
  {journal} {Commun. Phys.}\ }\textbf {\bibinfo {volume} {2}},\ \bibinfo
  {pages} {3} (\bibinfo {year} {2019})}\BibitemShut {NoStop}%
\bibitem [{\citenamefont {Greco}\ \emph {et~al.}(2020)\citenamefont {Greco},
  \citenamefont {Yamase},\ and\ \citenamefont {Bejas}}]{greco20a}%
  \BibitemOpen
  \bibfield  {author} {\bibinfo {author} {\bibfnamefont {A.}~\bibnamefont
  {Greco}}, \bibinfo {author} {\bibfnamefont {H.}~\bibnamefont {Yamase}},\ and\
  \bibinfo {author} {\bibfnamefont {M.}~\bibnamefont {Bejas}},\ }\bibfield
  {title} {\bibinfo {title} {Close inspection of plasmon excitations in cuprate
  superconductors},\ }\href
  {https://link.aps.org/doi/10.1103/PhysRevB.102.024509} {\bibfield  {journal}
  {\bibinfo  {journal} {Phys. Rev. B}\ }\textbf {\bibinfo {volume} {102}},\
  \bibinfo {pages} {024509} (\bibinfo {year} {2020})}\BibitemShut {NoStop}%
\bibitem [{\citenamefont {Prelov\ifmmode~\check{s}\else \v{s}\fi{}ek}\ and\
  \citenamefont {Horsch}(1999)}]{prelovsek99a}%
  \BibitemOpen
  \bibfield  {author} {\bibinfo {author} {\bibfnamefont {P.}~\bibnamefont
  {Prelov\ifmmode~\check{s}\else \v{s}\fi{}ek}}\ and\ \bibinfo {author}
  {\bibfnamefont {P.}~\bibnamefont {Horsch}},\ }\bibfield  {title} {\bibinfo
  {title} {Electron-energy loss spectra and plasmon resonance in cuprates},\
  }\href {https://doi.org/10.1103/PhysRevB.60.R3735} {\bibfield  {journal}
  {\bibinfo  {journal} {Phys. Rev. B}\ }\textbf {\bibinfo {volume} {60}},\
  \bibinfo {pages} {R3735} (\bibinfo {year} {1999})}\BibitemShut {NoStop}%
\bibitem [{\citenamefont {Bejas}\ \emph {et~al.}(2017)\citenamefont {Bejas},
  \citenamefont {Yamase},\ and\ \citenamefont {Greco}}]{bejas17}%
  \BibitemOpen
  \bibfield  {author} {\bibinfo {author} {\bibfnamefont {M.}~\bibnamefont
  {Bejas}}, \bibinfo {author} {\bibfnamefont {H.}~\bibnamefont {Yamase}},\ and\
  \bibinfo {author} {\bibfnamefont {A.}~\bibnamefont {Greco}},\ }\bibfield
  {title} {\bibinfo {title} {Dual structure in the charge excitation spectrum
  of electron-doped cuprates},\ }\href
  {https://doi.org/10.1103/PhysRevB.96.214513} {\bibfield  {journal} {\bibinfo
  {journal} {Phys. Rev. B}\ }\textbf {\bibinfo {volume} {96}},\ \bibinfo
  {pages} {214513} (\bibinfo {year} {2017})}\BibitemShut {NoStop}%
\bibitem [{\citenamefont {Klenner}\ \emph {et~al.}(1994)\citenamefont
  {Klenner}, \citenamefont {Falter},\ and\ \citenamefont {Chen}}]{klenner94a}%
  \BibitemOpen
  \bibfield  {author} {\bibinfo {author} {\bibfnamefont {M.}~\bibnamefont
  {Klenner}}, \bibinfo {author} {\bibfnamefont {C.}~\bibnamefont {Falter}},\
  and\ \bibinfo {author} {\bibfnamefont {Q.}~\bibnamefont {Chen}},\ }\bibfield
  {title} {\bibinfo {title} {Calculated phonon dispersion of infinite-layer
  compounds and the effects of charge fluctuations},\ }\href
  {https://doi.org/10.1007/BF01313348} {\bibfield  {journal} {\bibinfo
  {journal} {Z. Phys. B}\ }\textbf {\bibinfo {volume} {95}},\ \bibinfo {pages}
  {417} (\bibinfo {year} {1994})}\BibitemShut {NoStop}%
\bibitem [{\citenamefont {Hybertsen}\ \emph {et~al.}(1990)\citenamefont
  {Hybertsen}, \citenamefont {Stechel}, \citenamefont {Schluter},\ and\
  \citenamefont {Jennison}}]{hybertsen90}%
  \BibitemOpen
  \bibfield  {author} {\bibinfo {author} {\bibfnamefont {M.~S.}\ \bibnamefont
  {Hybertsen}}, \bibinfo {author} {\bibfnamefont {E.~B.}\ \bibnamefont
  {Stechel}}, \bibinfo {author} {\bibfnamefont {M.}~\bibnamefont {Schluter}},\
  and\ \bibinfo {author} {\bibfnamefont {D.~R.}\ \bibnamefont {Jennison}},\
  }\bibfield  {title} {\bibinfo {title} {{Renormalization from
  density-functional theory to strong-coupling models for electronic states in
  Cu-O materials}},\ }\href
  {https://link.aps.org/doi/10.1103/PhysRevB.41.11068} {\bibfield  {journal}
  {\bibinfo  {journal} {Phys. Rev. B}\ }\textbf {\bibinfo {volume} {41}},\
  \bibinfo {pages} {11068} (\bibinfo {year} {1990})}\BibitemShut {NoStop}%
\bibitem [{\citenamefont {Grecu}(1973)}]{grecu73a}%
  \BibitemOpen
  \bibfield  {author} {\bibinfo {author} {\bibfnamefont {D.}~\bibnamefont
  {Grecu}},\ }\bibfield  {title} {\bibinfo {title} {Plasma frequency of the
  electron gas in layered structures},\ }\href
  {https://link.aps.org/doi/10.1103/PhysRevB.8.1958} {\bibfield  {journal}
  {\bibinfo  {journal} {Phys. Rev. B}\ }\textbf {\bibinfo {volume} {8}},\
  \bibinfo {pages} {1958} (\bibinfo {year} {1973})}\BibitemShut {NoStop}%
\bibitem [{\citenamefont {Grecu}(1975)}]{grecu75a}%
  \BibitemOpen
  \bibfield  {author} {\bibinfo {author} {\bibfnamefont {D.}~\bibnamefont
  {Grecu}},\ }\bibfield  {title} {\bibinfo {title} {Self-consistent field
  approximation for the plasma frequencies of an electron gas in a layered thin
  film},\ }\href {https://doi.org/10.1088/0022-3719/8/16/014} {\bibfield
  {journal} {\bibinfo  {journal} {J. Phys. C Solid State}\ }\textbf {\bibinfo
  {volume} {8}},\ \bibinfo {pages} {2627} (\bibinfo {year} {1975})}\BibitemShut
  {NoStop}%
\bibitem [{\citenamefont {Greco}\ \emph {et~al.}(2016)\citenamefont {Greco},
  \citenamefont {Yamase},\ and\ \citenamefont {Bejas}}]{greco16a}%
  \BibitemOpen
  \bibfield  {author} {\bibinfo {author} {\bibfnamefont {A.}~\bibnamefont
  {Greco}}, \bibinfo {author} {\bibfnamefont {H.}~\bibnamefont {Yamase}},\ and\
  \bibinfo {author} {\bibfnamefont {M.}~\bibnamefont {Bejas}},\ }\bibfield
  {title} {\bibinfo {title} {{Plasmon excitations in layered high-${T}_{c}$
  cuprates}},\ }\href {https://doi.org/10.1103/PhysRevB.94.075139} {\bibfield
  {journal} {\bibinfo  {journal} {Phys. Rev. B}\ }\textbf {\bibinfo {volume}
  {94}},\ \bibinfo {pages} {075139} (\bibinfo {year} {2016})}\BibitemShut
  {NoStop}%
\bibitem [{\citenamefont {Singley}\ \emph {et~al.}(2001)\citenamefont
  {Singley}, \citenamefont {Basov}, \citenamefont {Kurahashi}, \citenamefont
  {Uefuji},\ and\ \citenamefont {Yamada}}]{singley01}%
  \BibitemOpen
  \bibfield  {author} {\bibinfo {author} {\bibfnamefont {E.~J.}\ \bibnamefont
  {Singley}}, \bibinfo {author} {\bibfnamefont {D.~N.}\ \bibnamefont {Basov}},
  \bibinfo {author} {\bibfnamefont {K.}~\bibnamefont {Kurahashi}}, \bibinfo
  {author} {\bibfnamefont {T.}~\bibnamefont {Uefuji}},\ and\ \bibinfo {author}
  {\bibfnamefont {K.}~\bibnamefont {Yamada}},\ }\bibfield  {title} {\bibinfo
  {title} {Electron dynamics in
  ${\mathrm{Nd}}_{1.85}{\mathrm{Ce}}_{0.15}{\mathrm{CuO}}_{4+\ensuremath{\delta}}:$
  evidence for the pseudogap state and unconventional c-axis response},\ }\href
  {https://doi.org/10.1103/PhysRevB.64.224503} {\bibfield  {journal} {\bibinfo
  {journal} {Phys. Rev. B}\ }\textbf {\bibinfo {volume} {64}},\ \bibinfo
  {pages} {224503} (\bibinfo {year} {2001})}\BibitemShut {NoStop}%
\bibitem [{\citenamefont {Pavarini}\ \emph {et~al.}(2001)\citenamefont
  {Pavarini}, \citenamefont {Dasgupta}, \citenamefont {Saha-Dasgupta},
  \citenamefont {Jepsen},\ and\ \citenamefont {Andersen}}]{pavarini01a}%
  \BibitemOpen
  \bibfield  {author} {\bibinfo {author} {\bibfnamefont {E.}~\bibnamefont
  {Pavarini}}, \bibinfo {author} {\bibfnamefont {I.}~\bibnamefont {Dasgupta}},
  \bibinfo {author} {\bibfnamefont {T.}~\bibnamefont {Saha-Dasgupta}}, \bibinfo
  {author} {\bibfnamefont {O.}~\bibnamefont {Jepsen}},\ and\ \bibinfo {author}
  {\bibfnamefont {O.~K.}\ \bibnamefont {Andersen}},\ }\bibfield  {title}
  {\bibinfo {title} {Band-structure trend in hole-doped cuprates and
  correlation with ${\mathit{T}}_{\mathit{c}\mathrm{max}}$},\ }\href
  {https://link.aps.org/doi/10.1103/PhysRevLett.87.047003} {\bibfield
  {journal} {\bibinfo  {journal} {Phys. Rev. Lett.}\ }\textbf {\bibinfo
  {volume} {87}},\ \bibinfo {pages} {047003} (\bibinfo {year}
  {2001})}\BibitemShut {NoStop}%
\bibitem [{\citenamefont {Suzuki}(1989)}]{suzuki89}%
  \BibitemOpen
  \bibfield  {author} {\bibinfo {author} {\bibfnamefont {M.}~\bibnamefont
  {Suzuki}},\ }\bibfield  {title} {\bibinfo {title} {Hall coefficients and
  optical properties of
  ${\mathrm{La}}_{2\ensuremath{-}x}{\mathrm{Sr}}_{x}\mathrm{Cu}{\mathrm{O}}_{4}$
  single-crystal thin films},\ }\href
  {https://doi.org/10.1103/PhysRevB.39.2312} {\bibfield  {journal} {\bibinfo
  {journal} {Phys. Rev. B}\ }\textbf {\bibinfo {volume} {39}},\ \bibinfo
  {pages} {2312} (\bibinfo {year} {1989})}\BibitemShut {NoStop}%
\bibitem [{\citenamefont {Uchida}\ \emph {et~al.}(1991)\citenamefont {Uchida},
  \citenamefont {Ido}, \citenamefont {Takagi}, \citenamefont {Arima},
  \citenamefont {Tokura},\ and\ \citenamefont {Tajima}}]{uchida91}%
  \BibitemOpen
  \bibfield  {author} {\bibinfo {author} {\bibfnamefont {S.}~\bibnamefont
  {Uchida}}, \bibinfo {author} {\bibfnamefont {T.}~\bibnamefont {Ido}},
  \bibinfo {author} {\bibfnamefont {H.}~\bibnamefont {Takagi}}, \bibinfo
  {author} {\bibfnamefont {T.}~\bibnamefont {Arima}}, \bibinfo {author}
  {\bibfnamefont {Y.}~\bibnamefont {Tokura}},\ and\ \bibinfo {author}
  {\bibfnamefont {S.}~\bibnamefont {Tajima}},\ }\bibfield  {title} {\bibinfo
  {title} {Optical spectra of
  ${\mathrm{La}}_{2\mathrm{\ensuremath{-}}\mathit{x}}$${\mathrm{Sr}}_{\mathit{x}}$${\mathrm{CuO}}_{4}$:
  Effect of carrier doping on the electronic structure of the
  ${\mathrm{CuO}}_{2}$ plane},\ }\href
  {https://link.aps.org/doi/10.1103/PhysRevB.43.7942} {\bibfield  {journal}
  {\bibinfo  {journal} {Phys. Rev. B}\ }\textbf {\bibinfo {volume} {43}},\
  \bibinfo {pages} {7942} (\bibinfo {year} {1991})}\BibitemShut {NoStop}%
\bibitem [{\citenamefont {Bauer}\ and\ \citenamefont {Falter}(2009)}]{bauer09a}%
  \BibitemOpen
  \bibfield  {author} {\bibinfo {author} {\bibfnamefont {T.}~\bibnamefont
  {Bauer}}\ and\ \bibinfo {author} {\bibfnamefont {C.}~\bibnamefont {Falter}},\
  }\bibfield  {title} {\bibinfo {title} {Impact of dynamical screening on the
  phonon dynamics of metallic ${\text{La}}_{2}{\text{CuO}}_{4}$},\ }\href
  {https://doi.org/10.1103/PhysRevB.80.094525} {\bibfield  {journal} {\bibinfo
  {journal} {Phys. Rev. B}\ }\textbf {\bibinfo {volume} {80}},\ \bibinfo
  {pages} {094525} (\bibinfo {year} {2009})}\BibitemShut {NoStop}%
\bibitem [{\citenamefont {Falter}\ and\ \citenamefont
  {Klenner}(1994)}]{falter94a}%
  \BibitemOpen
  \bibfield  {author} {\bibinfo {author} {\bibfnamefont {C.}~\bibnamefont
  {Falter}}\ and\ \bibinfo {author} {\bibfnamefont {M.}~\bibnamefont
  {Klenner}},\ }\bibfield  {title} {\bibinfo {title} {Nonadiabatic and nonlocal
  electron-phonon interaction and phonon-plasmon mixing in the high-temperature
  superconductors},\ }\href {https://doi.org/10.1103/PhysRevB.50.9426}
  {\bibfield  {journal} {\bibinfo  {journal} {Phys. Rev. B}\ }\textbf {\bibinfo
  {volume} {50}},\ \bibinfo {pages} {9426} (\bibinfo {year}
  {1994})}\BibitemShut {NoStop}%
\bibitem [{\citenamefont {Falter}\ \emph {et~al.}(1998)\citenamefont {Falter},
  \citenamefont {Klenner},\ and\ \citenamefont {Hoffmann}}]{falter98}%
  \BibitemOpen
  \bibfield  {author} {\bibinfo {author} {\bibfnamefont {C.}~\bibnamefont
  {Falter}}, \bibinfo {author} {\bibfnamefont {M.}~\bibnamefont {Klenner}},\
  and\ \bibinfo {author} {\bibfnamefont {G.}~\bibnamefont {Hoffmann}},\
  }\bibfield  {title} {\bibinfo {title} {{Screening and Phonon-Plasmon Scenario
  as Calculated from a Realistic Electronic Bandstructure Based on LDA for
  La$_2$CuO$_4$}},\ }\href
  {https://doi.org/10.1002/(SICI)1521-3951(199810)209:2<235::AID-PSSB235>3.0.CO;2-X}
  {\bibfield  {journal} {\bibinfo  {journal} {Phys. Status Solidi B}\ }\textbf
  {\bibinfo {volume} {209}},\ \bibinfo {pages} {235} (\bibinfo {year}
  {1998})}\BibitemShut {NoStop}%
\bibitem [{\citenamefont {Pintschovius}(2005)}]{pintschovius05}%
  \BibitemOpen
  \bibfield  {author} {\bibinfo {author} {\bibfnamefont {L.}~\bibnamefont
  {Pintschovius}},\ }\bibfield  {title} {\bibinfo {title} {{Electron-phonon
  coupling effects explored by inelastic neutron scattering}},\ }\href
  {https://doi.org/10.1002/pssb.200404951} {\bibfield  {journal} {\bibinfo
  {journal} {Phys. Status Solidi B}\ }\textbf {\bibinfo {volume} {242}},\
  \bibinfo {pages} {30} (\bibinfo {year} {2005})}\BibitemShut {NoStop}%
\bibitem [{\citenamefont {Brar}\ \emph {et~al.}(2014)\citenamefont {Brar},
  \citenamefont {Jang}, \citenamefont {Sherrott}, \citenamefont {Kim},
  \citenamefont {Lopez}, \citenamefont {Kim}, \citenamefont {Choi},\ and\
  \citenamefont {Atwater}}]{brar14}%
  \BibitemOpen
  \bibfield  {author} {\bibinfo {author} {\bibfnamefont {V.~W.}\ \bibnamefont
  {Brar}}, \bibinfo {author} {\bibfnamefont {M.~S.}\ \bibnamefont {Jang}},
  \bibinfo {author} {\bibfnamefont {M.}~\bibnamefont {Sherrott}}, \bibinfo
  {author} {\bibfnamefont {S.}~\bibnamefont {Kim}}, \bibinfo {author}
  {\bibfnamefont {J.~J.}\ \bibnamefont {Lopez}}, \bibinfo {author}
  {\bibfnamefont {L.~B.}\ \bibnamefont {Kim}}, \bibinfo {author} {\bibfnamefont
  {M.}~\bibnamefont {Choi}},\ and\ \bibinfo {author} {\bibfnamefont
  {H.}~\bibnamefont {Atwater}},\ }\bibfield  {title} {\bibinfo {title} {Hybrid
  surface-phonon-plasmon polariton modes in graphene/monolayer h-BN
  heterostructures},\ }\href {https://doi.org/10.1021/nl501096s} {\bibfield
  {journal} {\bibinfo  {journal} {Nano Lett.}\ }\textbf {\bibinfo {volume}
  {14}},\ \bibinfo {pages} {3876} (\bibinfo {year} {2014})}\BibitemShut
  {NoStop}%
\bibitem [{\citenamefont {Koch}\ \emph {et~al.}(2010)\citenamefont {Koch},
  \citenamefont {Seyller},\ and\ \citenamefont {Schaefer}}]{koch10}%
  \BibitemOpen
  \bibfield  {author} {\bibinfo {author} {\bibfnamefont {R.~J.}\ \bibnamefont
  {Koch}}, \bibinfo {author} {\bibfnamefont {T.}~\bibnamefont {Seyller}},\ and\
  \bibinfo {author} {\bibfnamefont {J.~A.}\ \bibnamefont {Schaefer}},\
  }\bibfield  {title} {\bibinfo {title} {Strong phonon-plasmon coupled modes in
  the graphene/silicon carbide heterosystem},\ }\href
  {https://link.aps.org/doi/10.1103/PhysRevB.82.201413} {\bibfield  {journal}
  {\bibinfo  {journal} {Phys. Rev. B}\ }\textbf {\bibinfo {volume} {82}},\
  \bibinfo {pages} {201413} (\bibinfo {year} {2010})}\BibitemShut {NoStop}%
\bibitem [{\citenamefont {Bezares}\ \emph {et~al.}(2017)\citenamefont
  {Bezares}, \citenamefont {Sanctis}, \citenamefont {Saavedra}, \citenamefont
  {Woessner}, \citenamefont {Alonso-Gonz\'alez}, \citenamefont {Amenabar},
  \citenamefont {Chen}, \citenamefont {Bointon}, \citenamefont {Dai},
  \citenamefont {Fogler}, \citenamefont {Basov}, \citenamefont {Hillenbrand},
  \citenamefont {Craciun}, \citenamefont {Garc\'{\i}a~de Abajo}, \citenamefont
  {Russo},\ and\ \citenamefont {Koppens}}]{bezares17}%
  \BibitemOpen
  \bibfield  {author} {\bibinfo {author} {\bibfnamefont {F.~J.}\ \bibnamefont
  {Bezares}}, \bibinfo {author} {\bibfnamefont {A.~D.}\ \bibnamefont
  {Sanctis}}, \bibinfo {author} {\bibfnamefont {J.~R.~M.}\ \bibnamefont
  {Saavedra}}, \bibinfo {author} {\bibfnamefont {A.}~\bibnamefont {Woessner}},
  \bibinfo {author} {\bibfnamefont {P.}~\bibnamefont {Alonso-Gonz\'alez}},
  \bibinfo {author} {\bibfnamefont {I.}~\bibnamefont {Amenabar}}, \bibinfo
  {author} {\bibfnamefont {J.}~\bibnamefont {Chen}}, \bibinfo {author}
  {\bibfnamefont {T.~H.}\ \bibnamefont {Bointon}}, \bibinfo {author}
  {\bibfnamefont {S.}~\bibnamefont {Dai}}, \bibinfo {author} {\bibfnamefont
  {M.~M.}\ \bibnamefont {Fogler}}, \bibinfo {author} {\bibfnamefont {D.~N.}\
  \bibnamefont {Basov}}, \bibinfo {author} {\bibfnamefont {R.}~\bibnamefont
  {Hillenbrand}}, \bibinfo {author} {\bibfnamefont {M.~F.}\ \bibnamefont
  {Craciun}}, \bibinfo {author} {\bibfnamefont {F.~J.}\ \bibnamefont
  {Garc\'{\i}a~de Abajo}}, \bibinfo {author} {\bibfnamefont {S.}~\bibnamefont
  {Russo}},\ and\ \bibinfo {author} {\bibfnamefont {F.~H.~L.}\ \bibnamefont
  {Koppens}},\ }\bibfield  {title} {\bibinfo {title} {Intrinsic plasmon-phonon
  interactions in highly doped graphene: A near-field imaging study},\ }\href
  {https://doi.org/10.1021/acs.nanolett.7b01603} {\bibfield  {journal}
  {\bibinfo  {journal} {Nano Lett.}\ }\textbf {\bibinfo {volume} {17}},\
  \bibinfo {pages} {5908} (\bibinfo {year} {2017})}\BibitemShut {NoStop}%
\end{thebibliography}
\end{document}